\documentclass[twocolumn,tighten]{aastex631}
\usepackage{tabto}
\usepackage{capt-of}
\usepackage{upgreek}
\usepackage{makecell}
\usepackage{comment}
\usepackage{amsmath}
\usepackage{mathrsfs}
\usepackage{colortbl}

\shorttitle{BigMAC Targets with the RFC}
\shortauthors{Schwartzman et al.}

\graphicspath{{./}{Figures/}}

\begin{document}
\title{A New Sample of Closely Separated Dual and Binary AGN Candidates Revealed with the Radio Fundamental Catalog}

\author[0000-0002-6454-861X]{Emma Schwartzman}
\affiliation{U.S. Naval Research Laboratory, 4555 Overlook Ave SW, Washington, DC 20375, USA}

\author[0000-0001-8640-8522]{Ryan W. Pfeifle}
\affiliation{U.S. Naval Observatory, 3450 Massachusetts Avenue NW, Washington, DC 20392, USA}

\author[0000-0001-6812-7938]{Tracy E. Clarke}
\affiliation{U.S. Naval Research Laboratory, 4555 Overlook Ave SW, Washington, DC 20375, USA}

\author[0000-0002-4902-8077]{Nathan J. Secrest}
\affiliation{U.S. Naval Observatory, 3450 Massachusetts Avenue NW, Washington, DC 20392, USA}

\author[0000-0003-2450-3246]{Henrique Schmitt}
\affiliation{U.S. Naval Research Laboratory, 4555 Overlook Ave SW, Washington, DC 20375, USA}

\author[0000-0003-2283-2185]{Barry Rothberg}
\affiliation{Department of Physics and Astronomy, George Mason University, 4400 University Drive, MSN 3F3, Fairfax, VA 22030, USA}
\affiliation{U.S. Naval Observatory, 3450 Massachusetts Avenue NW, Washington, DC 20392, USA}

\begin{abstract} 

The Big Multi-AGN Catalog (BigMAC) provides the first literature-complete compilation of confirmed and candidate dual, binary, and recoiling active galactic nuclei (AGN). The BigMAC provides a basis for the first population-wide radio study of multiple AGN (multi-AGN). The Radio Fundamental Catalog (RFC) is composed of four decades of Very Long Baseline Interferometry (VLBI) observations capable of resolving parsec-scale structure. All BigMAC candidate and confirmed multi-AGN were cross-matched to the RFC, highlighting a total of 241 BigMAC sources with available RFC imaging. We present the first detailed analysis of the ten systems exhibiting multiple compact components, most likely to be radio multi-AGN. Using multi-frequency radio and supporting multiwavelength information, we assess whether each system is more consistent with dual/binary AGN or intrinsic jet structure. The dual/binary nature of six systems is further supported by our radio analysis, while the four remaining systems are more plausibly explained by compact jet activity. Several candidate multi-AGN systems exhibit projected separations of only a few to tens of parsecs. In many cases, the parsec-scale radio separations are orders of magnitude smaller than the separations inferred from the observations that originally motivated their selection, emphasizing the importance of VLBI follow-up. Because all ten sources are members of the International Celestial Reference Frame (ICRF) and display significant intrinsic radio structure, they also represent potential sources of astrometric error. This work establishes the RFC as an archival resource for characterizing the closest SMBH pairs while improving future ICRF realizations.

\end{abstract}


\section{Introduction} \label{sec:intro_intro}

Galaxy evolution is widely understood to proceed hierarchically, as larger galaxies assemble through the mergers of smaller systems, alongside the continued accretion of gas and dark matter \cite[e.g.,][]{toomre1972, schweizer1996, rothberg2004}. Because the majority of massive galaxies harbor central supermassive black holes \cite[SMBHs;][]{kormendy1995}, with masses spanning roughly $10^6$ to $10^{10}~M_{\odot}$, galaxy mergers are expected to produce systems containing two SMBHs. As the merger progresses, the SMBHs lose orbital energy through dynamical friction and migrate toward the center of the host galaxy, ultimately forming a bound pair and eventually coalescing \cite[][]{barnes1992, hopkins2008, volonteri2016}. Observational evidence further suggests a connection between SMBH growth and host galaxy evolution, including the well-established correlations between SMBH mass and host galaxy properties \cite[][]{ferrassee2000, gebhardt2000, heckman2014}. Numerical simulations indicate that tidal and hydrodynamical torques generated during mergers can efficiently funnel gas toward the nuclear regions \cite[e.g.,][]{barnes1996, mihos1996, capelo2017, blumenthal2018}, enhancing accretion onto the SMBHs and potentially triggering active galactic nucleus \cite[AGN;][]{hopkins2008, hopkins2010, blecha2018} activity in one or both nuclei.

Galaxy mergers occur over timescales of hundreds of millions to billions of years \cite[][]{Callegari_2009, tremmel2018}, during which AGN pairs are thought to evolve through multiple stages. Systems in the earlier stages of interaction are commonly referred to as dual AGN, generally describing AGN pairs whose separations remain outside their mutual gravitational spheres of influence, typically at projected separations $\lesssim110$ kpc \cite[][]{ellison2011, liu2011, bigmacreal}. At later stages, the SMBHs become gravitationally bound as a binary AGN system, with separations shrinking to tens of parsecs or below \cite[][]{rodriguez2006, liu2018, pfeifle2023inprep}. The subsequent orbital evolution from parsec to sub-parsec scales remains an open theoretical problem, often referred to as the ``final parsec problem'' \cite[][]{burkespolaor2018,kelley2017,amaro2023}, with some recent theoretical solutions \citep[][]{alonso2024,koo2024}. Once the binary reaches sufficiently small separations ($\ll1$ pc), gravitational-wave emission becomes the dominant mechanism driving inspiral and eventual coalescence \cite[][]{burkespolaor2018,khan2018,fang2025,agazie2023,kelley2017,begelman1980}.

Merger-driven SMBH growth is believed to represent an important phase of SMBH and galaxy evolution \citep[][]{satyapal2014, ellison2013, barrows2013, weston2017, chen2023astrid, hopkins2008, hopkins2010}. However, confirmed AGN pair systems remain comparatively rare, particularly at small separations and higher redshifts, limiting the statistical power of current samples \citep[][]{bigmacreal}. Developing a complete picture of the growth and evolution of AGN pair populations across the merger sequence (and across cosmic time) requires continued efforts in finding and confirming larger and more statistically robust samples of genuine AGN pairs. 

\subsection{The Big Multi-AGN Catalog} \label{subsec:bigmac_intro}

The Big Multi-AGN Catalog \citep[the BigMAC,][]{bigmacreal} is the first literature-complete catalog of all known (confirmed or candidate) multi-AGN systems up to 2020, including dual, binary, and recoiling AGN. Among thousands of candidates, it includes 156 confirmed dual AGN and one confirmed binary AGN. The BigMAC includes systems drawn from all selection methods, redshifts, and galaxy mass ratios, providing a comprehensive census of multi-AGN searches in the literature.

The BigMAC has made clear that the vast majority of dual and binary AGN and candidates have been selected and confirmed via optical/infrared spectroscopic diagnostics \cite[including multiple velocity peaks;][]{comerford2009, comerford2013, wang2009, liu10b, lyu2016, barrows2013, liu2011}, spatially-resolved imaging in the X-ray, optical, infrared, and radio regimes \cite[][]{liu2013, liu2010, komossa2003, koss2011, bianchi2008, piconcelli2010, fu2015, u2013}, and mid-infrared colors \cite[][]{pfeifle2019, pfeifle2019b, satyapal2017, ellison2019, barrows2023}.

The dual AGN population is biased towards local redshifts (z $<$ 0.1) and larger physical separations \citep[$1 < r_p < 100$, where $r_p$ is the projected separation in kiloparsecs;][]{bigmacreal} with very few sub-kpc dual AGNs known to date \citep{bigmacreal,komossa2003,koss2023,mullersanchez2015}. The search is significantly more challenging for binary AGN ($\lesssim30$\,pc separations). The very small spatial scales (pc and sub-pc) of binary AGN require extraordinarily high angular resolution instruments. The only known binary AGN system, CSO 0402+379, has a separation of 7.3 pc \citep[][]{rodriguez2006} and was confirmed via very long baseline interferometry (VLBI). The challenge of pre-selection criteria and limited resolutions and sensitivities of current instruments has resulted in an observational gap at high redshifts and small separations \citep[see Figure 26 in][]{bigmacreal}. To probe populations of sub-kpc dual AGN and pc- or sub-pc scale binary AGNs, we must rely on the higher resolutions of radio interferometry, current space-based imaging, or ground-based imaging with large apertures or adaptive optics. For the realm of binary AGN, only VLBI observations will reach the exceedingly high resolutions required for a direct detection (sub-milliarcsecond scales and smaller: a milliarcsecond corresponds to $\sim$ 6.3 parsecs at a redshift of 0.5, and $\sim$ 8.6 parsecs at a redshift of 2).

\subsection{The Radio Fundamental Catalog} \label{subsec:rfc_intro}

Archival radio VLBI observations probing (sub-)milliarcsecond scales are one important tool that can be used in the search for previously unidentified dual and binary AGNs. The Radio Fundamental Catalog (RFC) is an all-sky VLBI catalog containing almost 22,000 compact radio sources \citep[][]{rfcreal}. This recently published catalog incorporates nearly all VLBI observations between 2 and 23 GHz from 73 programs over the course of more than three decades. The goals of the RFC are focused on its use as an absolute astrometry catalog. However, all calibrated images are available, making it an excellent starting point for dual and binary AGN searches. It is particularly useful alongside a pre-selected sample such as the BigMAC repository.

The purpose of this paper is to (1) identify all BigMAC confirmed/candidate sources with images available in the RFC, (2) present the ten BigMAC-RFC sources with two radio cores, and (3) confirm or refute their designation as dual or binary AGN. An analysis of the full BigMAC-RFC catalog (241 targets) is left to a later paper. In Section \ref{sec:radioanalysis}, we present the archival radio analysis methods, the results of which are presented in Section \ref{sec:results}, along with multiwavelength results, where available. In Section \ref{sec:discussion}, we discuss the likelihood of AGN pair versus jet activity designations, how this will impact the evolution of the BigMAC sample, and significant future work. Throughout this paper, all physical separations are the projected separation ($r_p$). A flat $\Lambda$CDM cosmology is adopted, with a $\Omega_{\Lambda}$ = 0.69, $\Omega_{m}$ = 0.31, and $H_{0}$ = 67.7 km s$^{-1}$ Mpc$^{-1}$ \cite[][]{planck2020}.

\section{Radio Observations} \label{sec:radioanalysis}

A BigMAC-RFC sample was derived from a cross-match of the BigMAC Data Release 1.0 repository \citep[confirmed and candidate dual, binary, and recoiling AGN;][]{bigmacreal} with the Radio Fundamental Catalog 2025D \citep[][]{rfcreal}. All 5,742 confirmed and candidate BigMAC sources were included, and their primary components were matched against the RFC 2025D's 21,949 catalog sources. The resulting cross-matched catalog includes 241 candidate and confirmed dual, binary, and recoiling AGN with RFC images. The RFC images are available at S, C, X, and U-bands (central frequencies of 2.3, 5.9, 8.4, and 13.7 GHz, respectively), though nominal frequencies vary with observations. The sources presented in this paper have observations from seven of the 73 large programs that make up the RFC: VLBA Calibrator Surveys III \cite[VCS-III;][]{vcs3}, VI \cite[VCS-VI;][]{vcs6}, IX \cite[VCS-IX;][]{vcs9}, and XII \cite[VCS-XII;][]{rfcreal}, as well as the VLBA Imaging and Polarimetry Survey \citep[VIPS;][]{helmboldt2007}, the VLBA survey of unassociated $\gamma$-ray objects in the seven-year Fermi/LAT catalog \citep[Fermi-LAT;][]{fermivlbi}, and the Monitoring of Jets in Active Galactic Nuclei with VLBA Experiments \citep[MOJAVE;][]{mojave2021} survey. Table \ref{tab:rfcinfo} presents the observational bands and beginning and end dates for all RFC projects used in this paper. Table \ref{tab:obsinfo} lists which programs and frequencies were available for each target, in addition to the responsible RFC data contributor. 

\begin{deluxetable}{ccc}
\tablenum{1}
\tablecaption{RFC Project Information\label{tab:rfcinfo}}
\tablewidth{0pt}
\tablehead{
\colhead{Project} & \colhead{Band/Frequency} & \colhead{Dates}\\
\colhead{(1)} & \colhead{(2)} & \colhead{(3)}
}
\startdata
VCS-III & S/X & 01-16-2017/10-21-2017\\
VCS-VI & S/X & 09-18-2020/12-12-2022\\
VCS-IX & S/X & 08-07-2015/09-07-2016\\
VCS-XII & C/X & 09-21-2021/12-02-2022\\
VIPS & C & 01-03-2006/08-12-2006\\
Fermi-VLBI & C/X & 08-25-2018/02-17-2019\\
MOJAVE & U & 08-31-1994/Present\\
\enddata
\caption{Column 1: RFC project acronym. Column 2: Bands utilized in RFC project. Column 3: Beginning/end dates of observations for RFC project.}
\end{deluxetable}

\begin{deluxetable}{cccc}
\tablenum{2}
\tablecaption{Observation Information\label{tab:obsinfo}}
\tablewidth{0pt}
\tablehead{
\colhead{Source} & \colhead{Band/Frequency} & \colhead{Contributor} & \colhead{Project}\\
\colhead{} & \colhead{GHz} & \colhead{} & \colhead{}\\
\colhead{(1)} & \colhead{(2)} & \colhead{(3)} &
\colhead{(4)}
}
\startdata
J0216-0105 & S/2.27 & Petrov & VCS-III\\
- & X/8.65 & Petrov & VCS-III\\
J0729-1320 & S/2.25 & Petrov & VCS-III\\
- & X/8.65 & Petrov & VCS-III\\
J0745+3142 & S/2.28 & Petrov & VCS-III\\
- & C/4.84 & Taylor & VIPS\\
- & X/8.65 & Petrov & VCS-III\\
J0843+4537 & S/2.25 & Petrov & VCS-III\\
- & C/4.84 & Taylor & VIPS\\
- & X/8.65 & Petrov & VCS-III\\
J1305-1033 & S/2.28 & Petrov & VCS-VI\\
- & C/4.35 & Petrov & VCS-XII\\
- & X/7.61 & Petrov & VCS-XII\\
- & U/15.36 & Lister & MOJAVE\\
J1414+4554 & S/2.28 & Petrov & VCS-III\\
- & C/4.84 & Taylor & VIPS\\
- & X/8.65 & Petrov & VCS-III\\
J1451+1343 & S/2.25 & Petrov & VCS-III\\
- & C/4.34 & Petrov & Fermi-VLBI\\
- & X/8.65 & Petrov & VCS-III\\
J1503+0917 & S/2.25 & Petrov & VCS-III\\
- & X/8.65 & Petrov & VCS-III\\
J2346+3011 & S/2.28 & Petrov & VCS-VI\\
- & C/4.34 & Petrov & VCS-IX\\
- & X/8.65 & Petrov & VCS-VI\\
J2347-1856 & S/2.25 & Petrov & VCS-III\\
- & C/8.65 & Petrov & VCS-III\\
\enddata
\caption{Column 1: Source names. Column 2: Named band and central frequency in GHz of observations. Column 3: RFC observation contributor. Column 4: Program of observation origin.}
\end{deluxetable}

The images in the RFC archive represent a synthesis of calibration inherited from the original observing programs and subsequent uniform processing, providing a homogeneous set of radio products suitable for structural studies. All RFC observations used in this paper are derived from calibrated Very Long Baseline Array (VLBA) visibility data collected across the observing programs described above. Calibration and imaging details are provided in the first RFC paper \citep[][]{petrov2024}. 

\section{Radio Results \& Analysis} \label{sec:results}

The results of the radio and multiwavelength analysis are presented below. In Section \ref{subsec:radiomorphs}, initial radio morphology designations are assigned. In Section \ref{subsec:radioparams}, the results from the Radio Fundamental Catalog observations are shown, including a tentative determination of the likelihood of AGN versus jet activity. 

\subsection{Radio Morphologies} \label{subsec:radiomorphs}

Each target has multiple observations available at each frequency, spanning timescales from months to decades. Because the RFC includes all observations obtained through the projects described above, not every image is suitable for scientific analysis. Observations were excluded if they were compromised by poor weather, radio frequency interference, insufficient \textit{uv}-coverage, prematurely terminated observing sessions, or other unrecoverable calibration issues. After removing these unusable datasets, the most recent observation at each available frequency was selected for analysis. This approach minimized the time separation between observations of the same source at different frequencies, thereby reducing the effects of intrinsic source variability and ensuring that derived quantities, such as spectral indices, were as close to quasi-simultaneous as possible.

The first consideration was radio morphology. The full 241 BigMAC-RFC targets were visually inspected. $\sim$58\% were found to exhibit extended emission, in many cases significant enough to be attributable to jet activity (e.g., radio lobes, collimated jet emission, etc.). However, it is possible for jet activity to exist in one or two radio cores in the same system \citep[e.g., CSO 0402, in which one of the components of the binary AGN is jetted][]{rodriguez2006}. Thus, a full analysis of the extended sources is necessary, but is left for future work. $\sim$38\% of the BigMAC-RFC targets were identified as point sources, in which the RFC source appears compact and Gaussian in nature, with no extension, secondary cores, or indicators of jet activity. While these systems may be multi-AGN with only one radio-emitting component, further study of them is also deferred for future work. 

Ten of the 241 BigMAC-RFC targets ($\sim$4\%) were identified as possible multi-AGN, based on the presence of two (or more) cores in the RFC image. The visual inspection does not reveal clear jet emission, and there is no positional and/or flux evolution of the secondary core that might indicate a jet hotspot. We measured flux densities for every core with the Python Blob Detector and Source Finder \citep[PyBDSF;][]{pybdsf}. PyBDSF identifies statistically significant emission islands directly from the image using thresholding and local noise estimation, then decomposes those islands into Gaussian source components without requiring any user-supplied positions. Brightness temperature and compactness parameters were calculated using the highest available frequency (and thus the highest available resolution). Finally, a radio spectrum was built for each component in each system, and fit with a standard (and, in some cases, curved) power law. 

\subsection{Radio Parameters} \label{subsec:radioparams}

Image and morphology parameters are presented in Table \ref{tab:imageinfo}. This includes target redshifts (where available), observational frequencies, image beam sizes, and the tentative morphological identifier for each source. The morphologies were initially identified via only the visual examination. A more in-depth radio analysis was used to confirm or reject this initial designation, as described in Sections \ref{subsec:pairs} and \ref{subsec:jets}. The designations in Table \ref{tab:imageinfo} reflect this analysis. 

\begin{deluxetable*}{ccccccc}
\tablenum{3}
\tablecaption{Image and Morphology Information\label{tab:imageinfo}}
\tablewidth{0pt}
\tablehead{
\colhead{Source} & \colhead{Redshift} & \colhead{Scale} & \colhead{Band} & \colhead{Beam} & \colhead{Morphology} & \colhead{Source Name}\\
\colhead{} & \colhead{} & \colhead{kpc/arcsec} & \colhead{} & \colhead{[mas, mas, deg]} & \colhead{} & \colhead{[ICRF]}\\
\colhead{(1)} & \colhead{(2)} & \colhead{(3)} &
\colhead{(4)} & \colhead{(5)} & \colhead{(6)} & \colhead{(7)}
}
\startdata
J0216-0105 & 1.49 & 8.4 & S & 9, 4, -2.2 & Candidate Dual & J021612.2-010518\\ 
- & - & - & X & 2, 1, 1.8 & - \\ 
J0729-1320 & 0.34 & 6.1 & S & 8, 3, -2.9 & Candidate Dual & J072917.8-132002\\ 
- & - & - & X & 2, 1, -2.8 & - \\ 
J0745+3142 & 0.46 & 5.8 & S & 7, 3, 6.4 & Candidate Dual & J074541.6+314256\\ 
- & - & - & C & 3, 2, 0.0 & - \\ 
- & - & - & X & 2, 1, 6.1 & - \\
J0843+4537 & 0.19 & 3.2 & S & 5, 4, 0.1 & Candidate Jet & J084307.0+453742\\ 
- & - & - & C & 3, 2, 0.0 & - \\ 
- & - & - & X & 1, 1, 5.8 & - \\
J1305-1033 & 0.28 & 4.2 & S & 11, 4, -5.3 & Candidate Dual & J130533.0-103319\\ 
- & - & - & C & 4, 1, 6.0 & - \\ 
- & - & - & X & 2, 1, 6.3 & - \\
- & - & - & U & 1, 0.5, -6.9 & - \\
J1414+4554 & 0.19 & 3.2 & S & 6, 4, -10 & Candidate Jet & J141414.8+455448\\ 
- & - & - & C & 3, 2, 0.0 & - \\
- & - & - & X & 2, 2, -12 & - \\
J1451+1343 & 0.5* & 6.1 & S & 8, 4, -3.9 & Candidate Jet & J145131.4+134324\\ 
- & - & - & C & 4, 3, -0.9 & - \\ 
- & - & - & X & 2, 1, -2.4 & - \\
J1503+0917 & 0.5* & 6.1 & S & 8, 4, -8.4 & Candidate Jet & J150300.8+091758\\ 
- & - & - & X & 2, 1, -5.2 & - \\
J2346+3011 & 0.5* & 6.1 & S & 7, 4, -8.2 & Candidate Dual & J234646.2+301159\\ 
- & - & - & C & 4, 2, -9.2 & - \\ 
- & - & - & X & 2, 1, -2.1 & - \\
J2347-1856 & 0.5* & 4.3 & S & 8, 3, -3.5 & Candidate Dual & J234708.6-185618\\ 
- & - & - & X & 2, 1, -2.3 & - \\ 
\enddata
\caption{Column 1: Source names; Column 2: Redshift (* indicates redshift is not available; z = 0.5 has been used as placeholder for all following analysis); Column 3: scale in kiloparsecs per arcsecond; Column 4: Observation band; Column 5: Image beam in milliarcseconds, position angle in degrees. Column 6: Tentative morphological designation; Column 7: International Celestial Reference Frame source identifier \citep[][]{charlot2020}.}
\end{deluxetable*}

The \texttt{PyBDSF} results are presented in Tables \ref{tab:radioparamsduals} and \ref{tab:radioparamsjets}. The results tables are split into the ``likely dual AGN'' and ``likely jet activity'' categories, respectively, across the two tables. The radio parameters for each source are sorted by both observational frequency and Component. Each radio source in each system was assigned a component label (e.g., North, South, West, or East). In cases where the resolution was not high enough to separate the components, the Component is labeled as ``Total.''

Total flux density and errors, peak flux density and errors, and image $1\sigma$ root-mean-square (RMS) are all drawn from the \texttt{PyBDSF} measurements. Error estimates are based on the work of \cite{condon1997}. Flux density scale errors were also taken into account. Following the VLBA Observing Guide\footnote{http://science.nrao.edu/facilities/vlba/docs/manuals/ \\propvlba/calibration-considerations}, a flux density scale calibration accuracy of 10\% was assumed for all frequencies \citep[][]{midderlberg2011}. Thus, the flux errors presented in this paper reflect both the errors in the Gaussian models and the flux density scaling errors, added in quadrature. 

The presented luminosity was calculated from the peak flux density. Finally, both compactness and brightness temperature were calculated. Compactness (C) is calculated as the fraction of the total emission from a source that is effectively within the peak, e.g.\ $\mathrm{C}=S_{\rm integrated}/I_{\rm peak}$ (the ratio of the flux density integrated over the beam to its peak value). A value of $\mathrm{C}\sim1$ would be indicative of a compact core, as is expected in the case of AGN activity. Extended emission such as that associated with jet activity might exhibit a value of $\mathrm{C}>2$. Compactness is only calculated for the high resolution observations available for each source in order to take into account the smallest scales of extended activity available in these observations.

Brightness temperature is similarly useful for identifying AGN activity, and requires the milliarcsecond-scale spatial resolutions provided by the VLBA observations. In the case of the RFC targets, there is little question as to the existence of at least one AGN within each source. However, the limit separating non-thermal AGN from star formation or otherwise weaker emission is $T_{b} = 10^5$ K \citep[][]{condon1991}, and it is useful to compare the BigMAC-RFC targets to that limit. Brightness temperature is calculated as:

\begin{equation}
    T_b = \frac{S}{\Omega_\mathrm{beam}}\frac{c^2}{2k\nu^2},
\end{equation}

where $\nu$ is the observing frequency, $S$ is the integrated flux density, and $\Omega_\mathrm{beam}$ is the beam solid angle. All sources within the BigMAC-RFC sample exhibit brightness temperatures significantly higher than the non-thermal AGN limit, making it likely that they represent accretion onto a central SMBH. 

\begin{deluxetable*}{cccccccccc}
\tablenum{4}
\tablecaption{Radio Parameters: Likely Dual AGN\label{tab:radioparamsduals}}
\tablewidth{0pt}
\tablehead{
\colhead{Source} & \colhead{Band} & \colhead{Component} & \colhead{$S_{total}$} & \colhead{$S_{Peak}$} & \colhead{$Log(L_{Peak})$} & \colhead{C} & \colhead{$T_{b}$} & \colhead{$\sigma$}\\
\colhead{} & \colhead{} & \colhead{} & \colhead{[mJy]} & \colhead{[mJy/bm]} & \colhead{[W Hz$^{-1}$]} & \colhead{} & \colhead{K} & \colhead{$\upmu$Jy}\\
\colhead{(1)} & \colhead{(2)} & \colhead{(3)} &
\colhead{(4)} & \colhead{(5)} & \colhead{(6)} & \colhead{(7)} & \colhead{(8)} & \colhead{(9)}
}
\startdata
J0216-0105 & S & Total & 128.7$\pm$13.5 & 105.6$\pm$10.8 & 27.19 & - & - & 203\\
- & X & North & 9.63$\pm$1.22 & 6.12$\pm$0.71 & 25.95 & 1.57 & $5.42 \times 10^{10}$ & 116\\
- & X & South & 75.79$\pm$7.88 & 68.98$\pm$7.01 & 27.01 & 1.09 & $6.11 \times 10^{11}$ & 116\\
J0729-1320 & S & Total & 246.8$\pm$25.6 & 158.3$\pm$16.1 & 25.78 & - & - & 229\\
- & X & East & 20.80$\pm$2.40 & 17.55$\pm$1.88 & 24.83 & 1.18 & $1.55 \times 10^{11}$ & 121\\
- & X & West & 168.6$\pm$17.3 & 146.4$\pm$14.8 & 25.75 & 1.15 & $1.30 \times 10^{12}$ & 121\\
J0745+3142 & S & Total & 723.9$\pm$73.3 & 589.3$\pm$59.2 & 26.69 & - & - & 233\\
- & C & Total & 542.9$\pm$55.9 & 323.9$\pm$32.6 & 26.43 & - & - & 232\\
- & X & North & 343.5$\pm$35.1 & 235.3$\pm$23.7 & 26.29 & 1.46 & $2.08 \times 10^{12}$ & 135\\
- & X & South & 111.1$\pm$11.6 & 79.70$\pm$8.11 & 25.83 & 1.39 & $7.06 \times 10^{11}$ & 135\\
J1305-1033 & S & Total & 642.5$\pm$65.9 & 590.6$\pm$59.5 & 26.18 & - & - & 364\\
- & C & Total & 579.6$\pm$58.6 & 482.7$\pm$48.5 & 26.09 & - & - & 246\\
- & X & Total & 445.6$\pm$45.3 & 352.1$\pm$35.4 & 25.96 & - & - & 138\\
- & U & North & 199.9$\pm$20.2 & 146.1$\pm$14.7 & 25.58 & 1.37 & $1.52 \times 10^{12}$ & 84\\
- & U & South & 459.5$\pm$46.4 & 459.2$\pm$45.9 & 26.07 & 1.00 & $4.79 \times 10^{12}$ & 84\\
\enddata
\caption{Column 1: Source names; Column 2: Band; Column 3: Component; Column 4: Total flux density and error in mJy; Column 5: Peak flux density and error in mJy/beam; Column 6: Log of peak luminosity in $W/Hz$; Column 7: Source compactness as calculated from total flux to peak flux ratio; Column 8: Brightness temperature as calculated from the total flux; Column 9: 1-$\sigma$ sensitivity.}
\end{deluxetable*}

The compactness parameter results show a more significant spread. In trying to disambiguate dual AGN from jet activity, it is critical to recognize that systems could exhibit one of several setups: (1) two peaks, each with $\mathrm{C}\sim1$ (indicative of dual AGN), (2) two peaks, one with $\mathrm{C}\sim1$, one with $\mathrm{C}>2$ (indicative of jet activity), or (3) multiple peaks with $\mathrm{C}>2$. In all cases, the compactness parameters must be taken into context with the other radio results.

Angular separations for all sources were measured from the highest available resolution observations. Using the known redshifts (assumed to be $z = 0.5$ where unavailable), projected physical separations were calculated. The results for all sources are listed in Table \ref{tab:sepinfo}.

\begin{deluxetable*}{cccccccccc}
\tablenum{4}
\tablecaption{Radio Parameters: Likely Dual AGN}
\tablewidth{0pt}
\tablehead{
\colhead{Source} & \colhead{Band} & \colhead{Component} & \colhead{$S_{total}$} & \colhead{$S_{Peak}$} & \colhead{$Log(L_{Peak})$} & \colhead{C} & \colhead{$T_{b}$} & \colhead{$\sigma$}\\
\colhead{} & \colhead{} & \colhead{} & \colhead{[mJy]} & \colhead{[mJy/bm]} & \colhead{[W Hz$^{-1}$]} & \colhead{} & \colhead{K} & \colhead{$\upmu$Jy}\\
\colhead{(1)} & \colhead{(2)} & \colhead{(3)} &
\colhead{(4)} & \colhead{(5)} & \colhead{(6)} & \colhead{(7)} & \colhead{(8)} & \colhead{(9)}
}
\startdata
J2346+3011 & S & North & 105.4$\pm$11.4 & 101.9$\pm$10.6 & 26.02 & - & - & 439\\
- & S & Middle & 135.9$\pm$14.4 & 132.5$\pm$13.7 & 26.13 & - & - & 439\\
- & S & South & 76.28$\pm$8.85 & 56.09$\pm$6.08 & 25.75 & - & - & 439\\
- & C & North & 118.5$\pm$12.4 & 103.9$\pm$10.5 & 26.03 & - & - & 118\\
- & C & Middle & 83.96$\pm$8.71 & 78.76$\pm$7.98 & 25.90 & - & - & 118\\
- & C & South & 23.52$\pm$2.98 & 10.24$\pm$1.13 & 25.02 & - & $6.78 \times 10^{10}$ & 118\\
- & X & North & 82.47$\pm$9.03 & 65.92$\pm$6.78 & 25.83 & 1.25 & $5.84 \times 10^{11}$ & 187\\
- & X & Middle & 62.49$\pm$7.31 & 47.16$\pm$4.92 & 25.68 & 1.32 & $4.18 \times 10^{11}$ & 187\\
J2347-1856 & S & North & 311.8$\pm$32.1 & 276.1$\pm$27.9 & 25.90 & - & - & 256\\
- & S & South & 445.6$\pm$45.8 & 350.6$\pm$35.3 & 26.00 & - & - & 256\\
- & X & North & 142.2$\pm$14.8 & 89.44$\pm$9.06 & 25.41 & 1.59 & $7.92 \times 10^{11}$ & 119\\
- & X & South & 156.5$\pm$16.4 & 71.54$\pm$7.27 & 25.31 & 2.19 & $6.33 \times 10^{11}$ & 119\\
\enddata
\caption{cont.}
\end{deluxetable*}

\begin{deluxetable*}{ccccccccc}
\tablenum{5}
\tablecaption{Radio Parameters: Likely Jet Activity\label{tab:radioparamsjets}}
\tablewidth{0pt}
\tablehead{
\colhead{Source} & \colhead{Band} & \colhead{Component} & \colhead{$S_{total}$} & \colhead{$S_{Peak}$} & \colhead{$Log(L_{Peak})$} & \colhead{C} & \colhead{$T_{b}$} & \colhead{$\sigma$}\\
\colhead{} & \colhead{} & \colhead{} & \colhead{[mJy]} & \colhead{[mJy/bm]} & \colhead{[$W Hz^{-1}$]} & \colhead{} & \colhead{K} & \colhead{mJy}\\
\colhead{(1)} & \colhead{(2)} & \colhead{(3)} &
\colhead{(4)} & \colhead{(5)} & \colhead{(6)} & \colhead{(7)} & \colhead{(8)} & \colhead{(9)}
}
\startdata
J0843+4537 & S & Total & 183.4$\pm$19.2 & 140.6$\pm$14.3 & 25.19 & - & - & 211\\
- & C & North & 112.48$\pm$12.3 & 71.72$\pm$7.44 & 24.90 & - & - & 264\\
- & C & South & 17.19$\pm$2.28 & 14.03$\pm$1.68 & 24.19 & - & - & 264\\
- & X & North & 66.7$\pm$7.32 & 29.41$\pm$3.05 & 24.52 & 2.26 & $5.21 \times 10^{11}$ & 118\\
- & X & South & 20.22$\pm$2.39 & 12.81$\pm$1.39 & 24.16 & 2.58 & $2.27 \times 10^{11}$ & 118\\
J1414+4554 & S & North & 197.4$\pm$20.7 & 111.1$\pm$11.3 & 25.06 & - & - & 174\\
- & S & South & 133.2$\pm$14.0 & 94.01$\pm$9.57 & 24.99 & - & - & 174\\
- & C & North & 110.6$\pm$13.1 & 38.29$\pm$4.09 & 24.60 & - & - & 281\\
- & C & South & 87.67$\pm$11.5 & 41.63$\pm$4.43 & 24.64 & - & - & 281\\
- & X & North & 56.31$\pm$5.93 & 32.48$\pm$3.32 & 24.53 & 1.73 & $1.44 \times 10^{11}$ & 81\\
- & X & South & 40.54$\pm$4.45 & 13.61$\pm$1.43 & 24.15 & 2.98 & $6.03 \times 10^{10}$ & 81\\
J1451+1343 & S & East & 262.4$\pm$27.4 & 125.4$\pm$12.7 & 26.11 & - & - & 205\\
- & S & West & 255.5$\pm$26.8 & 195.4$\pm$19.7 & 26.29 & - & - & 205\\
- & C & East & 162.5$\pm$16.9 & 94.75$\pm$9.59 & 25.98 & - & - & 138\\
- & C & West & 137.9$\pm$14.4 & 120.8$\pm$12.2 & 26.09 & - & - & 138\\
- & X & East & 83.81$\pm$8.97 & 39.87$\pm$6.49 & 25.61 & 2.10 & $3.53 \times 10^{11}$ & 97.9\\
- & X & West & 79.34$\pm$8.37 & 63.98$\pm$4.09 & 25.81 & 1.24 & $5.67 \times 10^{11}$ & 97.9\\
J1503+3011 & S & North & 395.5$\pm$40.8 & 330.3$\pm$33.3 & 26.53 & - & - & 265\\
- & S & South & 66.89$\pm$7.47 & 47.28$\pm$4.97 & 25.68 & - & - & 265\\
- & X & North & 182.2$\pm$19.4 & 102.8$\pm$10.4 & 26.02 & 1.77 & $9.11 \times 10^{11}$ & 130\\
- & X & South & 17.45$\pm$2.59 & 7.18$\pm$0.75 & 24.79 & 2.83 & $5.47 \times 10^{10}$ & 130\\
\enddata
\caption{Column 1: Source names; Column 2: Band; Column 3: Component; Column 4: Total flux density and error in mJy; Column 5: Peak flux density and error in mJy/beam; Column 6: Log of peak luminosity in $W/Hz$; Column 7: Source compactness as calculated from total flux to peak flux ratio; Column 8: Brightness temperature as calculated from the total flux; Column 9: 1-$\sigma$ sensitivity.}
\end{deluxetable*}

\begin{deluxetable}{cccc}
\tablenum{6}
\tablecaption{Separation Information\label{tab:sepinfo}}
\tablewidth{0pt}
\tablehead{
\colhead{Source} & \colhead{$\theta$} & \colhead{$R_p$} & \colhead{$\Delta$}\\
\colhead{} & \colhead{mas} & \colhead{pc} & \colhead{mas}\\
\colhead{(1)} & \colhead{(2)} & \colhead{(3)} & \colhead{(4)}
}
\startdata
J0216-0105 & 5.45 & 45.8 & 0.19\\
J0729-1320 & 3.41 & 16.5 & 0.53\\
J0745+3142 & 2.96 & 17.2 & 3.67\\
J0843+4537 & 4.14 & 13.2 & - \\
J1305-1033 & 1.51 & 6.34 & 0.37\\
J1414+4554 & 28.5 & 91.2 & -\\
J1451+1343 & 26.7 & 162.9$^*$ & -\\
J1503+0917 & 13.1 & 79.9$^*$ & -\\
J2346+3011 & 18.3 & 111.6$^*$ & -\\
J2347-1856 & 33.5 & 204.5$^*$ & -\\
\enddata
\caption{Column 1: Source names. Column 2: Angular separation in milliarcseconds as measured from the highest available resolution RFC image. Column 3: Projected physical separation in parsecs. $^*$: redshift assumed to be 0.5. Column 4: Radio-optical separations calculated from Gaia DR3 optical and ICRF3 radio positions, following the method presented in \cite{orosz2013}.}
\end{deluxetable}

\begin{figure*}[ht!]
    \centering
    \includegraphics[width=16cm]{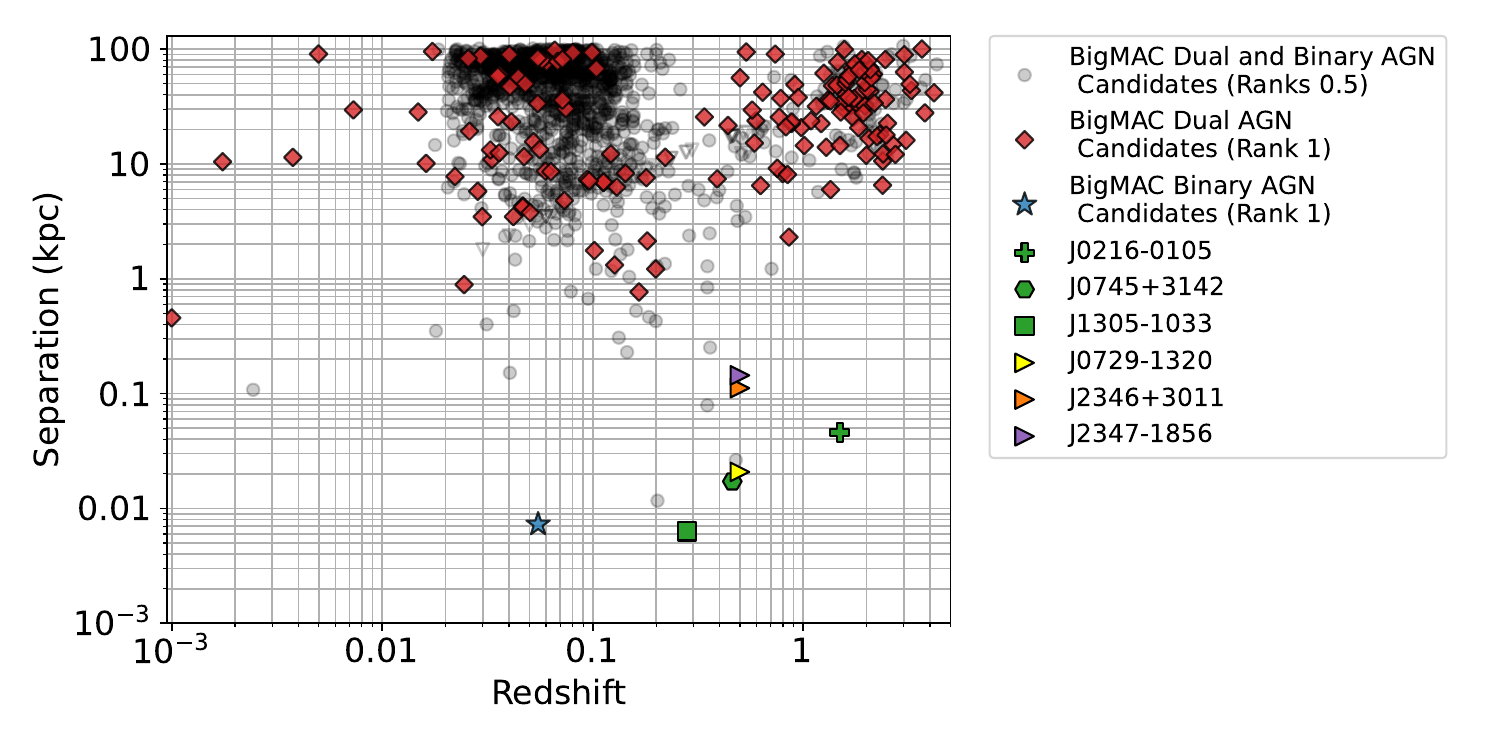}
        \vspace*{-3mm}
          \caption{Separation (in kiloparsecs) versus redshift plot for BigMAC-RFC candidate multi-AGN, plotted against the Rank 0.5 and 1 BigMAC confirmed and candidate dual and binary AGN. BigMAC rank 0.5 candidates are marked with grey circles. BigMAC rank 1 dual AGN candidates are marked with red diamonds. BigMAC rank 1 binary AGN candidates are marked with blue stars. BigMAC-RFC targets with known redshifts are marked in green, while those with assumed redshifts of 0.5 are marked with right-facing triangles denoting their possibly higher redshifts.}
   \label{fig:bigmac}
\end{figure*}

Given the measured separations for each BigMAC-RFC source, Figure \ref{fig:bigmac} plots the BigMAC-RFC candidate multi-AGN (six candidates) against the rank 0.5 and 1 BigMAC confirmed and candidate dual and binary AGN. Rank 1 candidates are sources in which the classification is established beyond a reasonable doubt, while rank 0.5 candidates are sources for which there is substantive evidence favoring the multiple interpretation. Grey circles, red diamonds, and blue stars correspond to BigMAC rank 0.5 dual and binary AGN candidates, BigMAC rank 1 dual AGN candidates, and BigMAC rank 1 binary AGN candidates, respectively. The three BigMAC-RFC multi-AGN candidates with known redshifts are marked in green, while the three candidates for which a redshift of 0.5 have been assumed are marked with right-facing triangles that denote their possibly higher redshifts. A distinct separation in the BigMAC versus the BigMAC-RFC samples is clear. As further discussed in Section \ref{subsec:bigmacevol}, the candidate multi-AGN in this paper probe a new redshift-separation regime. Finally, all radio images are shown in Appendix \ref{sec:radioims}, Figure \ref{fig:0216} - \ref{fig:2347}. The images shown were those used for the \texttt{PyBDSF} flux measurements and morphological identifications. 

\subsection{Radio Spectral Analysis} \label{subsec:radiospec}

Radio spectral shape and slope are other possible parameters for AGN identification. Radio spectra were built for all components in each BigMAC-RFC system, and are presented in Appendix \ref{sec:radiospec}. It is important to note that not all components in a source were resolved at all frequencies. In the cases where multiple high frequency components are not resolved in the lower frequency observations (due to their limited resolutions), an attempt was made to recover individual component fluxes by estimating the component ratios between the two cores as 10\%/90\%, 90\%/10\%, and 50\%/50\%. Some attempts produced spectral indices that were inconsistent with known physical emission mechanisms and were therefore deemed non-viable and discarded from further analysis. Each component was fit with two basic synchrotron models:

\begin{itemize}
    \item \textbf{Standard Power Law:} A standard power law model that is defined by $S_{\nu} = S_{o} \nu^{\alpha}$, where $S_{\nu}$ is the flux density at frequency $\nu$, $\alpha$ is the spectral index, and $S_{o}$ is the flux density at 1 GHz. The value and sign of the spectral index can be used to help distinguish between synchrotron and thermal emission \cite[e.g.,][]{patil2021,patil2022}.  
    \item \textbf{Curved Power Law:} A curved power law generated from the flux density at frequency $\nu$, $S_{\nu}$, the spectral index $\alpha$, and the flux density at 1 GHz $S_{o}$, defined as $S_{\nu} = S_{o} \nu^{\alpha} e^{q^{(\ln\nu)^2}}$, where $q$ represents the width of the peak, characterizes the degree of curvature, and is defined as $\nu_\mathrm{peak} = e^{-\alpha / 2q}$, where $\nu_\mathrm{peak}$ is the peak frequency. Significant spectral curvature is typically defined as $|q| \geq 0.2$ \cite[][]{duffy2012}, and is indicative of absorption of low frequency emission. Absorption may be due to a high synchrotron optical depth within the source \cite[i.e., synchrotron self-absorption (SSA);][]{odea2021} or free-free absorption from surrounding ionized gas \cite[][]{bicknell1998}. Understanding which process is in play can further constrain the radio source and environment properties. 
\end{itemize}

The radio spectra were fit using the \texttt{radio\_spectral\_fitting} package \citep{patil2022}, which performs weighted least-squares fits to both standard and curved power law spectral models. The \texttt{radio\_spectral\_fitting} package does not report parameter uncertainties when the number of measurements is equal to the number of free parameters in the fitted model. Consequently, uncertainties are not provided for standard power-law fits to two-frequency spectra or for curved power-law fits to three-frequency spectra. In these cases, the fitted model exactly reproduces the measured flux densities, leaving zero degrees of freedom and preventing estimation of model uncertainties from the residual variance in the data. While the best-fitting parameters remain well defined, their uncertainties must instead be estimated through propagation of the measured flux-density uncertainties. For standard power law fits, uncertainties in the spectral index were obtained by analytically propagating the flux density uncertainties, while treating the observing frequencies as exact. Flux density uncertainties include both the PyBDSF fitting uncertainties and an adopted 10\% absolute flux scale uncertainty, added in quadrature. For curved power law fits, parameter uncertainties were estimated using weighted least-squares covariance analysis in logarithmic space. 

In Figures \ref{fig:0216spec} - \ref{fig:2347spec}, all viable spectra are shown. All components and available observational frequencies are used in all spectra. Components are differentiated by shape (squares, triangles, and circles), while colors are attributable to observational frequencies: S-band in purple, C-band in green, X-band in brown, and U-band in grey. Error bars have been included, but for the majority of the flux densities, are too small to be visible. The standard power and curved power law fits are shown for each component. Non-physical fits have been excluded. Resultant spectral indices are included in the upper left hand corner. The superscript $C$ denotes a curved power law spectral index. The solid red, dot-dash blue, and dotted green lines each represent the standard and/or curved power laws for different components. 

Table \ref{tab:specinfo} presents the spectral index values. It includes all viable spectra with varied flux density percentages, and the spectral indices that result from both the standard power and curved power law fits. For each target, the resulting spectral indices from the best fit spectra are listed in bold. Given the resolution issues at the lower frequencies, these spectral indices must be treated with care. However, they are used in Section \ref{sec:discussion} to present a comprehensive radio understanding of all ten targets.

\begin{deluxetable*}{ccccccc}
\tablenum{7}
\tablecaption{Spectral Index Information\label{tab:specinfo}}
\tablewidth{0pt}
\tablehead{
\colhead{Source} & \colhead{Component} & \colhead{\%S$_{S}$} & \colhead{\%S$_{C}$} & \colhead{\%S$_{X}$} & \colhead{\%S$_{U}$} & \colhead{$\alpha \pm \sigma_{\alpha}$}\\
\colhead{(1)} & \colhead{(2)} & \colhead{(3)} &
\colhead{(4)} & \colhead{(4)} & \colhead{(4)} & \colhead{(7)}
}
\startdata
J0216-0105 & \makecell{North\\ South} & \makecell{90\\ 10} & - & Resolved & - & \makecell{-2.14$\pm$0.13\\ 1.46$\pm$0.82}\\
- & \makecell{North\\ South} & \makecell{10\\ 90} & - & Resolved & - & \makecell{\textbf{-0.42$\pm$0.80}\\ \textbf{-0.25$\pm$0.12}}\\
- & \makecell{North\\ South} & \makecell{50\\ 50} & - & Resolved & - & \makecell{-1.68$\pm$0.21\\ 0.21$\pm$0.19}\\
\hline
J0729-1320 & \makecell{East\\ West} & \makecell{10\\ 90} & - & Resolved & - & \makecell{\textbf{0.08$\pm$0.80}\\ \textbf{0.02$\pm$0.12}}\\
- & \makecell{East\\ West} & \makecell{90\\ 10} & - & Resolved & - & \makecell{-1.63$\pm$0.12\\ 1.73$\pm$0.79}\\
- & \makecell{East\\ West} & \makecell{50\\ 50} & - & Resolved & - & \makecell{-1.17$\pm$0.21\\ 0.48$\pm$0.18}\\
\hline
J0745+3142 & \makecell{North\\ South} & \makecell{50\\ 50} & \makecell{50\\ 50} & Resolved & - & \makecell{0.01$\pm$0.17\\ -1.04$\pm$0.13}\\
- & \makecell{North\\ South} & \makecell{90\\ 10} & \makecell{90\\ 10} & Resolved & - & \makecell{\textbf{-0.64$\pm$0.12}\\ \textbf{1.04$\pm$1.15}}\\
- & \makecell{North$_C$\\ South$_C$} & \makecell{90\\ 10} & \makecell{90\\ 10} & Resolved & - & \makecell{-1.60$\pm$0.99\\ -}\\
\hline
J0843+4537 & \makecell{North\\ South} & \makecell{90\\ 10} & Resolved & Resolved & - & \makecell{-0.35$\pm$0.10\\ -2.11$\pm$0.40}\\
- & \makecell{North\\ South} & \makecell{50\\ 50} & Resolved & Resolved & - & \makecell{\textbf{-0.79$\pm$0.14}\\ \textbf{-0.63$\pm$0.18}}\\
\hline
J1305-1033 & \makecell{North\\ South} & \makecell{50\\ 50} & \makecell{10\\ 90} & \makecell{10\\ 90} & Resolved & \makecell{0.01$\pm$0.20\\ 0.15$\pm$0.10}\\
- & \makecell{North$_C$\\ South$_C$} & \makecell{50\\ 50} & \makecell{10\\ 90} & \makecell{10\\ 90} & Resolved & \makecell{-\\ -0.23$\pm$0.59}\\
- & \makecell{North\\ South} & \makecell{50\\ 50} & \makecell{50\\ 50} & \makecell{50\\ 50} & Resolved & \makecell{\textbf{-0.38$\pm$0.16}\\ \textbf{0.47$\pm$0.18}}\\
- & \makecell{North$_C$\\ South$_C$} & \makecell{50\\ 50} & \makecell{50\\ 50} & \makecell{50\\ 50} & Resolved & \makecell{-0.34$\pm$0.82\\ -2.76$\pm$0.83}\\
\hline
J1414+4554 & \makecell{North\\ South} & Resolved & Resolved & Resolved & - & \makecell{\textbf{-1.01$\pm$0.14}\\ \textbf{-1.49$\pm$0.08}}\\
- & \makecell{North$_C$\\ South$_C$} & Resolved & Resolved & Resolved & - & \makecell{-\\ 0.64$\pm$0.94}\\
\hline
J1451+1343 & \makecell{West\\ East} & Resolved & Resolved & Resolved & - & \makecell{\textbf{-0.86$\pm$0.16}\\ \textbf{-0.87$\pm$0.12}}\\
- & \makecell{West$_C$\\ East$_C$} & Resolved & Resolved & Resolved & - & \makecell{-0.30$\pm$0.90\\ -}\\
\hline
J1503+0917 & \makecell{North\\ South} & Resolved & Resolved & - & - & \makecell{\textbf{-0.91$\pm$0.10}\\ \textbf{-1.59$\pm$0.15}}\\
\hline
J2346+3011 & \makecell{North\\ Middle\\ South} & Resolved & Resolved & Resolved & - & \makecell{\textbf{-0.33$\pm$0.12}\\ \textbf{-0.80$\pm$0.14}\\ \textbf{-3.36$\pm$0.13}}\\
- & \makecell{North$_C$\\ Middle$_C$\\ South$_C$} & Resolved & Resolved & Resolved & - & \makecell{1.63$\pm$0.91\\ -0.28$\pm$0.92\\ -}\\
\hline
J2347-1856 & \makecell{North\\ South} & Resolved & Resolved & - & - & \makecell{\textbf{-0.88$\pm$0.11}\\ \textbf{-1.24$\pm$0.12}}\\
\enddata
\caption{Column 1 + 2: Source and component names. A sub-script `C' denotes a curved power law spectra index. Column 3, 4, 5, + 6: Percentage of S-, C-, X-, and U-band flux attributed to component. `Resolved' indicates both components were resolved. Column 7: Spectral indices for all components, with error.}
\end{deluxetable*}

\subsection{Optical Analysis} \label{subsec:optical}

Another parameter that can be used to identify multiplicity in a system is evidence of extended structure indicating a recent merger in the optical images from the Dark Energy Camera Legacy Survey \citep[DESI LS;][]{dey2019}. J0729-1320 is outside out the DESI LS sky coverage, so PanSTARRS optical images were used in place. The optical images are presented in Appendix \ref{sec:decals}, Figure \ref{fig:decals}. Each image has a resolution of approximately 1''.

\begin{deluxetable}{ccc}
\tablenum{8}
\tablecaption{DESI LS Gaussian Models\label{tab:optical}}
\tablewidth{0pt}
\tablehead{
\colhead{Source} & \colhead{Gaussian Source Model} & \colhead{$e_{source}$}\\
\colhead{[SDSS]} & \colhead{[asec, asec, $^\circ$]} & \colhead{}\\
\colhead{[1]} & \colhead{[2]} & \colhead{[3]}
}
\startdata
J0216-0105 & 1.01, 1.08, 118 & 0.01\\
J0729-1320 & 1.52, 1.29, 57.3 & 0.15\\ 
J0745+3142 & 1.63, 1.48, 24.2 & 0.09\\
J0843+4537 & 5.43, 2.88, 23.4 & 0.47\\ 
J1305-1033 & 1.13, 1.07, 54.8 & 0.05\\
J1414+4554 & 3.13, 1.92, 169 & 0.38\\ 
J1451+1343 & 1.45, 1.36, 167 & 0.06\\ 
J1503+0917 & 1.72, 1.52, 142 & 0.12\\ 
J2346+3011 & 1.43, 1.25, 24.1 & 0.13\\ 
J2347-1856 & 2.44, 2.22, 180 & 0.09\\
\enddata
\caption{Column 1: Coordinate names. Column 2: DESI LS \textit{r}-band Gaussian target model: major axis, minor axis, position angle. Column 3: DESI LS \textit{r}-band target ellipticity.}
\end{deluxetable}

A careful examination of the DESI LS \textit{ugz} image for all then sources in this study has revealed that five of the ten optical hosts are slightly elliptical. The optical hosts in the \textit{r}-band were modeled with a single Gaussian using CASA's IMFIT. The results, including ellipticity, are presented in Table \ref{tab:optical}.

The five targets with optical ellipticity are J0729-1320, J0843+4537, J1414+4554, J1503+0917, and J2346+3011. For all five targets, the ellipticity might be evidence for extended structure that indicates a recent merger. Of these, J0729-1320, J1503+0917, and J2346+3011 are hosts for the candidate multi-AGN identified in this work. However, further optical observations at significantly higher resolutions are required to confirm the presence of any true extended, merger-induced structure. Even the best ground-based AO observations are still unable to match the VLBA resolutions.

\section{Discussion} \label{sec:discussion}

Given the above radio parameters, and significant existing multiwavelength data, the nature of the candidate BigMAC/RFC multi-AGN can now be assessed. In Sections \ref{subsec:pairs} and \ref{subsec:jets}, each of the ten BigMAC/RFC sources are discussed in the context of all above results. In Section \ref{sec:ICRFdisc}, the impact of these results on the International Celestial Reference Frame (ICRF) is discussed. Finally, in Section \ref{subsec:bigmacevol}, all sources are discussed in the broader context of the full BigMAC sample. Additionally, there is significant follow-up work possible with these sources, as well as the entire BigMAC/RFC sample, which is also discussed in Section \ref{subsec:bigmacevol}.

Given the range of redshifts and angular separations probed with this sample, each target must be considered from the perspective of both dual and binary AGN. The angular separations (or the target extent, in the case of jetted sources) range from 1.51 to 33.5 milliarcseconds, probing down into the binary regime at the redshifts of many of these sources. The only confirmed binary AGN is CSO 0402+379, with a separation of 7.3 parsecs, which was confirmed with VLBI observations \citep[][]{rodriguez2006}. While the targets presented in this paper must remain candidates given the existing observations, confirming even a single one would double the known binary AGN population. In the case of jetted sources, the resolutions of the RFC now allow examination of other multi-AGN pre-selection methods. In some cases, these observations allow validation of previously measured source separations. In others, these results highlight a disparity in inferred separation and directly measured separation. 

All sources must also be considered from the perspective of the International Celestial Reference Frame \citep[ICRF;][]{charlot2020}. The ICRF serves as the defining standard for astrometry, the quasi-inertial reference system against which positions and motions of celestial objects are measured. The ICRF is realized by VLBI observations of extragalactic radio sources, primarily quasars, which are assumed to be point-like and stable. In practice, intrinsic source structure and multiplicity can shift the apparent centroid of the emission observed by VLBI, inducing increased systematic uncertainty \citep[e.g.,][]{makarov2022,cigan2024}.

All ten of the sources analyzed in this paper are found in the current iteration of the ICRF \cite[ICRF3;][]{charlot2020}. Given that all ten sources also exhibit either intrinsic source multiplicity or other source structure on the milli- and sub-milliarcsecond scales that are used to define the ICRF, further discussion of these sources from an ICRF perspective is presented in Section \ref{sec:ICRFdisc}.

Finally, there also exists an optical reference frame, defined by observations from the Gaia spacecraft \citep[][]{gaiacrf3}, which reaches microarcsecond astrometric precision. Comparisons between the radio and optical reference frames have highlighted significant radio-optical offsets \citep[][]{Makarov_2017,petrov2017}. Aligning the multi-frequency reference frames has critical astrometric applications, and thus it is increasingly important to identify and understand the observed radio-optical offsets \citep[][]{gaiacrf3,charlot2020,mignard2016}. The RFC observations now allow us to compare measured source separations or extents to known radio-optical offset measurements. While source multiplicity is posited to be a driver of radio-optical offsets \citep[][]{orosz2013,kovalev2017,petrov2017,makarov2019}, these RFC observations can help further highlight this possibility.

Radio-optical positional offsets were included in the source-by-source discussion presented in Sections \ref{subsec:pairs} and \ref{subsec:jets}. Radio-optical offsets were calculated using the SDSS DR9 and ICRF2 positions \citep[][]{orosz2013}. For comparison, radio-optical offsets were calculated in this paper using the Gaia DR3 and ICRF3 positions. Separation information is presented in Table \ref{tab:sepinfo}. We note that Gaia DR3 positions were only available for four of the ten targets. The radio-optical offsets from both \cite{orosz2013} and this paper have been included in the source-by-source discussions.

\subsection{Likely Binary or Dual AGN} \label{subsec:pairs}

\subsubsection{RFC J0216-0105} \label{subsubsec:0126}

RFC J0216-0105 ($z = 1.49$) is a blazar \citep{massaro2009} and was originally selected by \cite{orosz2013} as a dual AGN candidate based upon an observed offset of 0.179\arcsec{} between the ICRF2 \citep{fey2015} radio position and the optical position given in SDSS DR9 \cite{ahn2012}. Within the BigMAC, this system is cross-listed as a dual AGN candidate and a binary AGN candidate, but at a redshift of 1.49, 0.179\arcsec{} corresponds to $\sim1.6$\,kpc. This source should therefore be considered a kpc-scale dual AGN candidate based upon the radio-optical offset \citep{orosz2013}. We selected RFC J0216-0105 as a candidate BigMAC-RFC multi-AGN based on the RFC X-band morphology. Both S- and X-band RFC observations were available, but the target is unresolved at S-band. Two radio peaks are visible at X-band, with a separation of 5.45 mas. Given the initial morphology, this source was labeled as a likely multi-AGN. This separation corresponds to a 2D projected separation of only $45.8$\,pc, placing the system close to the transition point between dual and gravitationally bound binary SMBHs \citep{bigmacreal}. 

The radio-optical offset calculated between ICRF3 and Gaia DR3 is 0.19 mas, which is smaller than the 5.45 mas measured separation between the radio peaks. This is most likely explained by the existence of a small-scale optical jet, and is not reflective of the potential multiple nature of the system.

At S-band, the unresolved core is radio-loud, with a $Log(L_{Peak})$ = 27.19 W Hz$^{-1}$ and a brightness temperature well in excess of the 10$^5$ K limit for star formation. At X-band, both resolved cores are also radio-loud, with $Log(L_{Peak})$ = 25.95 W Hz$^{-1}$ and 27.01 W Hz$^{-1}$ in the Northern and Southern components, respectively. Both cores are compact, with C values of 1.57 in the north and 1.09 in the south. Both resolved cores at X-band exhibit brightness temperatures well in excess of the limit for star formation.

Finally, the best fit radio spectra for J0216-0105 is the 50\%/50\% North/South flux ratio at S-band, which gives standard power law fit results of $\alpha_{North}$ = -0.42 and $\alpha_{South}$ = -0.25. Both spectral indices are within the range expected for optically thin synchrotron emission from an AGN. Thus, J0216-0105 is a likely dual AGN and could potentially be a binary AGN depending upon the masses of the SMBHs. It is noteworthy that the radio-optical offset observed by \citet{orosz2013} is two orders of magnitude larger than the separation of the two X-band sources in the RFC imaging, so the potential presence of a binary in this system is likely unrelated to the radio-optical offset.

\subsubsection{RFC J0729-1320} \label{subsubsec:0729}

RFC J0729-1320 ($z = 0.34$) is a BL Lac \citep{DAbrusco2014} and was another dual AGN candidate selected by \citet{orosz2013} based on an observed radio-optical offset of 0.4480\arcsec{} (or 2.2\,kpc at $z=0.34$). Within the BigMAC, this system is cross-listed as a dual AGN candidate and a binary AGN candidate, but at a redshift of 0.34, 0.4480\arcsec{} corresponds to $\sim2.2$\,kpc. This source should therefore be considered a kpc-scale dual AGN candidate based upon the radio-optical offset \citep{orosz2013}. We selected RFC J0729-1320 as a candidate BigMAC-RFC multi-AGN based on both the RFC S- and X-band morphologies. The S-band shows significant extension, and the X-band resolves that extension into two clear radio peaks with a separation of 3.41 mas (or $16.5$\,pc at $z=0.34)$. Given the initial morphology, this source was labeled as a likely multi-AGN, and the projected, 2D separation specifically places this system within the regime of gravitationally bound binary AGNs.

The radio-optical offset calculated between ICRF3 and Gaia DR3 is 0.53 mas, which is smaller than the 3.41 mas measured separation between the radio peaks. This is most likely explained by the existence of a small-scale optical jet, and is not reflective of the potential multiple nature of the system.

At S-band, the unresolved extended source is radio-loud, with a $Log(L_{Peak}$) = 26.21 W Hz$^{-1}$ and a brightness temperature well in excess of the 10$^5$ K limit for star formation. At X-band, both resolved cores are also radio-loud, with $Log(L_{Peak}$) = 25.25 W Hz$^{-1}$ and 26.17 W Hz$^{-1}$ in the East and West components, respectively. Both cores are compact, with C values of 1.18 in the Eastern component and 1.15 in the Western component. Both resolved cores at X-band exhibit brightness temperatures well in excess of the limit for star formation.

Finally, the best fit radio spectra for J0729-1320 is the 50\%/50\% East/West flux ratio at S-band, which gives standard power law fit results of $\alpha_{East}$ = 0.08 and $\alpha_{West}$ = 0.02. Both spectral indices are within the range expected for optically thin synchrotron emission from an AGN. Thus, J0729-1320 is a likely gravitationally bound binary AGN similar to CSO 0402+735 \citep{rodriguez2006}, but higher frequency VLBI imaging is required for confirmation. The radio-optical offset is $\sim2$ orders of magnitude larger than the separation of the two radio sources in the RFC imaging, suggesting that the two radio sources are unrelated to the observed radio-optical offset in \citet{orosz2013}.

\subsubsection{RFC J0745+3142} \label{subsubsec:0745}

RFC J0745+3142 ($z = 0.46$) is a Seyfert 1 that was originally selected as a binary AGN candidate in \citet{liu2014} based on velocity offset broad spectroscopic emission lines and evidence of a line-of-sight radial velocity emission line shift. Velocity offset broad emission lines and radial velocity shifts could potentially trace the orbital motion of one AGN in a gravitationally bound sub-pc binary. We selected RFC J0745+3142 as a candidate BigMAC-RFC multi-AGN based on the RFC S-, C-, and X-band morphologies. The S- and C-bands show significant extension, and the X-band resolves that extension into two clear radio peaks with a separation of 2.96 mas (or $17.2$\,pc at $z=0.46$). Given the initial morphology, this source was labeled as a likely multi-AGN, and the projected separation of the two cores places this system within the gravitationally bound binary regime.

The radio-optical offset calculated between ICRF3 and Gaia DR3 is 3.67 mas, making this is the system for which the radio-optical offset and the radio peak separation measurements are the closest in value. While it is possible that the radio-optical offset reflects a genuine detection of multiplicity, this is most likely explained by the existence of a small-scale optical jet.

At S-band, the unresolved extended source is radio-loud, with a $Log(L_{Peak}$) = 26.69 W Hz$^{-1}$ and a brightness temperature well in excess of the 10$^5$ K limit for star formation. At C-band, the same source is also radio-loud, with a $Log(L_{Peak}$) = 26.43 W Hz$^{-1}$. At X-band, both resolved cores are also radio-loud, with $Log(L_{Peak}$) = 26.29 W Hz$^{-1}$ and 25.83 W Hz$^{-1}$ in the Northern and Southern components, respectively. Both cores are compact, with C values of 1.46 in the North and 1.39 in the South. Both resolved cores at X-band exhibit brightness temperatures well in excess of the limit for star formation.

Finally, the best fit radio spectra for J0745+3142 is the 90\%/10\% North/South flux ratio at S-band with the 90\%/10\% North/South flux ratio at C-band. This gives a standard power law fit result of $\alpha_{North}$ = -0.65 and $\alpha_{South}$ = 0.40. Both spectral indices are within the range expected for optically thin synchrotron emission from an AGN. Thus, J0745+3142 is a likely binary AGN, but higher frequency VLBI imaging is needed for confirmation. If J0745+3142 is indeed a binary AGN, the separation is too large for it to explain the velocity offset broad lines and radial velocity shifts observed by \citet{liu2014}. These spectroscopic peculiarities would then more likely be attributable to gas motion within the accretion disk of one (or both) of the AGNs.

\subsubsection{RFC J1305-1033} \label{subsubsec:1305}

RFC J1305-1033 ($z = 0.28$) is a blazar \citep{massaro2009} and is also the well known binary AGN candidate PG 1302-102, which was selected by \citet{graham2015} based on a potential periodicity observed in the Catalina Real-time Transient Survey optical light curve and archival observations. J1305-1033 was selected in this work as a candidate BigMAC-RFC multi-AGN based on the RFC U-band morphology. RFC observations are available at S-, C-, X-, and U-bands, though the target is unresolved at S-, C-, and X-bands. At U-band, the source resolves into two peaks exhibiting a separation of 1.51 mas (which corresponds to $6.34$\,pc at $z=0.28$). Given the initial morphology, this source was labeled as a likely multi-AGN and specifically a candidate gravitationally bound binary AGN.

At S-band, the unresolved source is radio loud, with a $Log(L_{Peak}$) = 26.19 W Hz$^{-1}$ and a brightness temperature well in excess of the limit for star formation. At C- and X-bands, the unresolved source is also radio loud, with a $Log(L_{Peak}$) = 26.09 and 25.96 W Hz$^{-1}$, as well as brightness temperatures indicative of AGN activity. At U-band, the two peaks exhibit $Log(L_{Peak}$) = 25.58 and 26.07 W Hz$^{-1}$ in the Northern and Southern component, respectively. Both cores exhibit brightness temperatures in excess of the limit for star formation, and are compact, with C values of 1.37 and 1.00, respectively.

The radio-optical offset calculated between ICRF3 and Gaia DR3 is 0.37 mas, which is smaller than the 1.51 mas measured separation between the radio peaks. This is most likely explained by the existence of a small-scale optical jet, and is not reflective of the potential multiple nature of the system.

Finally, the best fit radio spectra for J1305-1033 is the S-band 50\%/50\% North/South, C-band 50\%/50\% North/South, and X-band 50\%/50\% North/South flux ratios. This gives a standard power law fit result of $\alpha_{North}$ = -0.39 and $\alpha_{South}$ = 0.77, and a curved power law fit result of $\alpha_{North}$ = -0.32 and $\alpha_{South}$ = -2.69. Both standard power law spectral indices are consistent with optically thin synchrotron emission expected from an AGN. The very steep curved spectral index of the southern source is likely a fitting error, or attributable to the estimated flux ratios used to generate the spectrum. Thus, J1305-1033 is likely a gravitationally bound binary AGN. If confirmed with additional high frequency radio imaging, J1305-1033 would be the second confirmed pc-scale binary AGN and would be the smallest separation binary known, just barely overtaking the well known 7\,pc hosted in CSO 0402+735 \citep{rodriguez2006}. The pc-scale separation of this binary AGN candidate is too large to explain the periodicity reported by \citet{graham2015}, which would have implied a $\sim0.01$\,pc binary instead (see also works by \citealp{tliu2018} demonstrating that the expected periodicity did not continue).

\subsubsection{RFC J2346+3011} \label{subsubsec:2346}

J2346+3011 (no published redshift; $z = 0.5$ assumed throughout) is another dual AGN candidate identified by \cite{orosz2013} based on an observed 0.285\arcsec{} radio-optical offset. Within the BigMAC, this system is cross-listed as a dual AGN candidate and a binary AGN candidate, but at an assumed redshift of 0.5, 0.285\arcsec{} corresponds to $\sim1.74$\,kpc. This source should therefore be considered a kpc-scale dual AGN candidate based upon the radio-optical offset \citep{orosz2013}. We selected J2346+3011 as a candidate BigMAC-RFC multi-AGN based on the RFC S-, C-, and X-band morphologies. RFC observations at all three frequencies are available, and two radio cores (northern and middle) are clearly visible at all frequencies. At S- and C-bands, a third radio source (southern) is visible. At X-band, the northern and middle components exhibit a separation of 18.3 mas, corresponding to a projected physical separation of 111.5 pc, with the assumed redshift of 0.5. A more in-depth visual examination reveals jet emission associated with only the middle source; the southern radio source is likely a jet hotspot. The northern core does not reveal jetted emission. Given the initial morphology, this source is labeled as a likely multi-AGN.

At S-band, all three cores are radio-loud, with a $Log(L_{Peak}$) = 26.02, 26.13, and 25.75 W Hz$^{-1}$ in the Northern, Middle, and Southern core, respectively. At C-band, all cores are radio-loud, with a $Log(L_{Peak}$) = 26.03, 25.90, and 25.02 W Hz$^{-1}$ in the Northern, Middle, and Southern core, respectively. At X-band, both cores are radio-loud, with a $Log(L_{Peak}$) = 25.83 and 25.68 W Hz$^{-1}$ in the Northern and Middle core, respectively. The Southern core is not visible at X-band. For the radio spectral analysis, the $1\sigma$ RMS value was assumed as an upper limit. At all frequencies, both cores exhibit brightness temperatures in excess of that expected for star formation. At X-band, the Northern and Southern cores exhibit C values of 1.25 and 1.32, respectively, indicating some extension.

Finally, the standard power law fit to the spectra for these three cores gives $\alpha_{North}$ = -0.40, $\alpha_{Middle}$ = -0.77, and $\alpha_{South}$ = -2.38, and curved power law spectral indices of $\alpha_{North}$ = 1.63 and $\alpha_{Middle}$ = -0.28. Note that the spectra for the Southern source includes the X-band image RMS as an upper limit on the non-detected source flux. The spectral indices of the Northern and Middle cores are within the range expected for optically thin synchrotron emission from an AGN. However, the Southern source exhibits a steeper spectral index, likely driven by jet activity. Putting all the radio parameters together, the Northern and Middle cores are likely both AGN, given their compactness parameters and spectral indices. The Southern core is likely a jet hotspot, driven by the middle core, judging by the compactness value and steep spectral index. We note that the curved power law fit for the middle core is consistent with an AGN. The significantly more positive fit measured with the curved power law in the northern source is likely attributable to fitting errors. Thus, J2346+3011 is potentially a dual AGN with separation 111.6 pc (assuming $z=0.5$); the separation would be smaller if the object resides at a redshift $z<0.5$, and it therefore could also represent a potential binary AGN. Follow-up spectroscopic observations are required to derive the host redshift and therefore the actual multi-AGN class of this system. The disparity between the RFC radio source separations and the observed radio-optical offset from \citet{orosz2013} suggests the two are unrelated.

\subsubsection{RFC J2347-1856} \label{subsubsec:2347}

J2347-1856 (no published redshift; $z = 0.5$ assumed throughout) is a radio galaxy and another dual AGN candidate identified by \citet{orosz2013} based on a radio-optical offset of 0.244\arcsec{}.  Within the BigMAC, this system is cross-listed as a dual AGN candidate and a binary AGN candidate, but at a redshift of 0.5, 0.244\arcsec{} corresponds to $\sim1.49$\,kpc; this source should therefore be considered a kpc-scale dual AGN candidate based upon the radio-optical offset \citep{orosz2013}. In this work, we selected J2347-1856 as a candidate BigMAC-RFC multi-AGN based on both the RFC S- and X-band morphologies. Both S- and X-band RFC observations were available, and two radio cores are clearly visible at both frequencies. At X-band, the components show an angular separation of 33.5 mas, which corresponds to a projected physical separation of 204.5 parsecs. Given the initial morphology, this source was labeled as a likely multi-AGN.

At S-band, both cores are radio-loud, with a $Log(L_{Peak}$) = 25.90 and 26.00 W Hz$^{-1}$ in the Northern and Southern core, respectively. At X-band, a similar result is seen, with $Log(L_{Peak}$) = 25.41 and 25.31 W Hz$^{-1}$ in the Northern and Southern core, respectively. At both frequencies, both cores exhibit brightness temperatures in excess of the limit for star formation. Finally, the Northern core exhibits a C value of 1.59, while the Southern core exhibits a C-value of 2.19. 

The standard power law fit to the spectra for these two cores gives $\alpha_{North}$ = -0.88 and $\alpha_{South}$ = -1.24, respectively. While the Northern spectral index is within the range expected for optically thin synchrotron emission from an AGN, the Southern spectral index is steeper, as might be exhibited by jet activity. Taken into account, along with the Southern C-value of 2.19, the southern source in J2347-1856 is likely exhibiting smaller scale jet activity associated with the core, and the northern source the AGN core. This morphology has been seen in the case of CSO 0402+379 \citep[][]{rodriguez2006}. Thus, J2347-1856 is likely a small-scale dual AGN; the radio-optical offset reported by \citet{orosz2013} is too large to be explained by the $\sim0.03$\arcsec{} separation seen in the RFC imaging, but larger scale radio jet emission not seen in the RFC imaging could conceivably explain the radio-optical offset.

\subsection{Likely Jet Activity} \label{subsec:jets}

\subsubsection{RFC J0843+4537} \label{subsubsec:0843}

RFC J0843+4537 ($z = 0.19$) is classified as a BL Lac \citep{DAbrusco2019} and a Seyfert 2 \citep{sexton2022} and was another dual AGN candidate selected by \citet{orosz2013} based on a 0.177\arcsec{} radio-optical positional offset. Here we selected RFC J0843+4537 as a candidate BigMAC-RFC multi-AGN based on the RFC S-, C-, and X-band morphologies. However, a more in-depth visual analysis of all three frequencies reveals emission consistent with jet activity. While S-band shows what appears to be two radio peaks, both C- and X-bands highlight extended, complex emission. The southern source is the radio core, while the northern source is the jet hotspot. The full extent of the source at X-band is measured to be 4.14 mas, which corresponds to a projected physical size of 13.2 parsecs. This is significantly smaller than the extent measured in \cite{orosz2013}. Given the initial morphology, this source was labeled as likely jet activity. 

At S-band, the extended source is radio-loud, with a $Log(L_{Peak}$) = 25.19 W Hz$^{-1}$ and a brightness temperature well in excess of the 10$^5$ K limit for star formation. At C-band, the two clear features appear to be the core and a jet hotspot, which exhibit $Log(L_{Peak}$) = 24.90 and 24.19 W Hz$^{-1}$, respectively. The same two features at X-band exhibit $Log(L_{Peak}$) = 24.52 and 24.16 W Hz$^{-1}$, respectively, with compactness values of 2.26 and 2.58. The compactness values in particular mark these sources as extended emission.

Finally, the best fit radio spectra for J0843+4537 is the 10\%/90\% South/North flux ratio at S-band. This gives a standard power law fit result of $\alpha_{Core}$ = -0.08 and $\alpha_{Hotspot}$ = -1.08, and a curved power law fit result of $\alpha_{Core}$ = 0.30 and $\alpha_{Hotspot}$ = 0.83. Both spectral indices are consistent with their designations: the core value matches optically thin synchrotron emission, and the hotspot value is somewhat steeper, as is expected of jet activity. Thus, J0843+4537 is likely jet activity. The observed radio-optical positional offset identified by \citet{orosz2013} may be attributable to larger-scale jet activity beyond the scales probed in this work.

\subsubsection{RFC J1414+4554} \label{subsubsec:1414}

J1414+4554 ($z = 0.458$) is a Seyfert 2 \citep{sexton2022} and is another dual AGN candidate identified by \citet{orosz2013} based on a 0.171\arcsec{} radio-optical positional offset. Here we selected J1414+4554 as a candidate BigMAC-RFC dual AGN based on the S-, C-, and X-band morphologies. RFC observations are available at all three bands, and two radio peaks are clearly visible. The full extent of the source at X-band is measured to be 28.5 mas, which corresponds to a projected physical separation of 91.2 parsecs. However, an in-depth visual examination, particularly at C-band, reveals a jet-like morphology, and thus the source was labeled as jet activity.

At S-band, both cores are radio-loud, with a $Log(L_{Peak}$) = 25.06 and 24.99 W Hz$^{-1}$ in the Northern and Southern core, respectively. At C-band, both cores are radio-loud, with a $Log(L_{Peak}$) = 24.60 and 24.64 W Hz$^{-1}$ in the Northern and Southern core, respectively. At X-band, both cores are radio-loud, with a $Log(L_{Peak}$) = 24.53 and 24.15 W Hz$^{-1}$ in the Northern and Southern core, respectively. At all frequencies, both cores exhibit brightness temperatures in excess of that expected for star formation. At X-band, the Northern and Southern cores exhibit C values of 1.73 and 2.98, respectively, indicating some extension.

Finally, the standard power law fit to the spectra for both cores gives $\alpha_{North}$ = -0.97 and $\alpha_{South}$ = -1.46, as well as a curved power law spectral index of $\alpha_{South}$ = 0.64. While the Northern spectral index is within the range expected for optically thin synchrotron emission from an AGN, the Southern spectral index is steeper, as might be exhibited by jet activity. Taken into account, along with the compactness values, the southern source is most likely driven by jet activity. It is possible that the northern source is either an AGN core or similarly driven by jet activity. The flatter curved power law spectral index is likely a fitting error, or attributable to the estimate flux ratios used to generate the spectrum. Thus, J1414+4554 is likely jet activity; the angular separations found between the components in the RFC imaging are roughly an order of magnitude smaller than the radio-optical offset reported by \citet{orosz2013}, but it is conceivable that larger scale jet emission could be contributing to the observed 0.171\arcsec{} offset.

\subsubsection{RFC J1451+1343} \label{subsubsec:1451}

J1451+1343 (no published redshift; $z = 0.5$ assumed throughout) is a radio galaxy and another dual AGN candidate selected by \citet{orosz2013} based on an observed radio-optical offset of 0.42\arcsec{}. We  selected J1451+1343 here as a candidate BigMAC-RFC dual AGN based on S-, C-, and X-band morphologies. RFC observations are available at all three bands, and two radio peaks are clearly visible. At X-band, the full source extent is measured to be 26.7 mas, which corresponds to a projected physical separation of 162.9 parsecs, at the assumed redshift of 0.5. However, an in-depth visual examination, particularly at S- and C-bands, reveals a jet-like morphology, and thus the source was labeled as jet activity.

At S-band, both cores are radio-loud, with a $Log(L_{Peak}$) = 26.11 and 26.29 W Hz$^{-1}$ in the East and West core, respectively. At C-band, both cores are radio-loud, with a $Log(L_{Peak}$) = 25.98 and 26.09 W Hz$^{-1}$ in the East and West core, respectively. At X-band, both cores are radio-loud, with a $Log(L_{Peak}$) = 25.61 and 25.81 W Hz$^{-1}$ in the East and West core, respectively. At all frequencies, both cores exhibit brightness temperatures in excess of that expected for star formation. At X-band, the East and West cores exhibit C values of 2.10 and 1.24, respectively, indicating some extension.

Finally, the standard power law fit to the spectra for both cores gives $\alpha_{East}$ = -0.78 and $\alpha_{West}$ = -0.82, and a curved power law spectral index of $\alpha_{West}$ = 0.30. Both East and West cores exhibit spectral indices expected for optically thin synchrotron emission from an AGN. This is particularly interesting in the context of the clear jet-like morphologies and the compactness parameters. It is possible that one of the peaks is a core, and the other a jet hotspot. It is also possible that both are AGN cores exhibiting jet activity, which would make this a dual AGN system with a separation of 162.9 pc between the cores (assuming $z=0.5$). However, given the lack of confirmation, J1451+1343 remains likely jet activity. Sufficiently deep, higher frequency VLBI imaging would shed further light on the nature of these radio sources. Furthermore, the projected separation of the cores will be smaller if the system resides at a redshift $z<0.5$, potentially making it a binary AGN candidate as well. Follow-up spectroscopic observations are required to derive the redshift of this system and properly classify it as a dual or binary AGN candidate. The observed separations in S, C, and X band are, however, over an order of magnitude too small to explain the radio-optical offset reported in \citet{orosz2013}; larger scale jet emission not observed in the RFC imaging could explain the radio-optical offset.

\subsubsection{RFC J1503+0917} \label{subsubsec:1503}

J1503+0917 (no published redshift; $z = 0.5$ assumed throughout) is a radio galaxy and another dual AGN candidate selected by \citet{orosz2013} based on an observed radio-optical positional offset of 0.291\arcsec{}. In this work, we selected J1503+0917 as a candidate BigMAC-RFC dual AGN based on both the S- and X-band morphologies. RFC observations are available at both frequencies. At X-band, the full extent of the source is measured to be 13.1 mas, which corresponds to a projected physical separation of 79.9 parsecs at the assumed redshift of 0.5. However, a more in-depth visual analysis of the source reveals jet-like structure, especially at X-band. Given the initial morphology, this source was labeled as likely jet activity.

At S-band, both Northern and Southern cores are radio-loud, with a $Log(L_{Peak}$) = 26.53 and 25.68 W Hz$^{-1}$, respectively. At X-band, both Northern and Southern cores are radio-loud, with a $Log(L_{Peak}$) = 26.02 and 24.79 W Hz$^{-1}$, respectively. At both frequencies, both cores exhibit brightness temperatures in excess of those expected for star formation. The Northern core exhibits a C value of 1.77, while the Southern core exhibits a C-value of 2.83. 

Finally, the standard power law fit to the spectra for these two cores gives $\alpha_{North}$ = -0.91 and $\alpha_{South}$ = -1.59. While the Northern spectral index is within the range expected for optically thin synchrotron emission from an AGN, the Southern spectral index is steeper, as might be exhibited by a jet hotspot. Given the compactness value of the Southern Core, it is likely that the emission is attributable to jet activity. Thus, J1503+0917 is likely jet activity. The observed separation at S- and X-band is over an order of magnitude too small to explain the radio-optical offset observed by \citet{orosz2013}, but larger-scale jet activity not observed in the RFC imaging could contribute to this offset.

\subsection{The ICRF Perspective} \label{sec:ICRFdisc}

As previously discussed, the ICRF is realized by VLBI observations of extragalactic radio sources, namely quasars. It depends on the critical assumption that these radio sources are compact, stable, and free from structural, temporal, or positional variability. However, in practice, intrinsic structure can shift the apparent centroid of emission as observed by VLBI, leading to positional offsets that evolve over time \citep[][]{charlot2020, makarov2022, titov2022}. This represents a significant source of error in the ICRF and has motivated ongoing efforts to identify, monitor, and maintain a subset of compact, structurally stable defining sources from which the celestial reference frame is realized \citep[][]{charlot2020, charlot2007, liu2022}.

Multi-AGN and AGN with milliarcsecond-scale structure have been singled out as an increasingly recognized source of error \citep[][]{shabala2015, anderson2018, titov2022, bourda2016}, producing distinct or blended radio emission components within a single host galaxy \citep[][]{rodriguez2006,fu2011, bondi2016}. When unresolved or partially resolved by VLBI, these components can induce apparent position shifts, astrometric jitter, and time-variable centroid motion as the relative brightnesses of the components change. Further, studies have shown that radio-optical positional offsets are often associated with complex intrinsic AGN structures \citep[][]{mignard2016, kovalev2017, petrov2019, secrest2022, petrov2017, liu2021, popkov2025, titov2022, lambert2024, xu2021}. These offsets can degrade the stability of the reference frame and introduce biases in the radio-optical frame tie. As a result, they represent a critical population to identify and exclude in future realizations of the ICRF, especially as astrometric precision approaches the microarcsecond regime.

All ten BigMAC-RFC sources presented in this paper are ICRF sources. All ten show either clear intrinsic multi-AGN structure or other complex intrinsic structure most likely associated with small-scale jet activity. We note in the cases of the jet activity that it is particularly bright, increasing the likelihood of apparent position shifts in VLBI observations. Thus, it is recommended that all ten BigMAC-RFC sources analyzed here be removed from future iterations of the ICRF.

We note that future studies will expand on this work to include BigMAC/RFC sources that are not likely to be multi-AGN. This will include a significant population of sources exhibiting intrinsic complex structure (in most cases, likely driven by jet activity), most of which will likely be identified for removal from future iterations of the ICRF. Though this work is left for future papers, the present work has already highlighted the number of sources in the current iteration of the ICRF that are likely to experience apparent positional shifts, even up to the higher frequencies. 

\subsection{The Evolving BigMAC Sample \& Future Work} \label{subsec:bigmacevol}

Of the ten BigMAC-RFC targets examined here, six exhibit compact radio morphologies consistent with likely multi-AGN. Confirmation will require additional dedicated follow-up observations capable of demonstrating that the detected radio components remain compact at higher frequencies, and are thus associated with AGN rather than alternative explanations. If confirmed, these targets would represent six new multi-AGN, including some likely binary AGN, adding to the 156 confirmed systems (including one confirmed binary AGN) in the BigMAC DR~1.0 repository and those identified in more recent studies \citep[e.g.,][]{chen_hst,Mannucci_2022,hwang2020varstrometry}. While some fraction of the remaining BigMAC candidates are also likely to represent genuine AGN pairs, this work further illustrates that confirming close SMBH systems remains observationally challenging.

\subsubsection{Implications for Close Multi-AGN Searches} \label{subsubsec:close}

One of the most notable results of this study is that the compact radio separations measured with VLBI are frequently much smaller than the separations inferred from the observations that originally motivated inclusion of these systems in BigMAC. While this possibly represents a contradiction, it also likely reflects the fundamentally different spatial scales probed by the respective selection techniques. The studies represented within BigMAC identify merger candidates using a wide range of diagnostics, including double-peaked emission lines, disturbed morphologies, infrared or optical excesses, multiple optical nuclei, and kpc-scale radio structures. These signatures are effective at identifying interacting galaxies and candidate SMBH pairs, but they should not be interpreted as direct predictors of the parsec-scale separation of compact radio cores. Instead, the VLBI observations presented here demonstrate that some of these systems harbor compact radio components on substantially smaller scales than previously recognized, emphasizing the importance of multi-scale observations for characterizing the late stages of SMBH mergers. It is similarly possible for both to be true; a system identified in BigMAC as a kiloparsec-scale multi-AGN might also host a parsec-scale binary AGN. 

Recent efforts have increasingly focused on identifying sub-kpc and parsec-scale SMBH pairs using complementary observational techniques. Targeted VLBI campaigns have identified several promising close dual AGN candidates \citep[][]{koss2023}, while high-resolution optical and near-infrared observations have revealed dual nuclei at comparable physical scales \citep[][]{voggel2022}. Likewise, systematic searches such as varstrometry-based searches \citep[][]{hwang_initial,schwartzman2024} continue to expand the population of candidate systems requiring detailed follow-up. The present work complements these efforts by leveraging the Radio Fundamental Catalog to perform a uniform VLBI investigation of candidates drawn from a literature-wide compilation rather than from a single selection strategy.

Unlike studies that primarily target systems already suspected to host sub-kpc nuclei, the BigMAC-RFC sample demonstrates that compact radio structures can emerge in systems originally selected using diagnostics sensitive to much larger merger scales. The radio separations measured here are often factors of several to orders of magnitude smaller than those inferred from the original discovery observations, illustrating that VLBI follow-up can reveal the final stages of SMBH pairing that remain inaccessible to lower-resolution observations.

Beyond expanding the overall census of confirmed multi-AGN, even confirmation of a subset of the systems presented here as genuine binary AGN would substantially increase the extremely small population currently known, which presently consists of a single confirmed binary AGN \citep[][]{rodriguez2006}. Such discoveries would provide valuable constraints on SMBH pairing efficiencies, merger timescales, and the evolution of AGN during the final stages preceding coalescence.

\subsubsection{The BigMAC-RFC Sample as a Resource} \label{subsubsec:resource}

Although confirming individual systems remains challenging, the BigMAC repository provides an unusually broad foundation for studying the multi-AGN population. As a literature-based compilation, it encompasses the full range of selection techniques employed prior to 2020, including optical, infrared, radio, and multiwavelength searches spanning a broad range of redshifts, merger stages, and galaxy mass ratios. The sample naturally inherits the selection biases of the studies from which it was assembled. Most notably, this is represented by the historical emphasis on optically selected systems. Nevertheless, it represents one of the most comprehensive collections of candidate and confirmed multi-AGN currently available. Continued characterization of the radio properties of these systems will provide an opportunity to evaluate how different selection techniques perform at identifying genuine close SMBH pairs.

As demonstrated throughout this work, the Radio Fundamental Catalog is a critical resource for extending these studies into the parsec regime. Characterizing the radio properties of candidate multi-AGN cannot rely solely on (sub-)arcsecond observations, as VLBI provides the angular resolution required to resolve the smallest separations currently accessible. For the closest binary AGN, depending on redshift, VLBI observations may provide the only direct probe capable of resolving distinct compact radio cores. At the same time, the RFC alone cannot fully characterize the BigMAC sample. A non-detection does not necessarily imply the absence of a multi-AGN, as VLBI observations require sufficiently compact radio emission and are inherently insensitive below a certain luminosity threshold, depending on the observatory in question. Furthermore, the relatively small VLBI field of view limits sensitivity to widely separated companions within a single pointing. For systems with larger expected separations, the most effective strategy is therefore to combine VLBI observations with high-resolution optical, infrared, and longer-baseline radio imaging capable of identifying distinct galactic nuclei over much larger fields of view. Together, these complementary observations provide sensitivity across the full range of separations represented within the BigMAC sample.

An additional consideration is the population of multi-AGN in which one or both SMBHs are radio quiet. For isolated AGN, only approximately $\sim10\%$ are traditionally classified as radio loud \citep[][]{osterbrock1993}, although the applicability of this fraction to merging AGN remains uncertain. It is therefore likely that a substantial fraction of the BigMAC sample will ultimately require confirmation through optical, infrared, or X-ray observations rather than radio observations alone. Multiwavelength follow-up of both the broader BigMAC repository and the BigMAC-RFC subsample will therefore remain an important component of future work.

While this paper has focused on the ten BigMAC-RFC targets judged most likely to host multiple AGN, the complete BigMAC-RFC sample contains 241 candidate and confirmed systems, all of which warrant detailed radio and multiwavelength investigation. The RFC observations of this larger sample are already available and will form the basis of a subsequent study examining the radio properties of the full BigMAC-RFC population. These analyses will be complemented by lower-resolution pointed and archival radio observations, together with optical spectroscopy to establish redshifts and confirm physical associations between candidate nuclei. They will also enable systematic searches for radio-emitting companions associated with AGN initially identified through optical or mid-infrared selection techniques, further expanding the parameter space over which close SMBH pairs can be identified and confirmed.

\section{Summary and Conclusions} \label{sec:summary}

This paper represents the first VLBI study of the BigMAC-RFC sample by combining the literature-complete Big Multi-AGN Catalog with archival observations from the Radio Fundamental Catalog. This work demonstrates that existing VLBI datasets provide an efficient means of identifying and characterizing parsec-scale AGN multiplicity while simultaneously evaluating the impact of intrinsic radio structure on precision astrometry.

The principal results of this work are summarized below:

\begin{itemize}
    \item Cross-matching the BigMAC with the RFC has produced 241 candidate and confirmed multi-AGN systems with available VLBI imaging, providing a large archival sample for future systematic study.
    \item Visual inspection of the RFC images identified ten sources exhibiting multiple compact radio components and no obvious large-scale morphological evidence requiring immediate rejection as multi-AGN candidates.
    \item Combining morphology, brightness temperature, compactness, spectral indices, luminosities, and existing multiwavelength information has highlighted six systems are most consistent with candidate dual or binary AGN, while four are more naturally explained by compact jet structure.
    \item Multiple candidates exhibit projected separations of only a few to a few tens of parsecs. In particular, J1305-1033 exhibits a projected separation of approximately 6.3 pc and, if confirmed, would become the smallest known binary AGN.
    \item For many targets, the directly measured radio component separations are factors of several to orders of magnitude smaller than the scales inferred from the observations that originally motivated their classification as dual or binary AGN candidates. This demonstrates that lower-resolution selection techniques can identify systems that ultimately host much more compact SMBH pairs.
    \item The RFC represents an underutilized resource for searching for close SMBH pairs, enabling parsec-scale investigations without requiring new observations for every candidate.
    \item All ten analyzed systems are ICRF sources and exhibit either intrinsic source multiplicity or significant compact radio structure capable of introducing astrometric centroid shifts. These objects should therefore be carefully evaluated before inclusion in future realizations of the celestial reference frame.
    \item This paper represents only the first analysis of the BigMAC-RFC sample. Extending the radio analysis to all 241 cross-matched sources, obtaining higher-frequency and multi-epoch VLBI observations, incorporating Gaia astrometric diagnostics, and performing targeted optical, infrared, and X-ray follow-up will substantially improve confirmation rates and refine the census of close SMBH pairs.
\end{itemize}

This work highlights the unique power of combining literature-based multi-AGN catalogs with archival VLBI observations. As the population of confirmed binary AGN remains extremely small, even confirmation of a subset of the candidates presented here would significantly advance our understanding of SMBH pairing, merger-driven galaxy evolution, and the final stages of black hole coalescence. At the same time, identifying structurally complex radio sources is increasingly important for maintaining the long-term stability of the International Celestial Reference Frame as astrometric precision approaches the microarcsecond regime.

\begin{acknowledgements}
This research made use of Astropy, a community-developed core Python package for Astronomy (\cite{astropy}), $\mathrm{TOPCAT}$ (\cite{topcat}), the Common Astronomy Software Application (\cite{casanew2022}), and the Python Blob Detector and Source Finder (\cite{pybdsf}). 
Funding for the Sloan Digital Sky Survey IV has been provided by the Alfred P. Sloan Foundation, the U.S. Department of Energy Office of Science, and the Participating Institutions.
The National Radio Astronomy Observatory is a facility of the National Science Foundation operated under cooperative agreement by Associated Universities, Inc. 
Basic research in radio astronomy at the U.S. Naval Research Laboratory is supported by 6.1 Base Funding. 
\end{acknowledgements}

\vspace{5mm}
\facilities{VLA (NRAO), Sloan, VLBA}

\software{Astropy (\cite{astropy}), CASA (\cite{casanew2022}), PyBDSF (\cite{pybdsf}), $\mathrm{TOPCAT}$ (\cite{topcat})}

\clearpage
\bibliography{citation.bib} 

\bibliographystyle{aasjournal}

\section{Appendix A: Radio Images} \label{sec:radioims}

Figures \ref{fig:0216} - \ref{fig:2347} are presented below. In all images, the observational frequencies are listed in the upper left, scale bars are shown in the lower right, beams are shown in the lower left, and color bars (units of Janskys) are show on the right. Contours are drawn in light grey. Positive contours are solid at levels: 3$\sigma$, 15$\sigma$, 27$\sigma$, 39$\sigma$, and 51$\sigma$. Negative contours are dashed at levels: -9$\sigma$, -6$\sigma$, and -3$\sigma$. The blue and green crosses note the positions of the two components at the highest available frequency (X- or U-bands). Axes show angular offsets in right ascension and declination relative to the reference position.

\begin{figure*}[ht!]
    \centering
    \includegraphics[width=8cm]{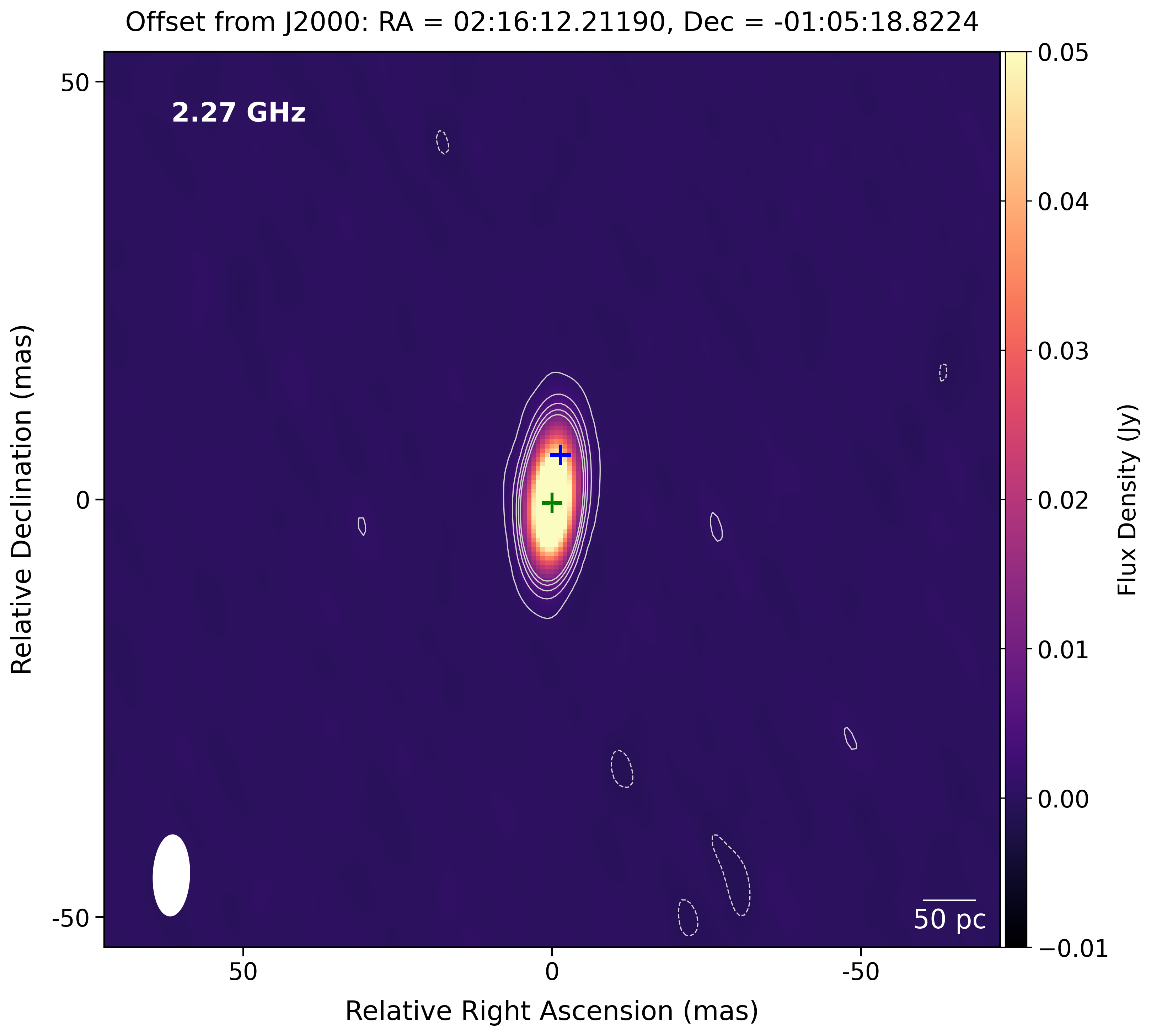}
     \includegraphics[width=7.80cm]{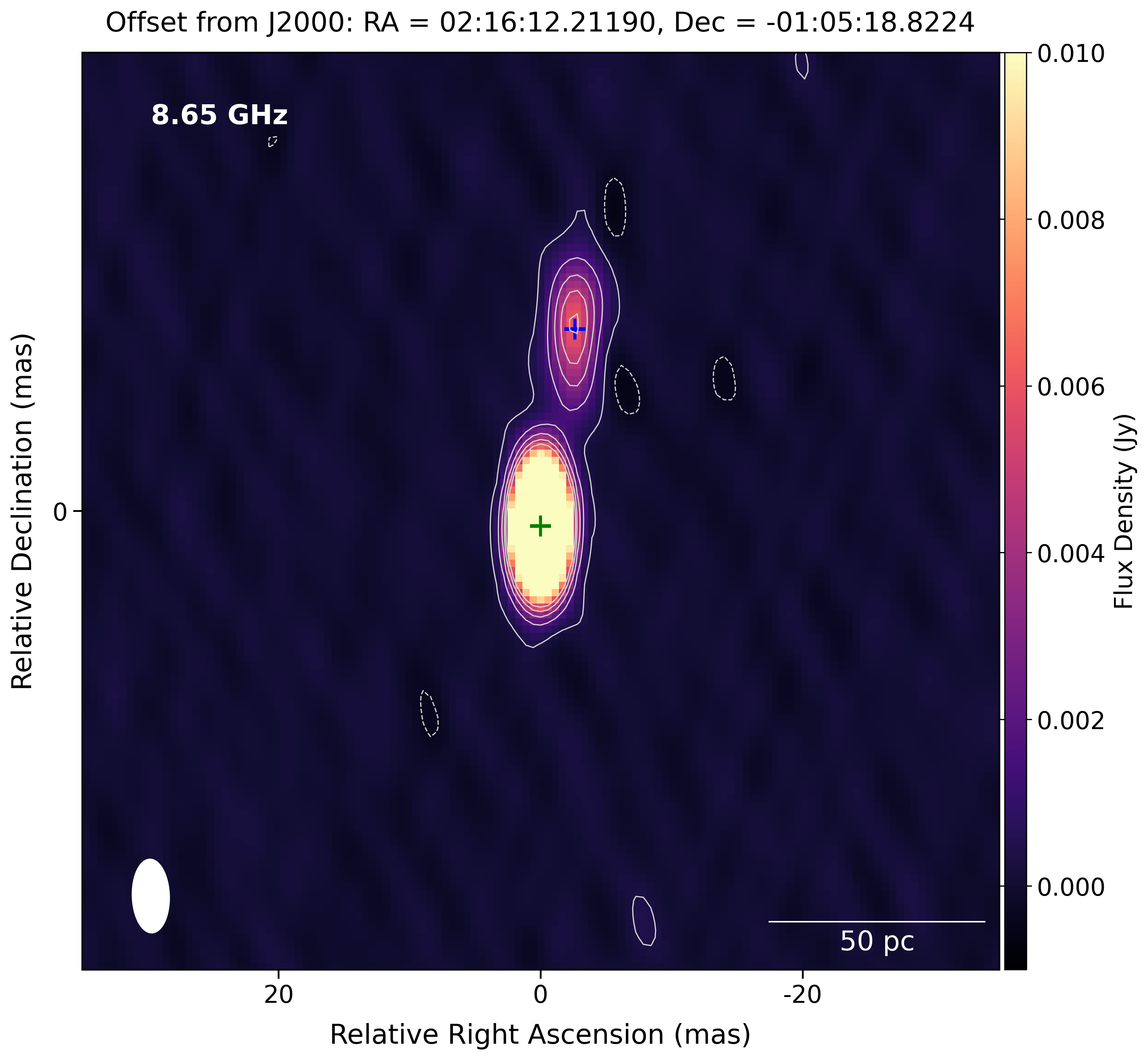}
        \vspace*{-3mm}
          \caption{VLBA 2.27 (Left) and 8.65 (Right) GHz images of J0216-0105. Scalebars are located in the lower left hand corners, beams are shown in the lower right hand corners. Colorbars in Jys are displayed to the right of each image. Contours are drawn in light grey. Positive contours are solid at levels: 3$\sigma$, 15$\sigma$, 27$\sigma$, 39$\sigma$, and 51$\sigma$. Negative contours are dashed at levels: -9$\sigma$, -6$\sigma$, and -3$\sigma$. The blue and green crosses note the positions of the two components at the highest available frequency (X- or U-bands). Axes show angular offsets in right ascension and declination relative to the reference position.}
   \label{fig:0216}
\end{figure*}

\begin{figure*}[ht!]
    \centering
    \includegraphics[width=8cm]{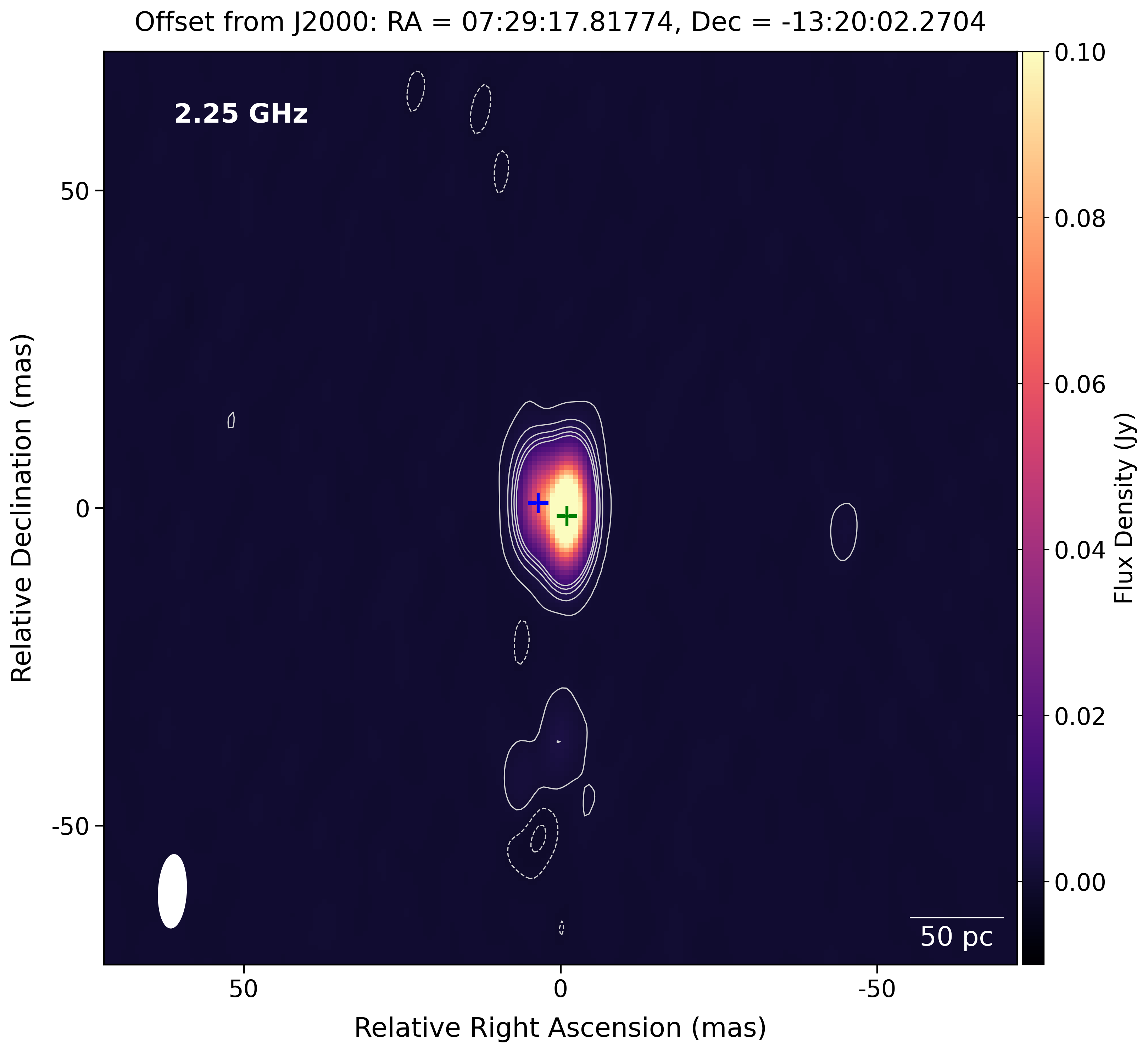}
     \includegraphics[width=8.05cm]{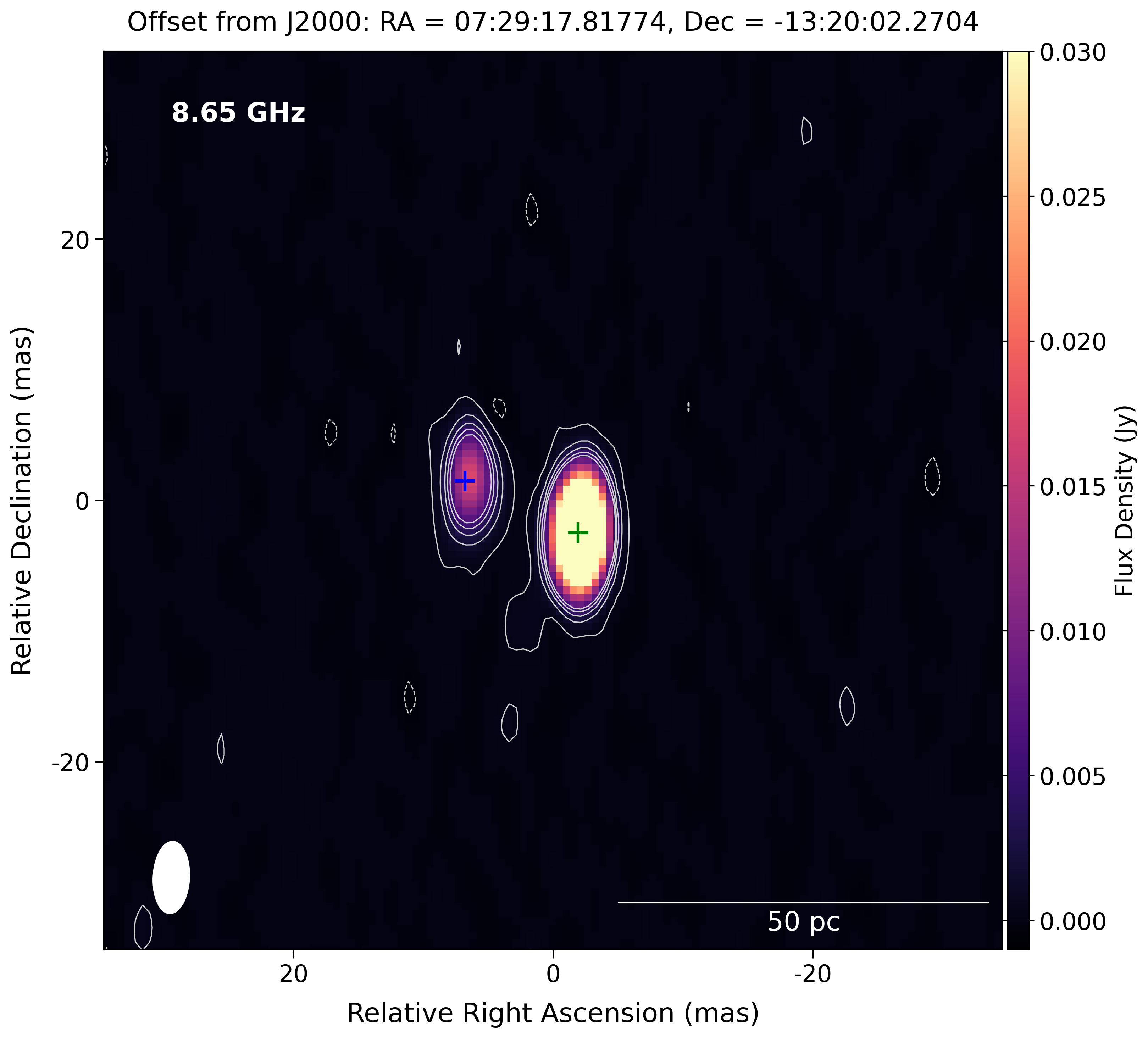}
        \vspace*{-3mm}
          \caption{VLBA 2.25 (Left) and 8.65 (Right) GHz images of J0729-1320. Scalebars are located in the lower left hand corners, beams are shown in the lower right hand corners. Colorbars in Jys are displayed to the right of each image. Contours are drawn in light grey. Positive contours are solid at levels: 3$\sigma$, 15$\sigma$, 27$\sigma$, 39$\sigma$, and 51$\sigma$. Negative contours are dashed at levels: -9$\sigma$, -6$\sigma$, and -3$\sigma$. The blue and green crosses note the positions of the two components at the highest available frequency (X- or U-bands). Axes show angular offsets in right ascension and declination relative to the reference position.}
   \label{fig:0729}
\end{figure*}

\begin{figure*}[ht!]
    \centering
    \includegraphics[width=8cm]{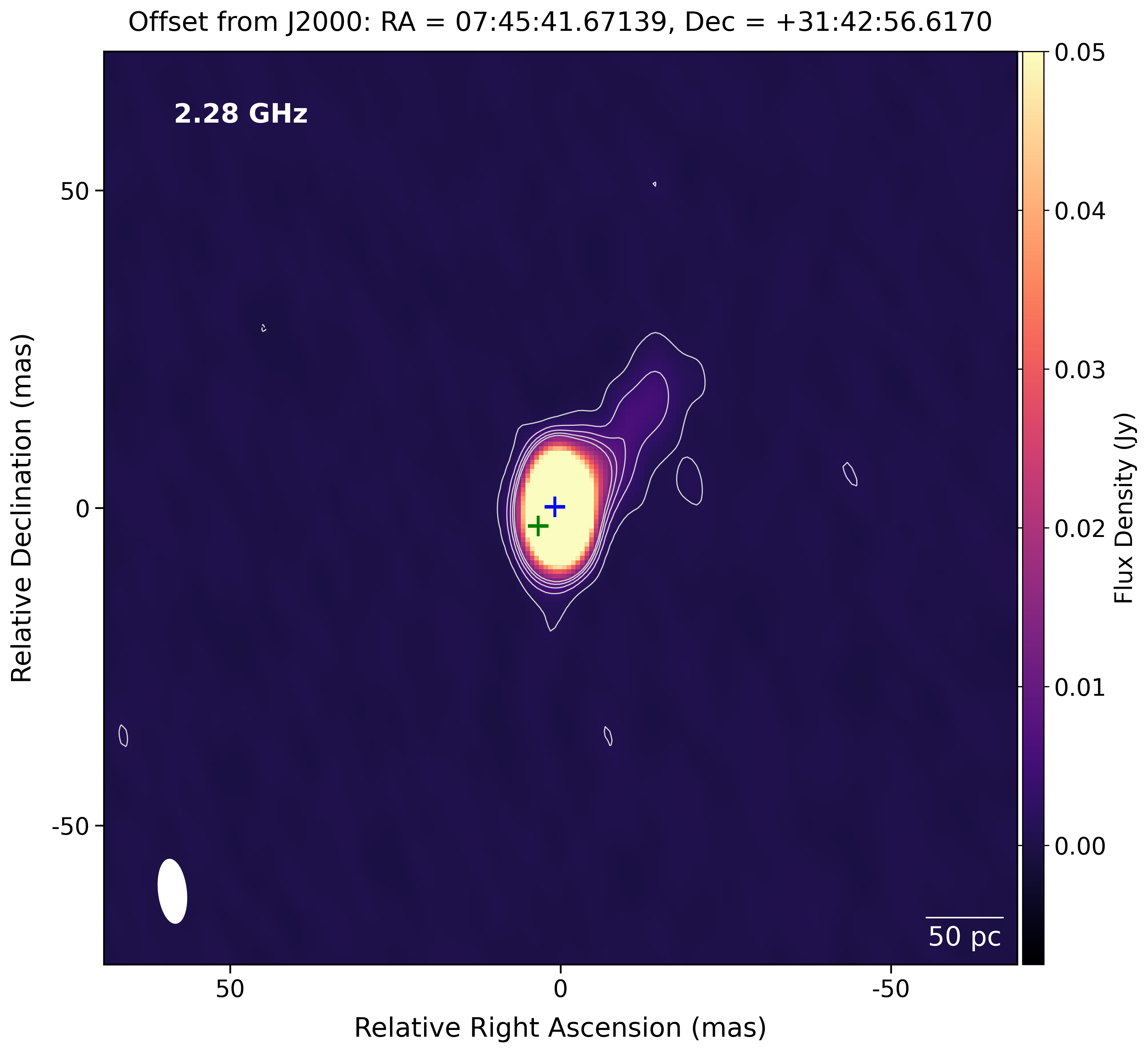}
     \includegraphics[width=8.20cm]{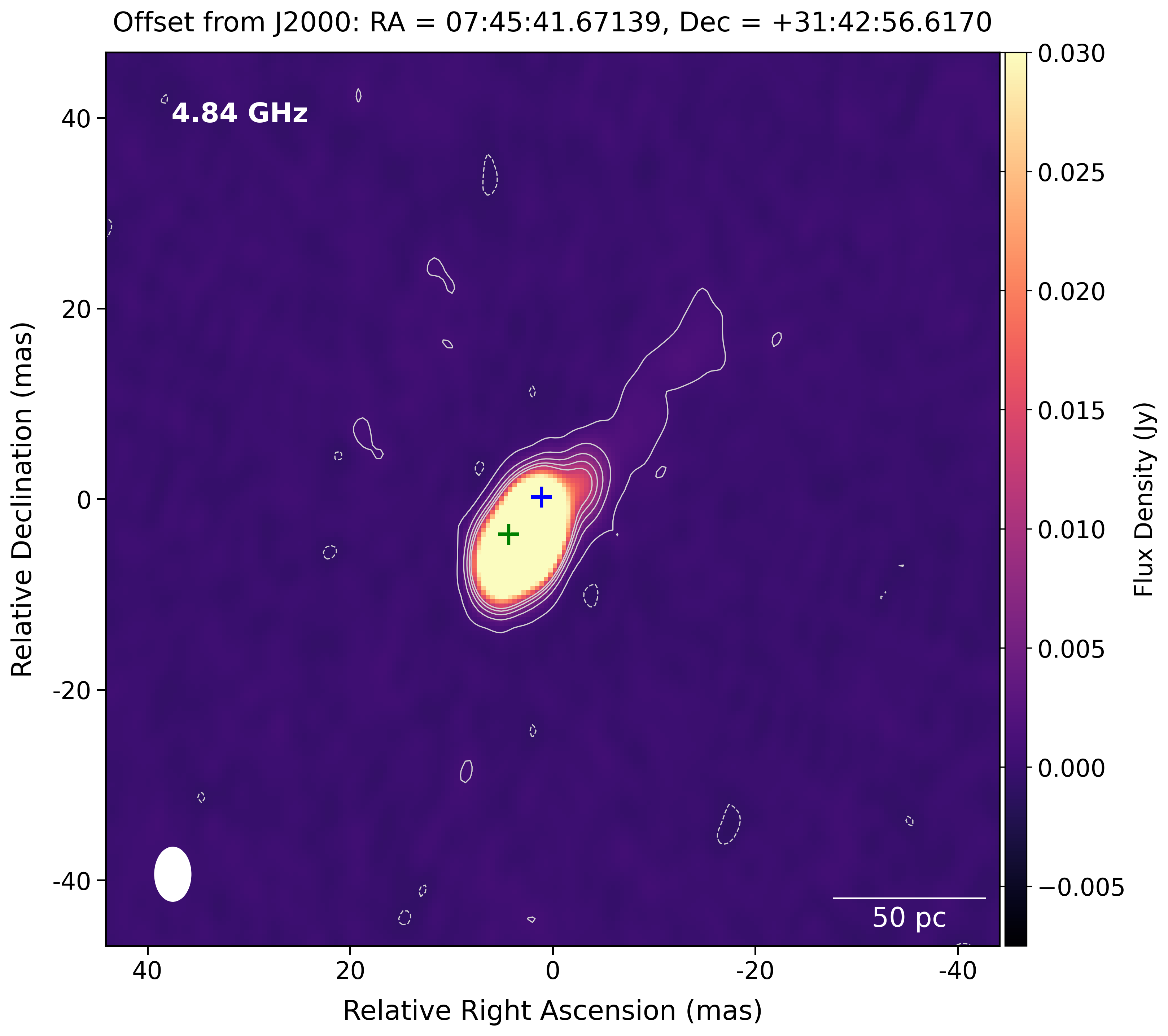}
     \includegraphics[width=8cm]{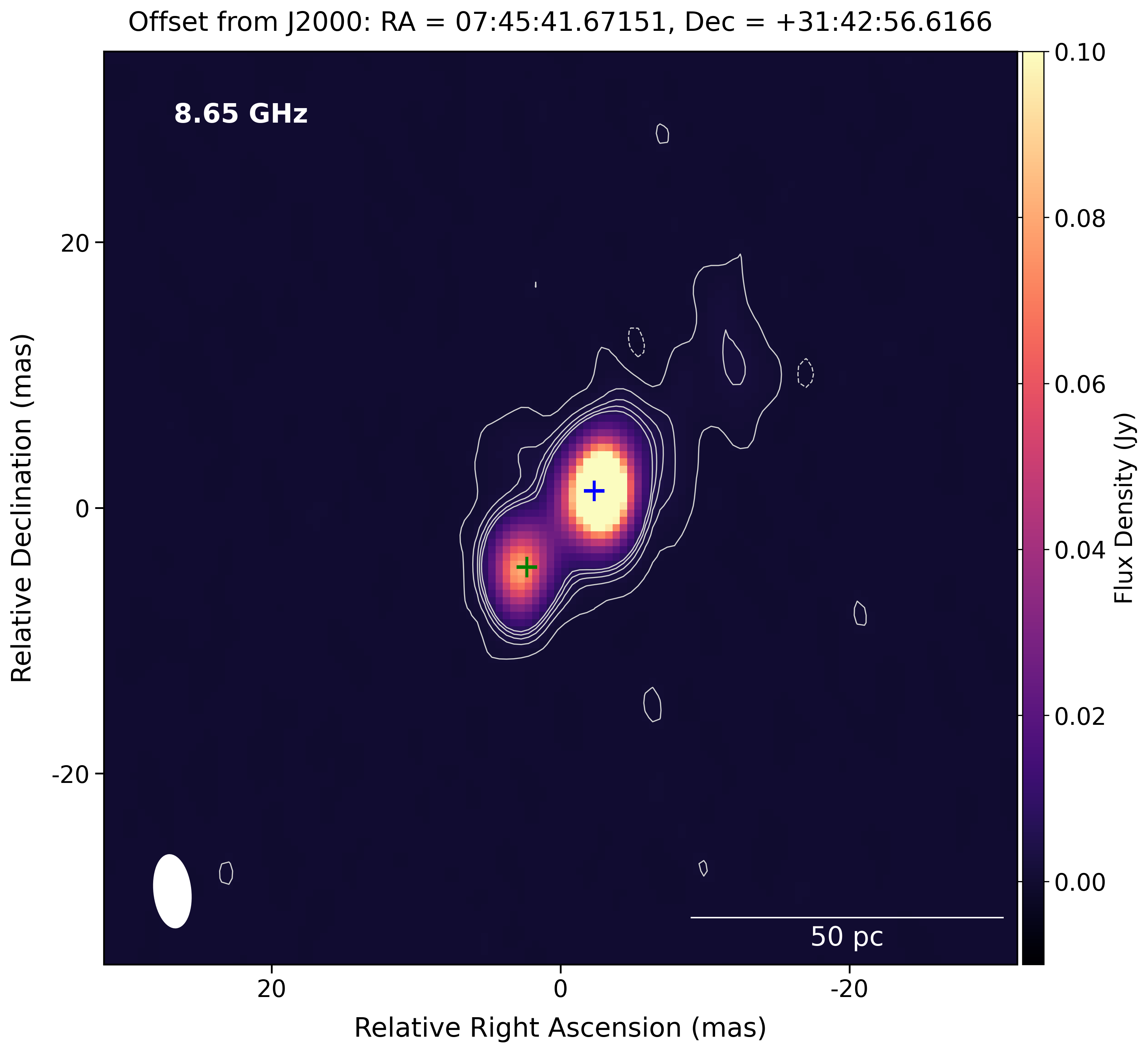}
        \vspace*{-3mm}
          \caption{VLBA 2.28 (Top Left), 4.84 (Top Right), and 8.65 (Bottom) GHz images of J0745+3142. For plot description, see Figure \ref{fig:0216}.}
   \label{fig:0745}
\end{figure*}

\begin{figure*}[ht!]
    \centering
    \includegraphics[width=8cm]{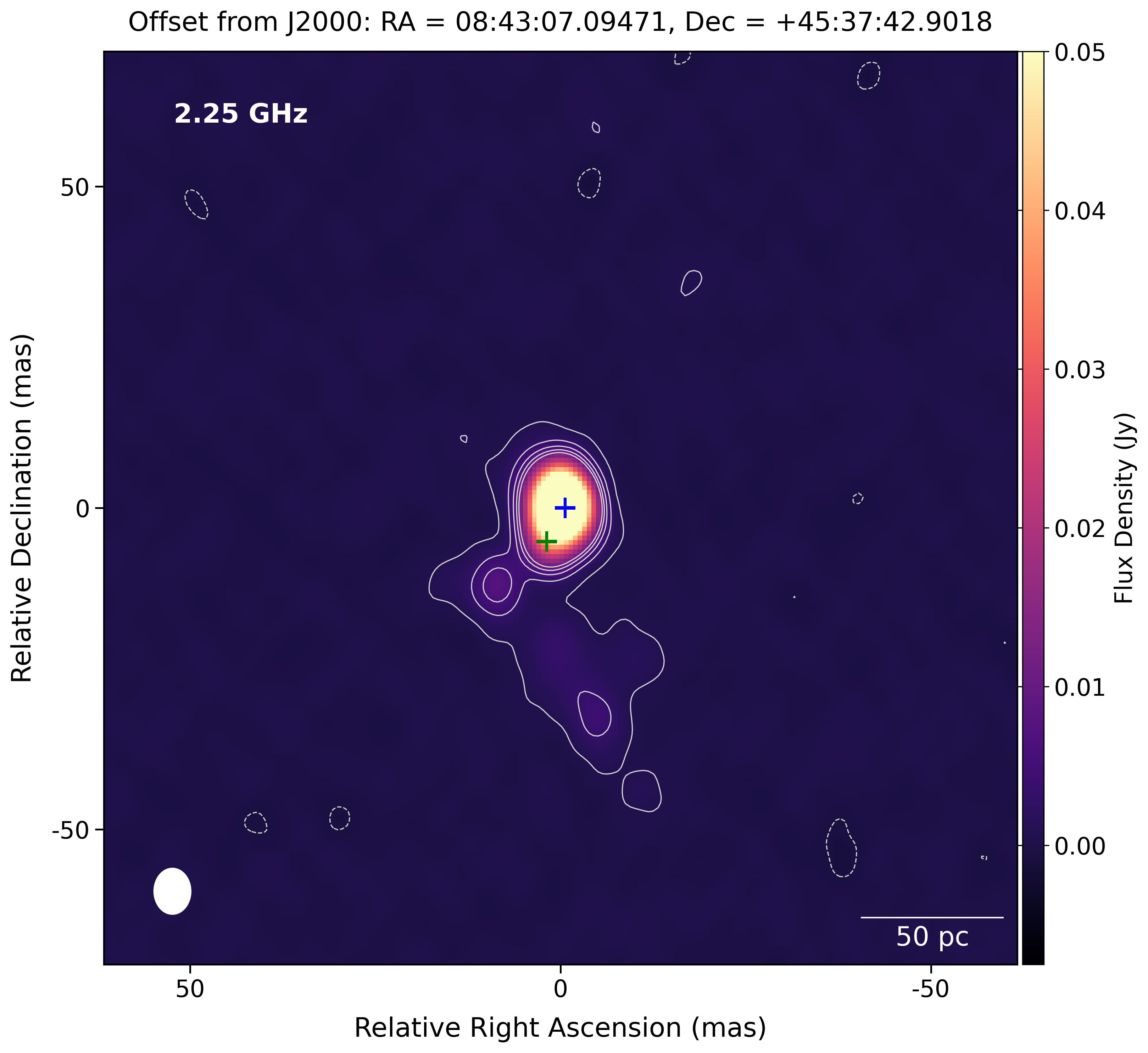}
     \includegraphics[width=8cm]{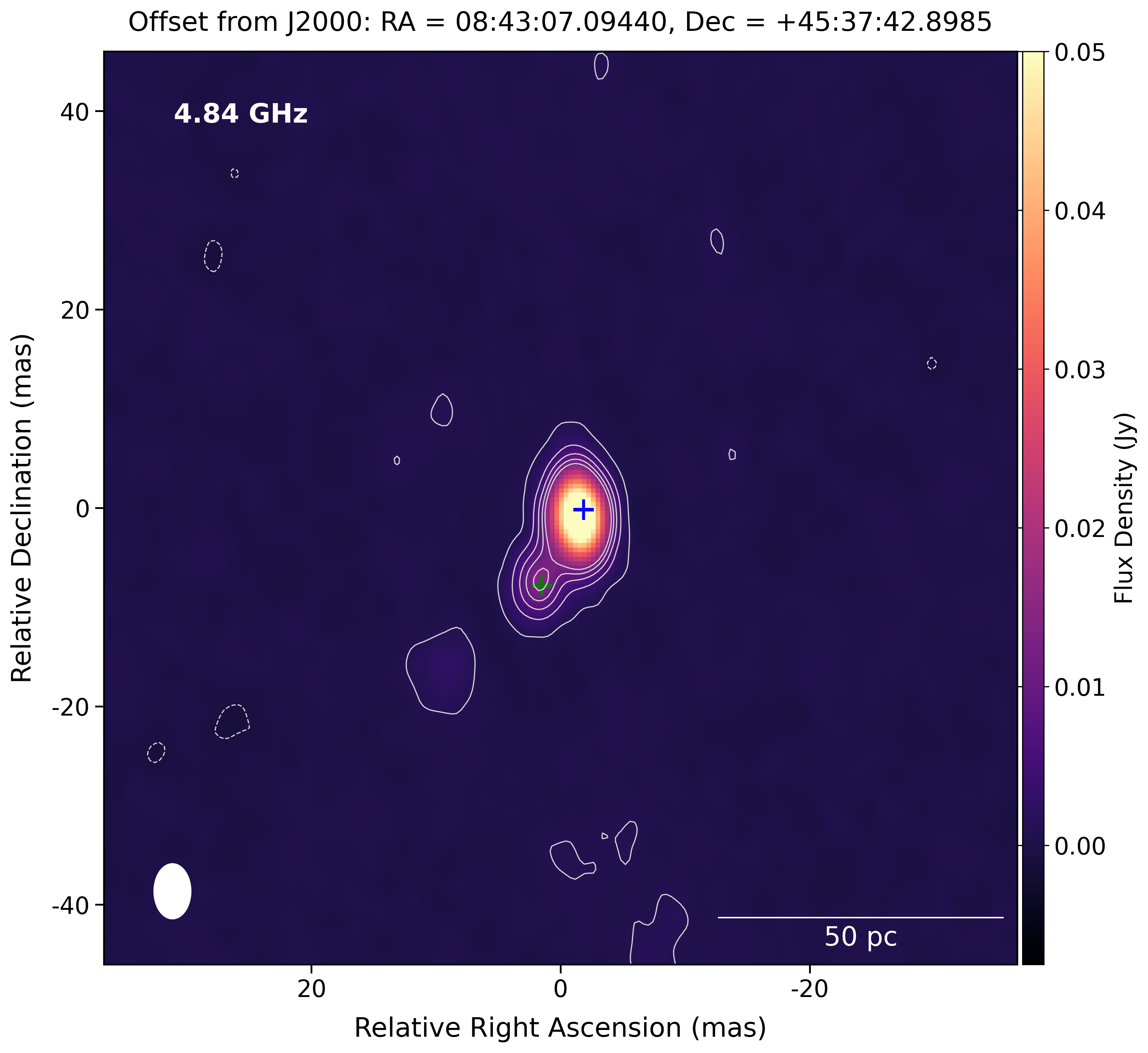}\\
     \includegraphics[width=8.2cm]{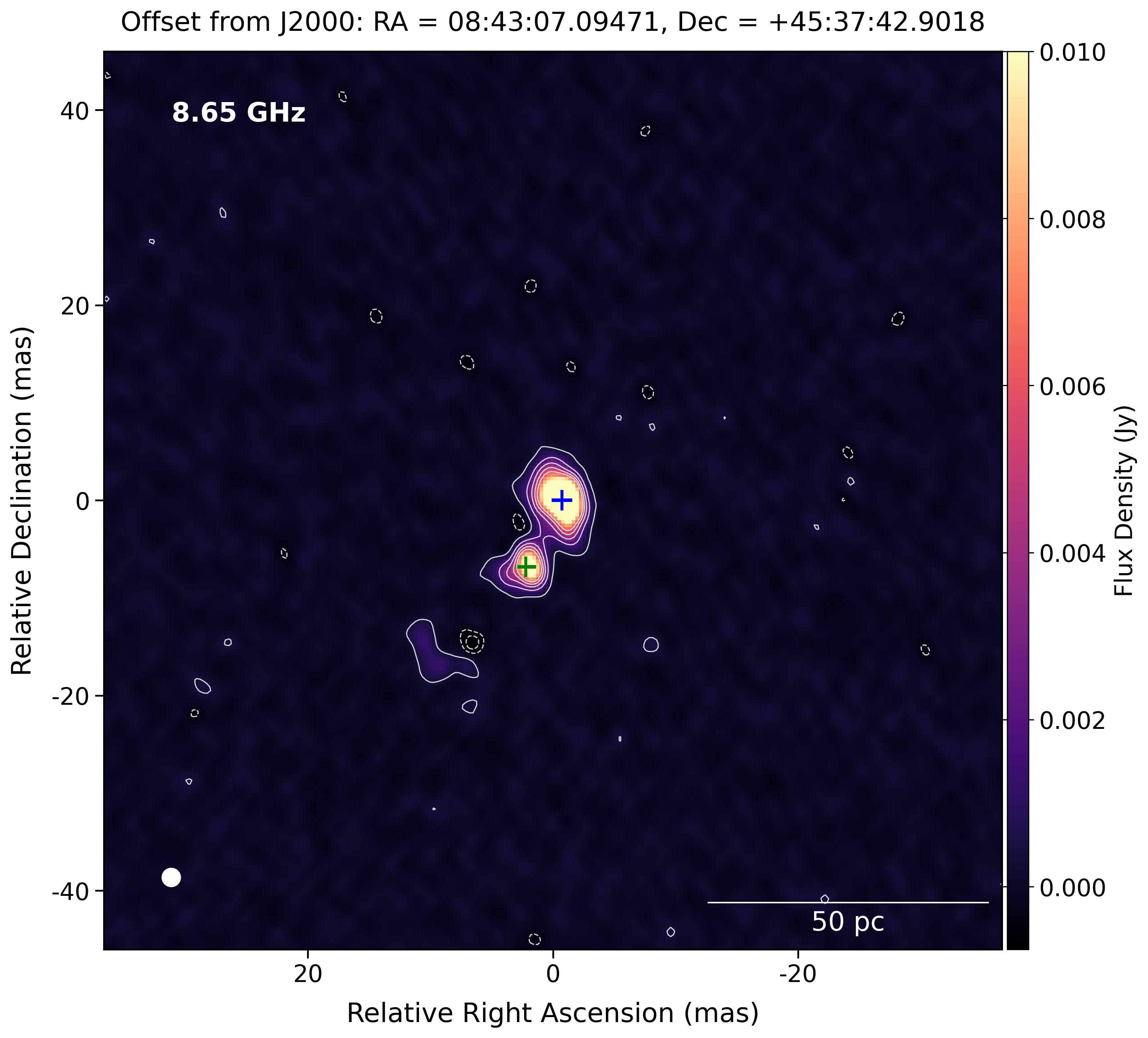}
        \vspace*{-3mm}
          \caption{VLBA 2.25 (Top Left), 4.84 (Top Right), and 8.65 (Bottom) GHz images of J0843+4537. For plot description, see Figure \ref{fig:0216}.}
   \label{fig:0843}
\end{figure*}

\begin{figure*}[ht!]
    \centering
    \includegraphics[width=8cm]{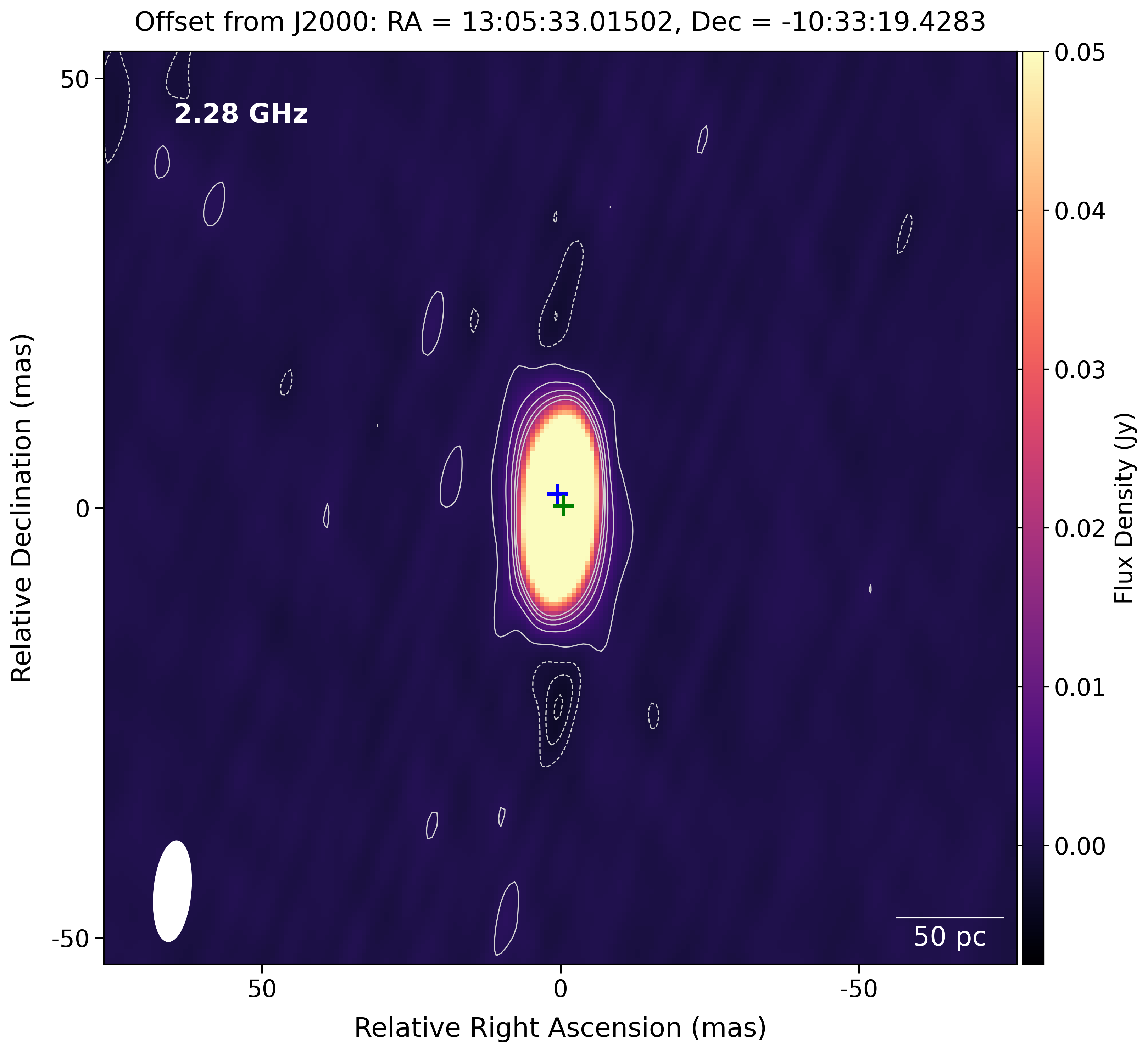}
     \includegraphics[width=8cm]{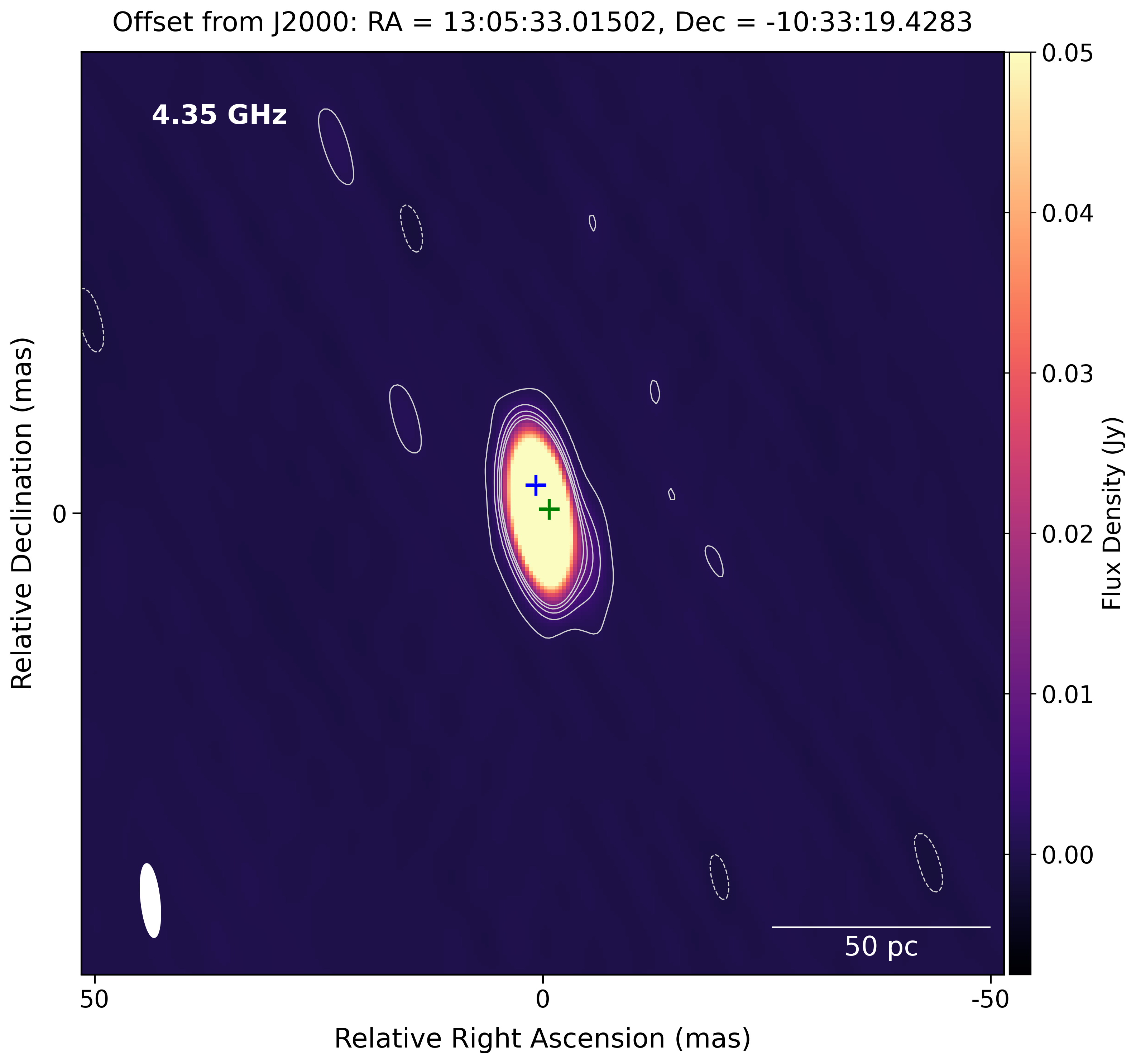}\\
     \includegraphics[width=7.85cm]{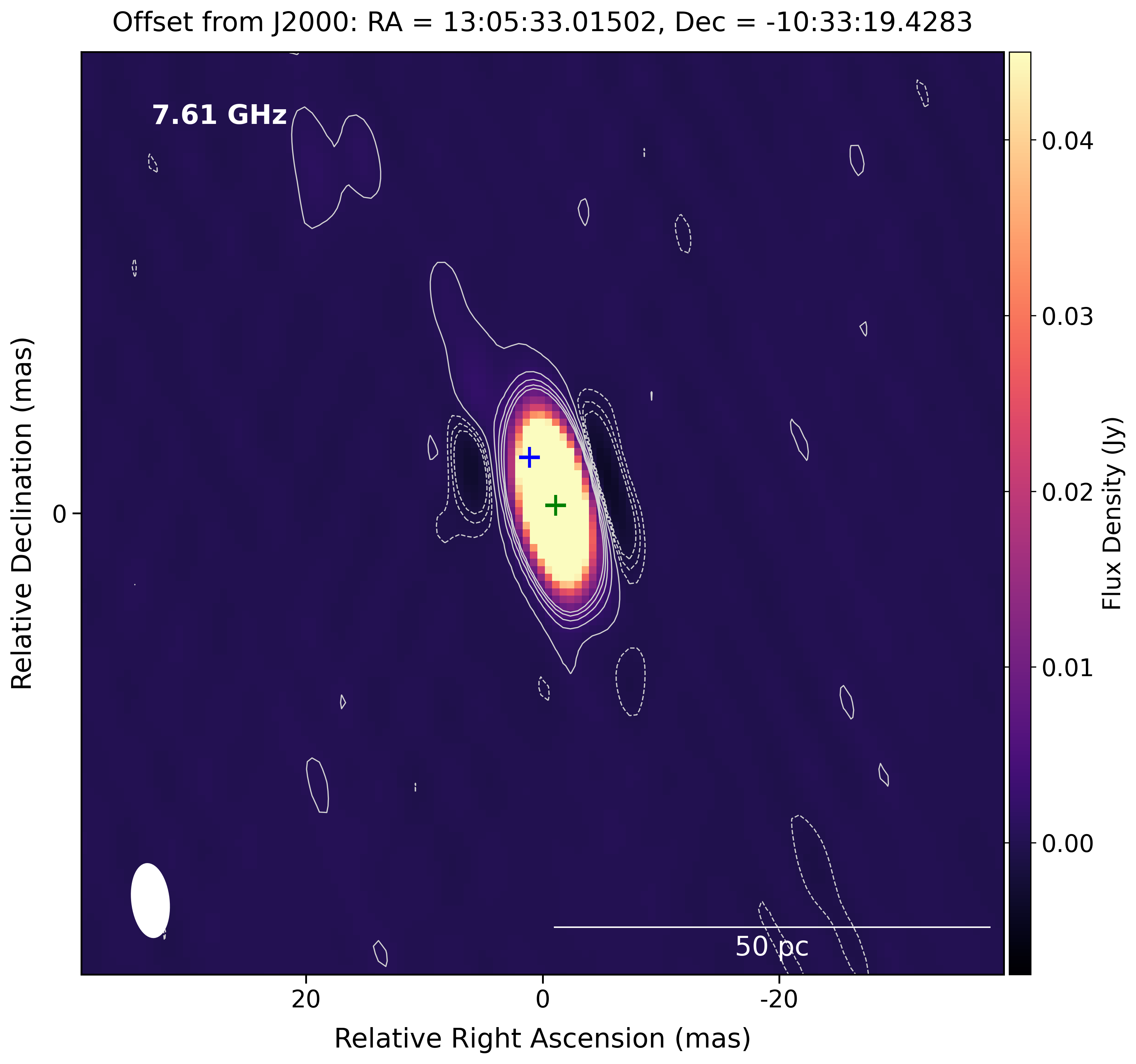}
     \includegraphics[width=8cm]{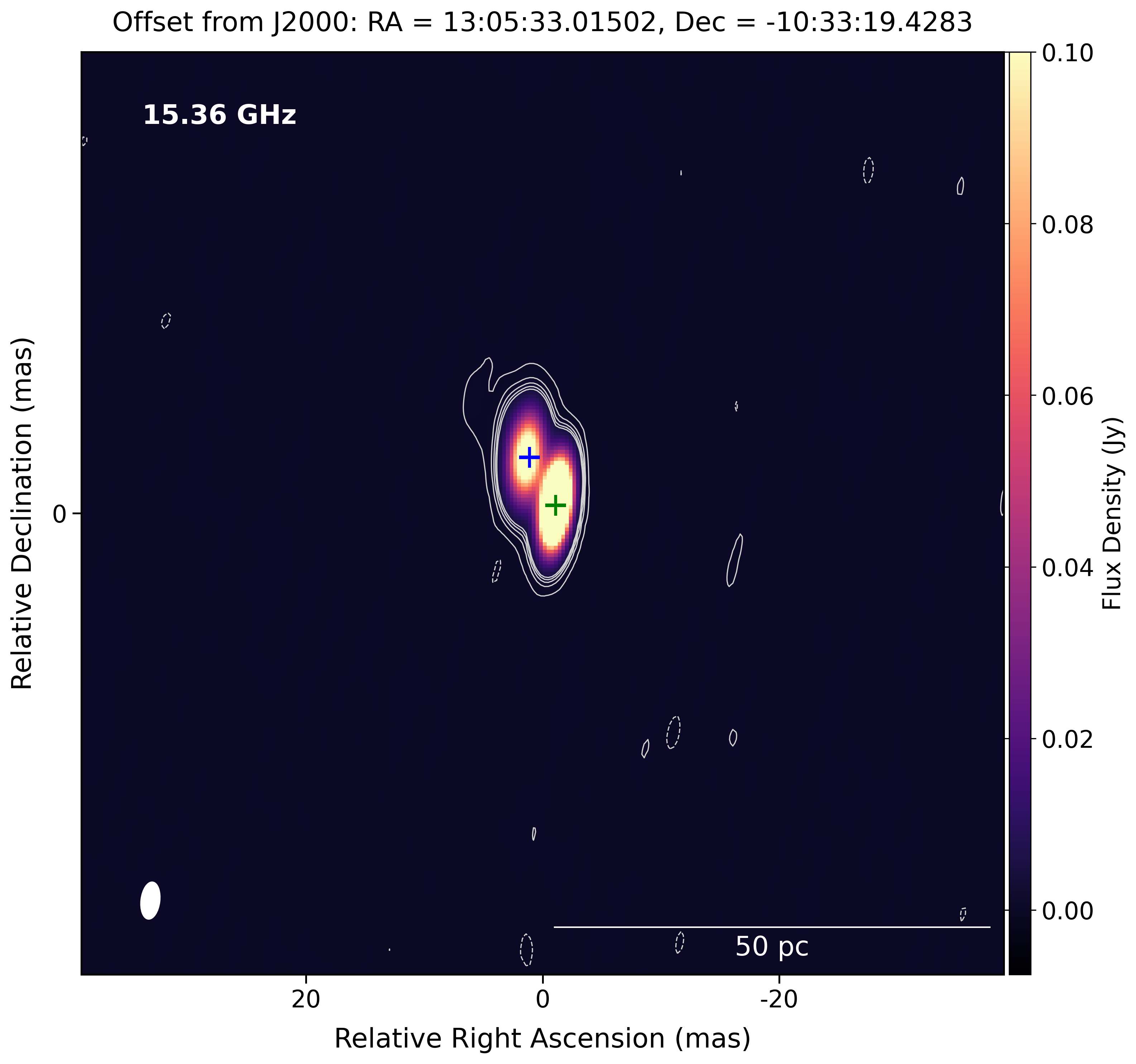}
        \vspace*{-3mm}
          \caption{VLBA 2.28 (Top Left), 4.35 (Top Right), 7.61 (Bottom Left), and 15.36 (Bottom Right) GHz images of J1305-1033. For plot description, see Figure \ref{fig:0216}.}
   \label{fig:1305}
\end{figure*}

\begin{figure*}[ht!]
    \centering
    \includegraphics[width=8cm]{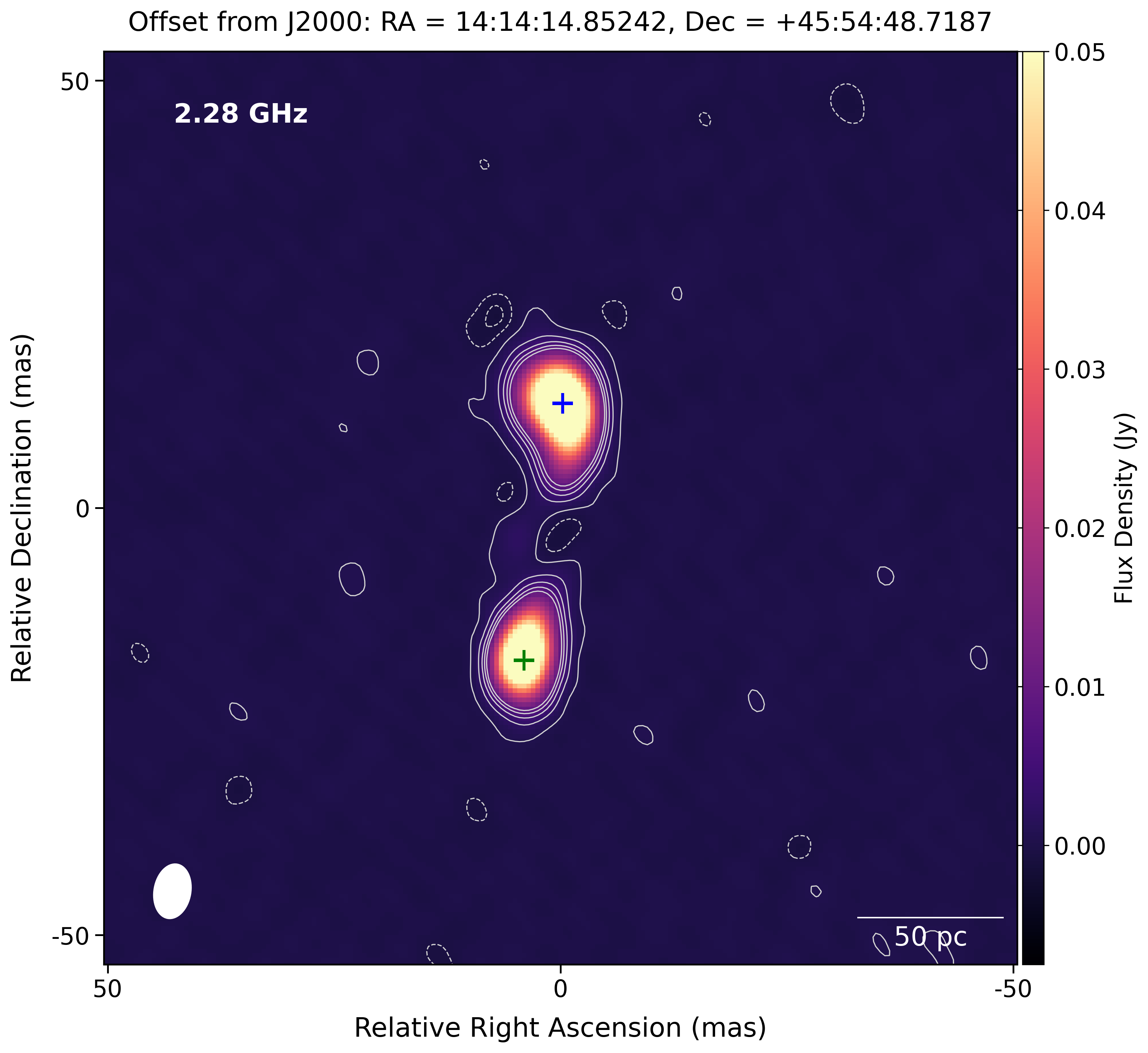}
     \includegraphics[width=8.1cm]{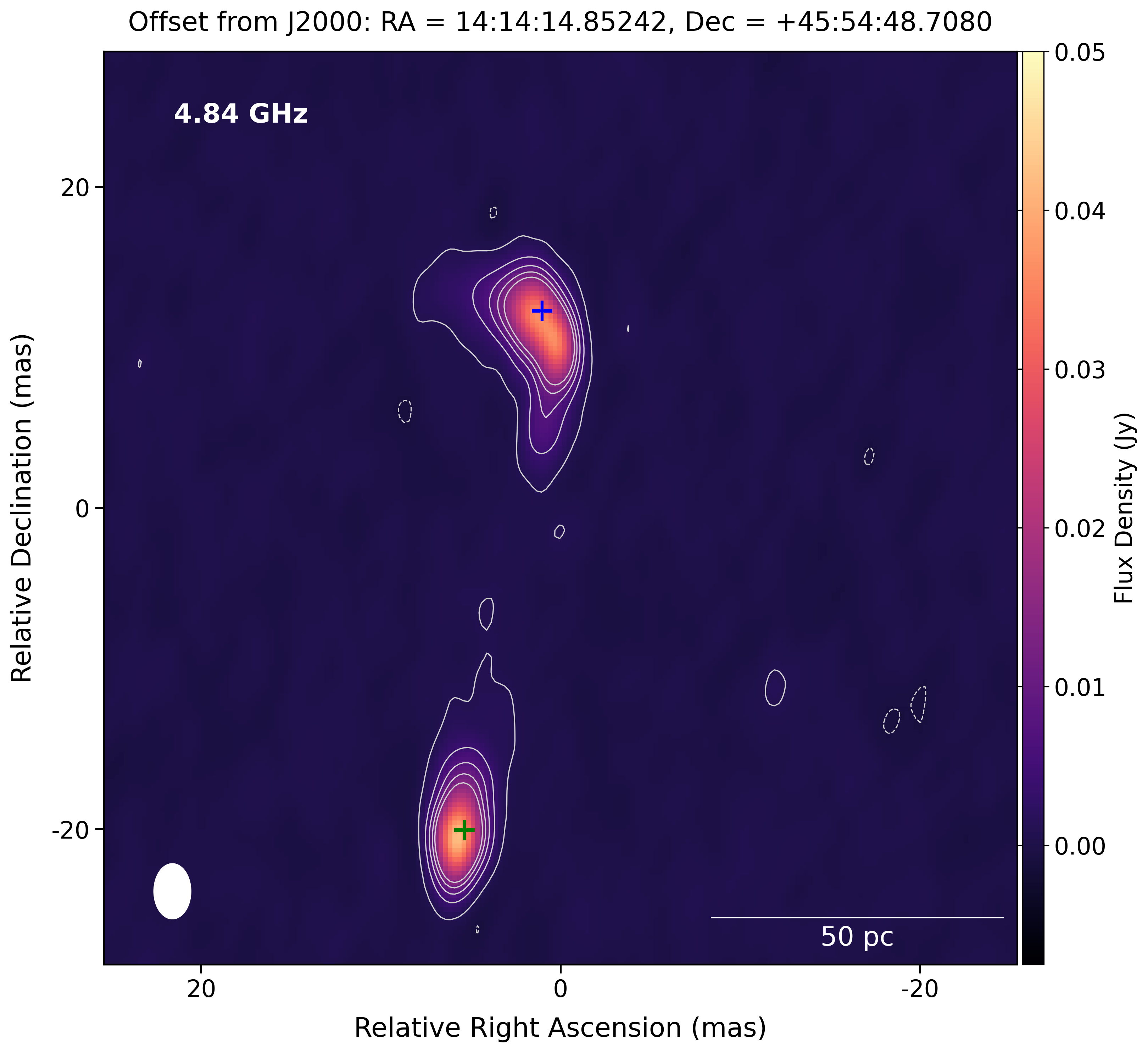}\\
     \includegraphics[width=8.2cm]{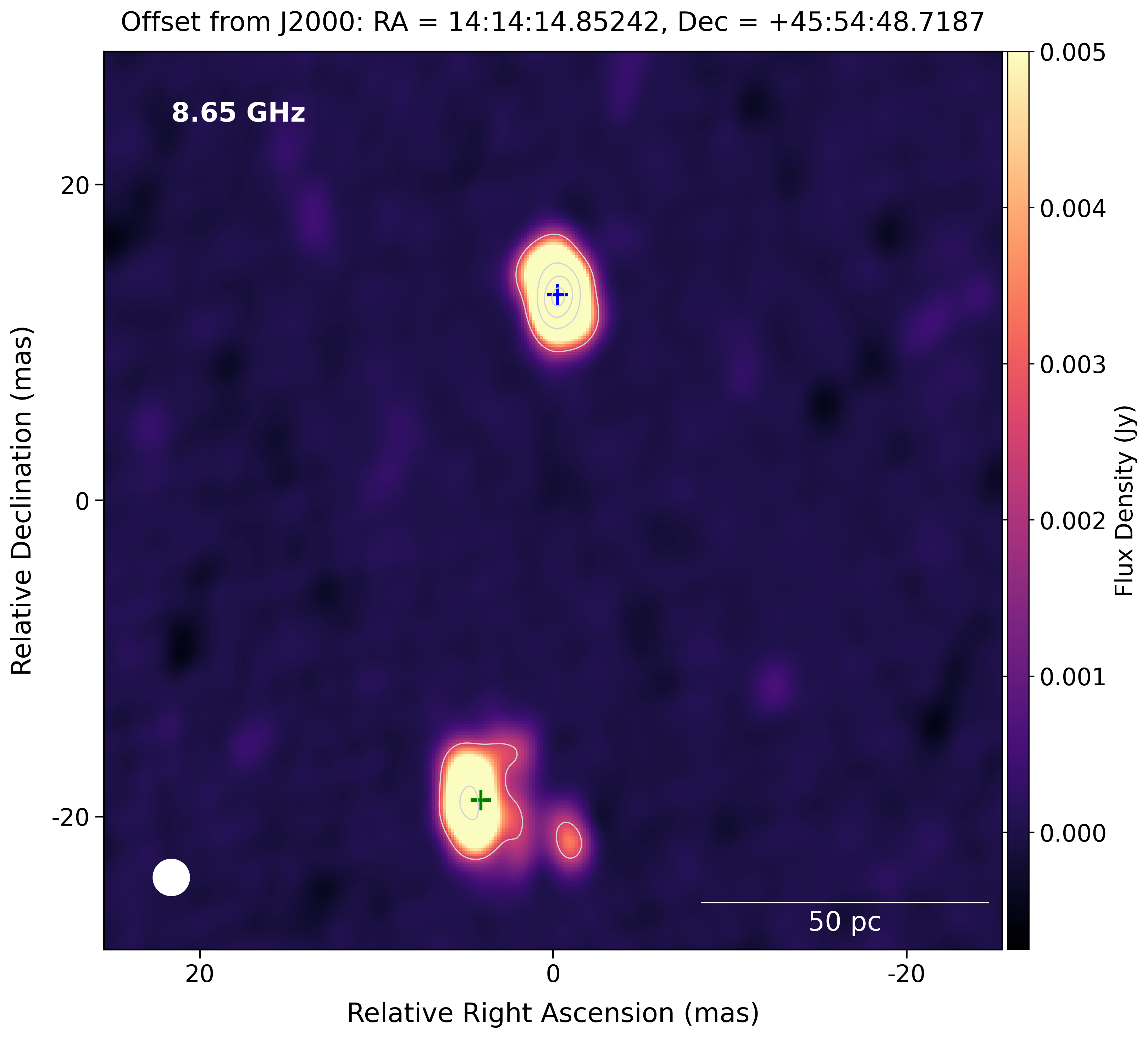}
        \vspace*{-3mm}
          \caption{VLBA 2.28 (Top Left), 4.84 (Top Right), and 8.65 (Bottom) GHz images of J1414+4554. For plot description, see Figure \ref{fig:0216}.}
   \label{fig:1414}
\end{figure*}

\begin{figure*}[ht!]
    \centering
    \includegraphics[width=8cm]{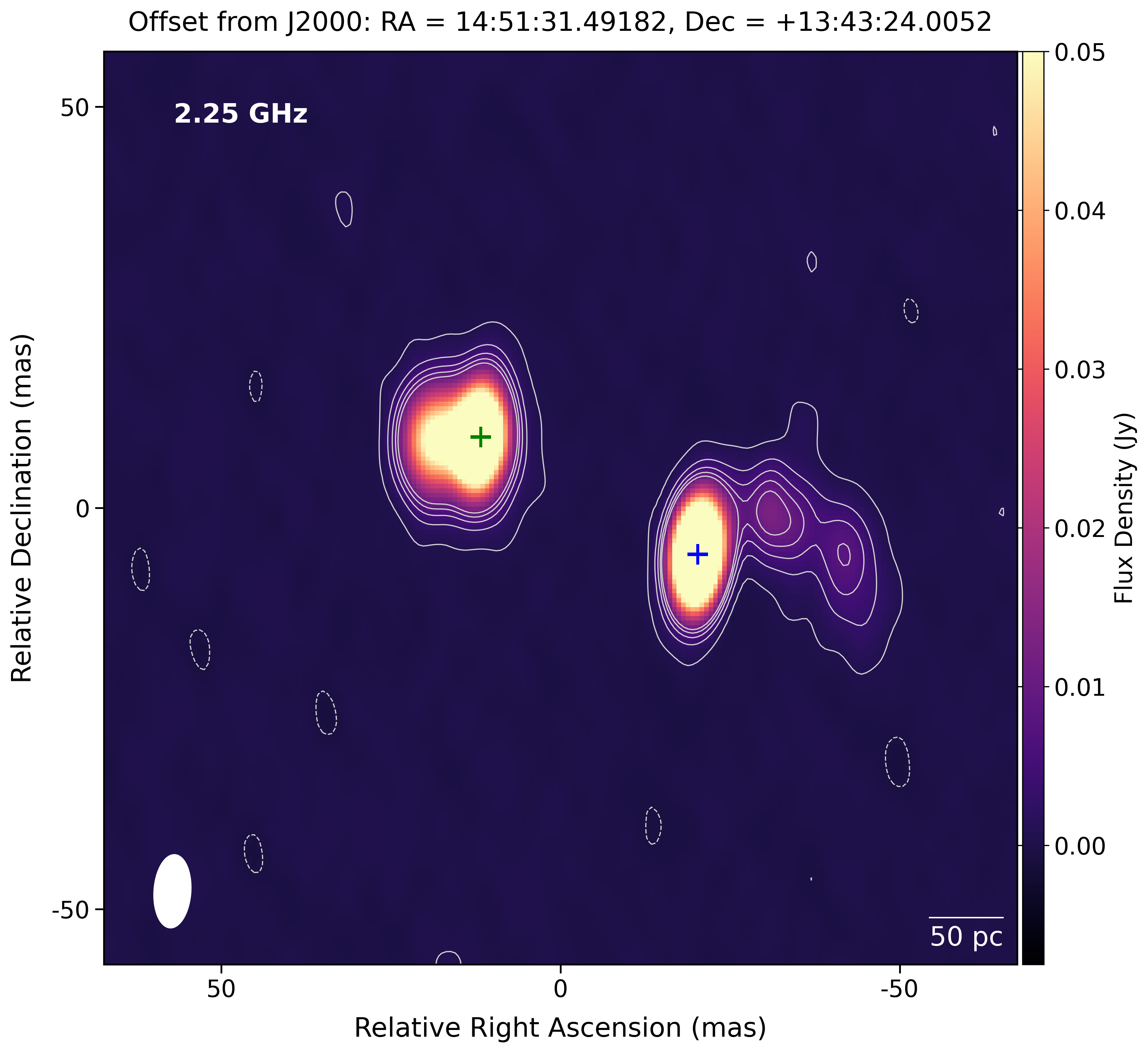}
     \includegraphics[width=7.9cm]{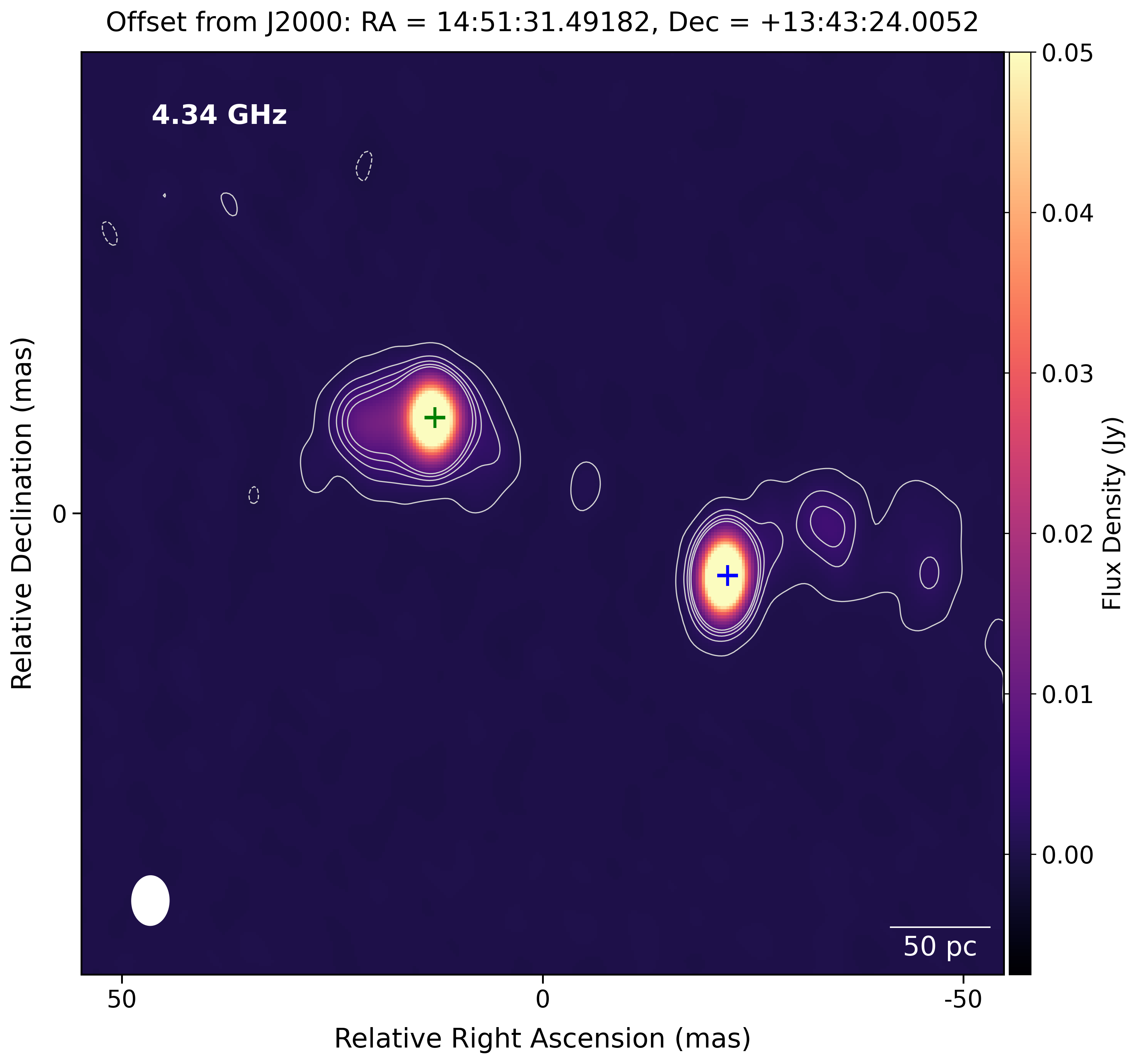}\\
     \includegraphics[width=8.2cm]{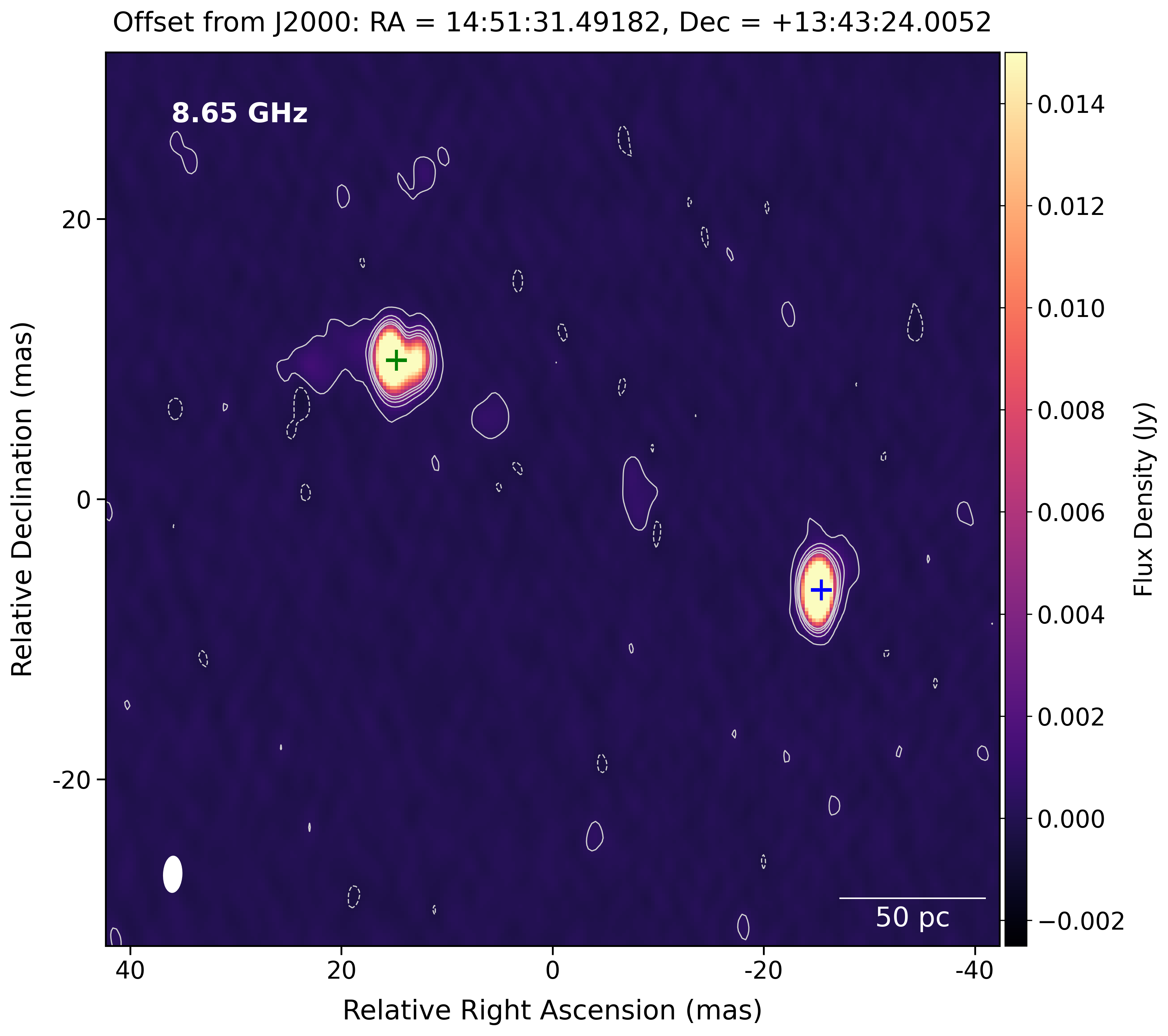}
        \vspace*{-3mm}
          \caption{VLBA 2.25 (Top Left), 4.34 (Top Right), and 8.65 (Bottom) GHz images of J1451+1343. For plot description, see Figure \ref{fig:0216}.}
   \label{fig:1451}
\end{figure*}

\begin{figure*}[ht!]
    \centering
    \includegraphics[width=8cm]{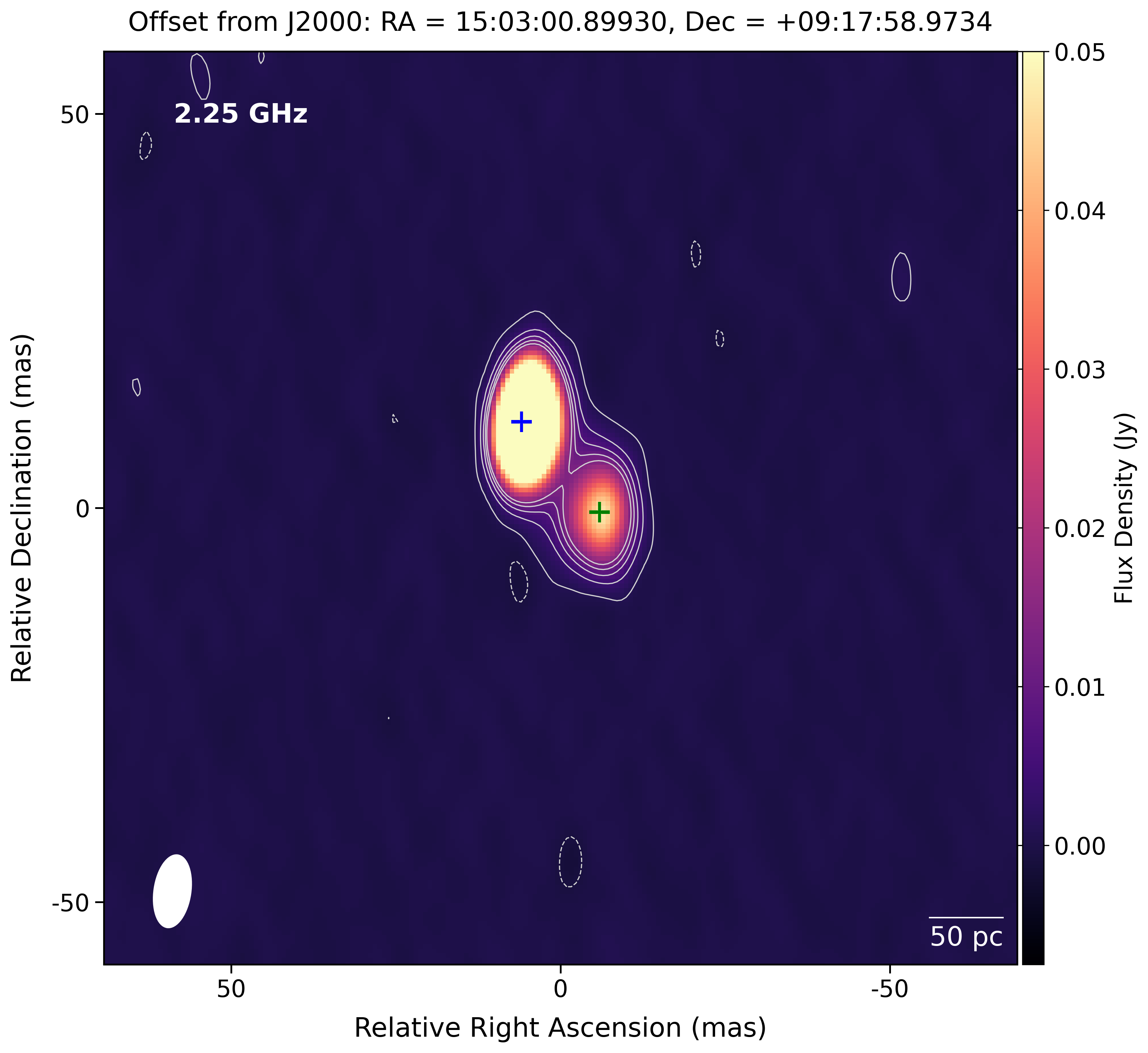}
     \includegraphics[width=8.2cm]{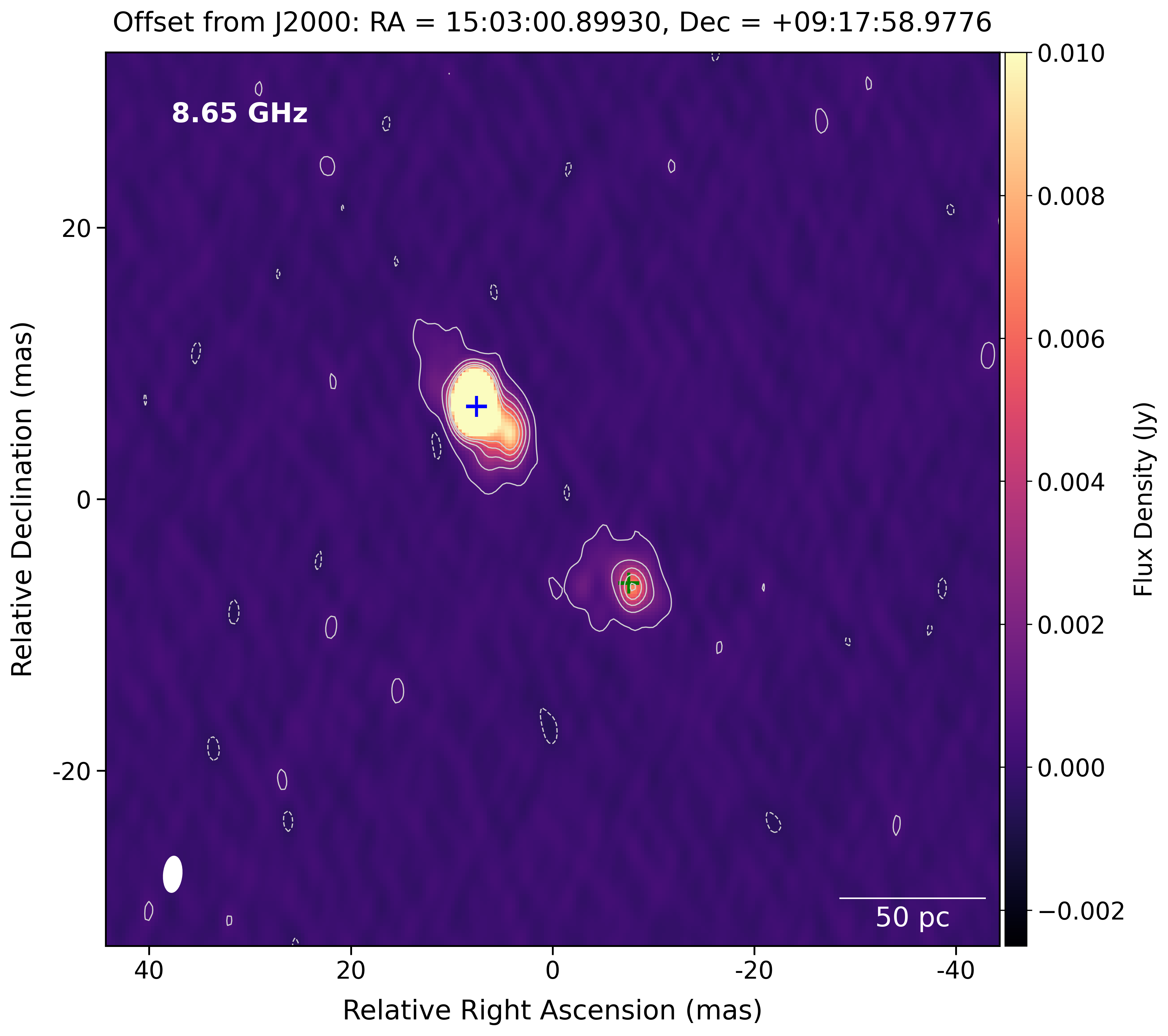}
        \vspace*{-3mm}
          \caption{VLBA 2.25 (Left) and 8.65 (Right) GHz images of J1503+0917. For plot description, see Figure \ref{fig:0216}.}
   \label{fig:1503}
\end{figure*}

\begin{figure*}[ht!]
    \centering
    \includegraphics[width=8cm]{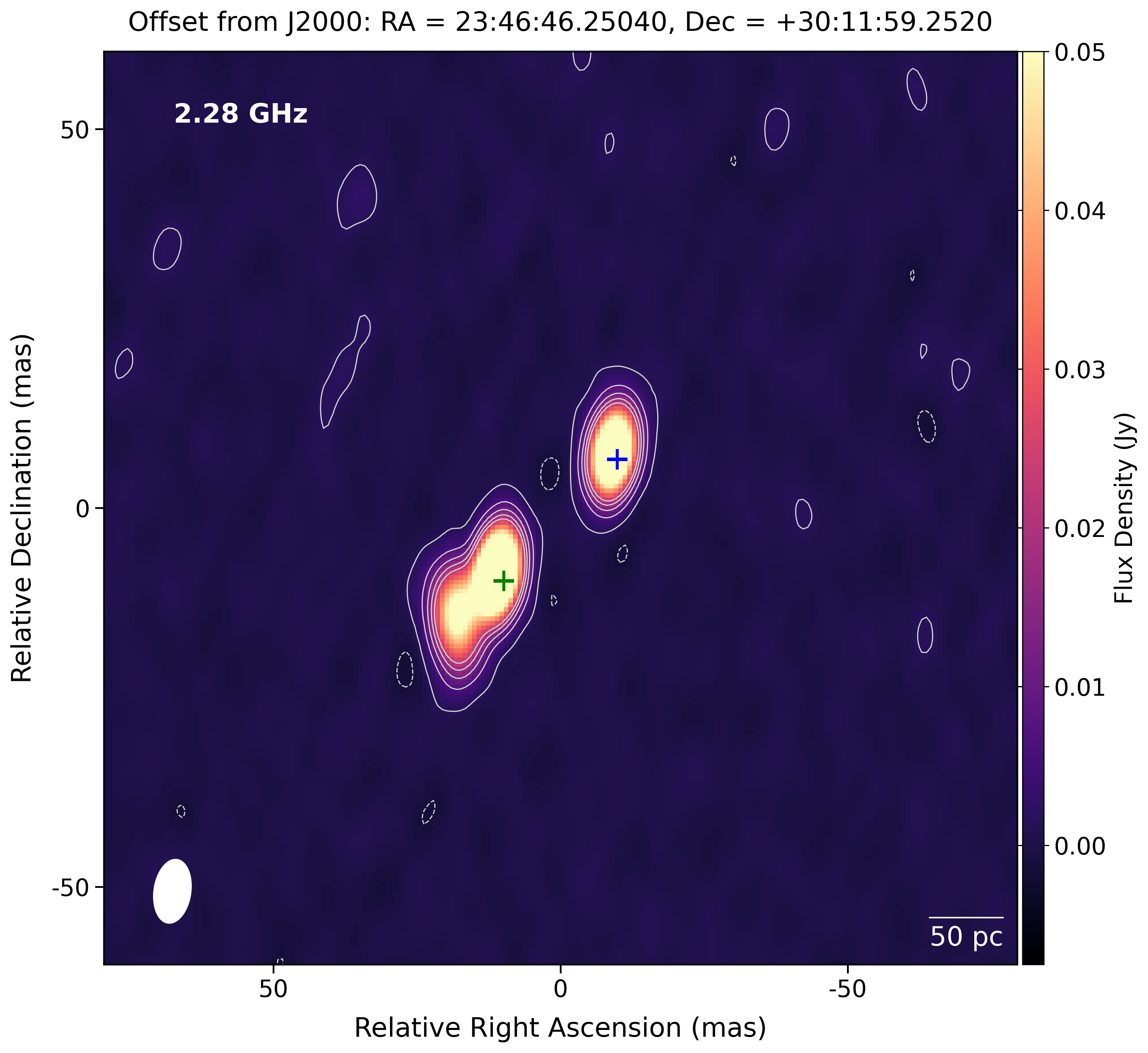}
     \includegraphics[width=7.9cm]{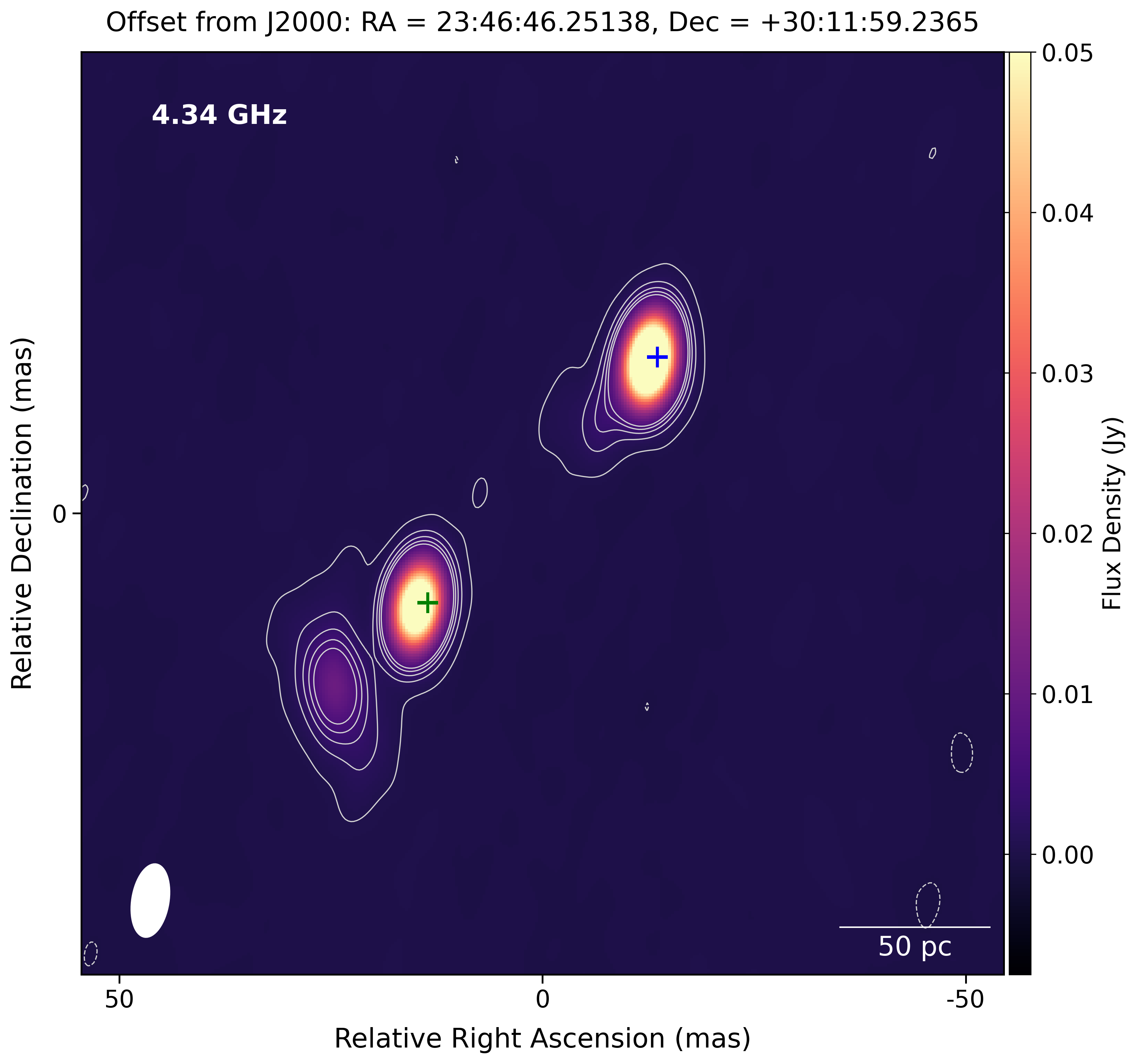}\\
     \includegraphics[width=8.2cm]{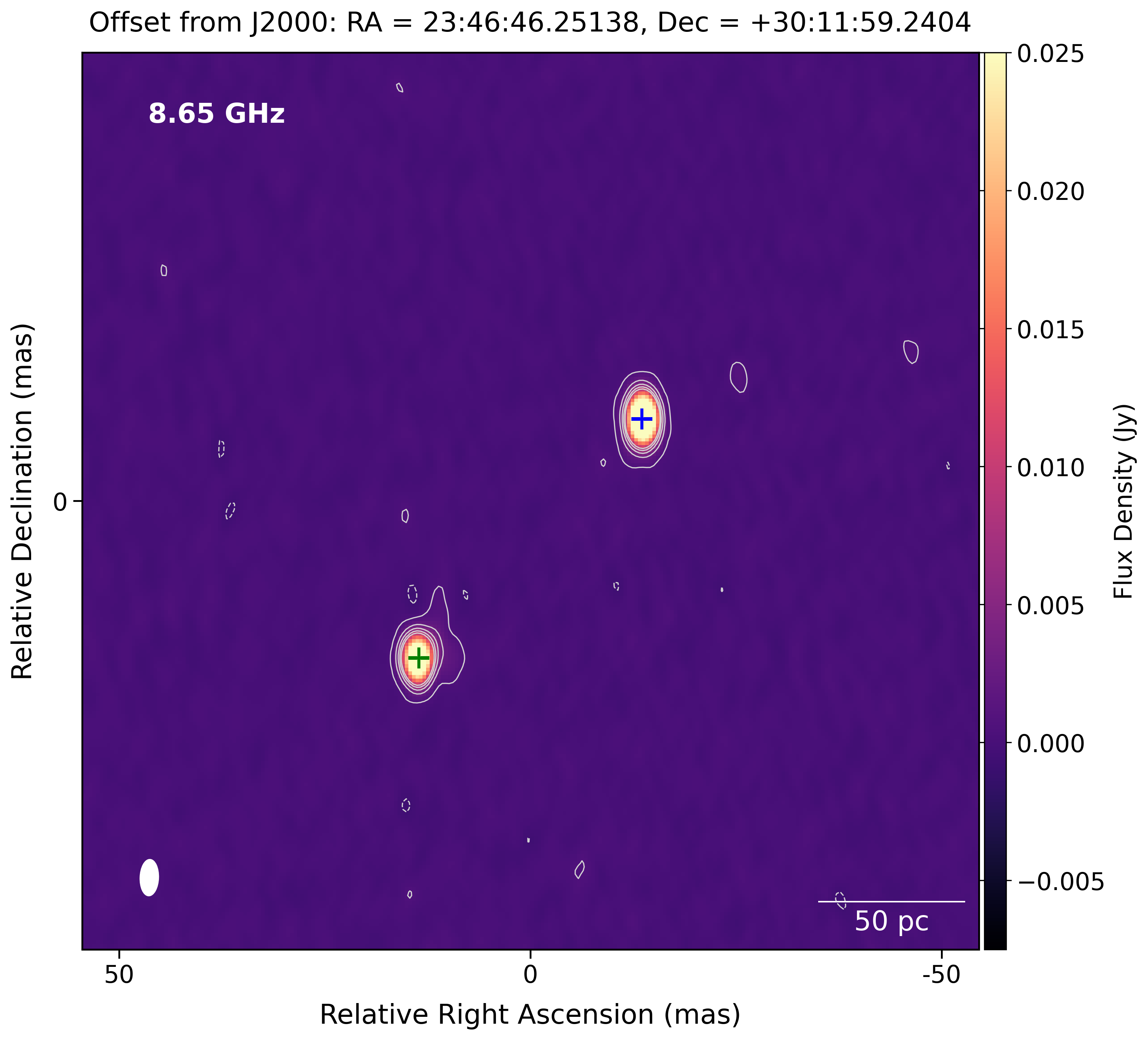}
        \vspace*{-3mm}
          \caption{VLBA 2.28 (Top Left), 4.34 (Top Right), and 8.65 (Bottom) GHz images of J2346+3011. For plot description, see Figure \ref{fig:0216}.}
   \label{fig:2346}
\end{figure*}

\begin{figure*}[ht!]
    \centering
    \includegraphics[width=8cm]{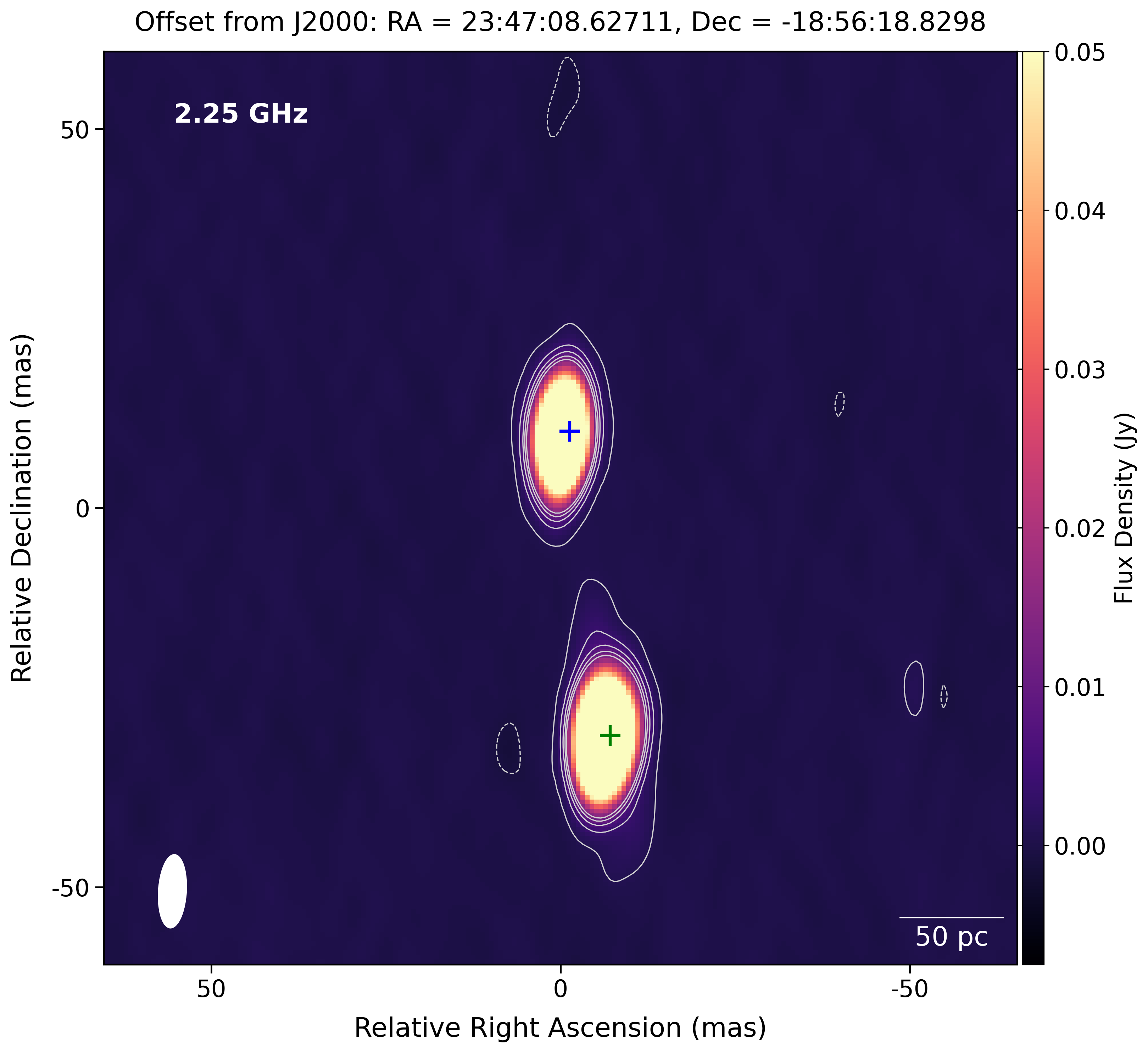}
     \includegraphics[width=8cm]{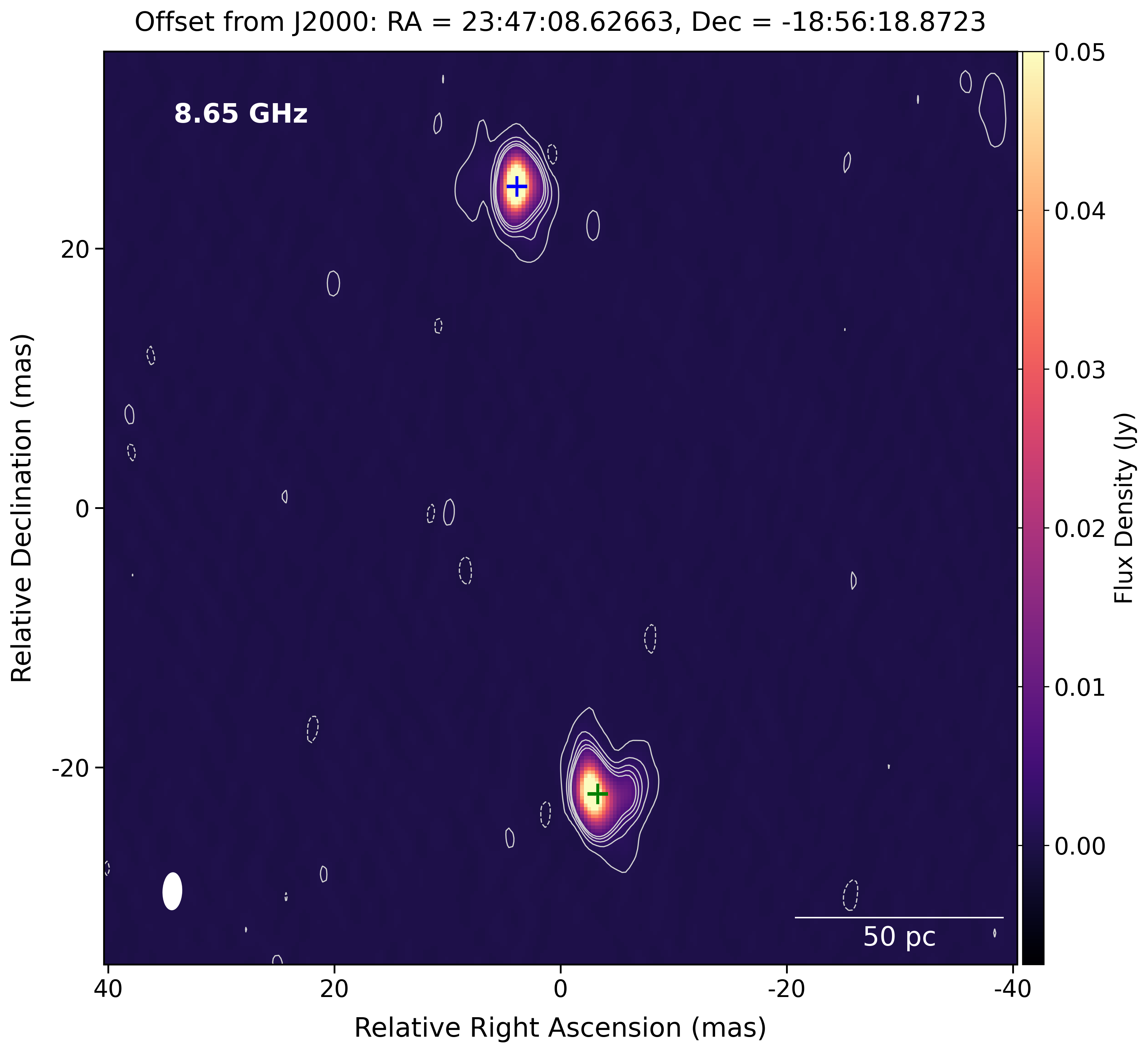}
        \vspace*{-3mm}
          \caption{VLBA 2.25 (Left) and 8.65 (Right) GHz images of J2347-1856. For plot description, see Figure \ref{fig:0216}.}
   \label{fig:2347}
\end{figure*}

\clearpage

\section{Appendix B: Radio Spectra} \label{sec:radiospec}

\begin{figure*}[ht!]
    \centering
    \includegraphics[width=8cm]{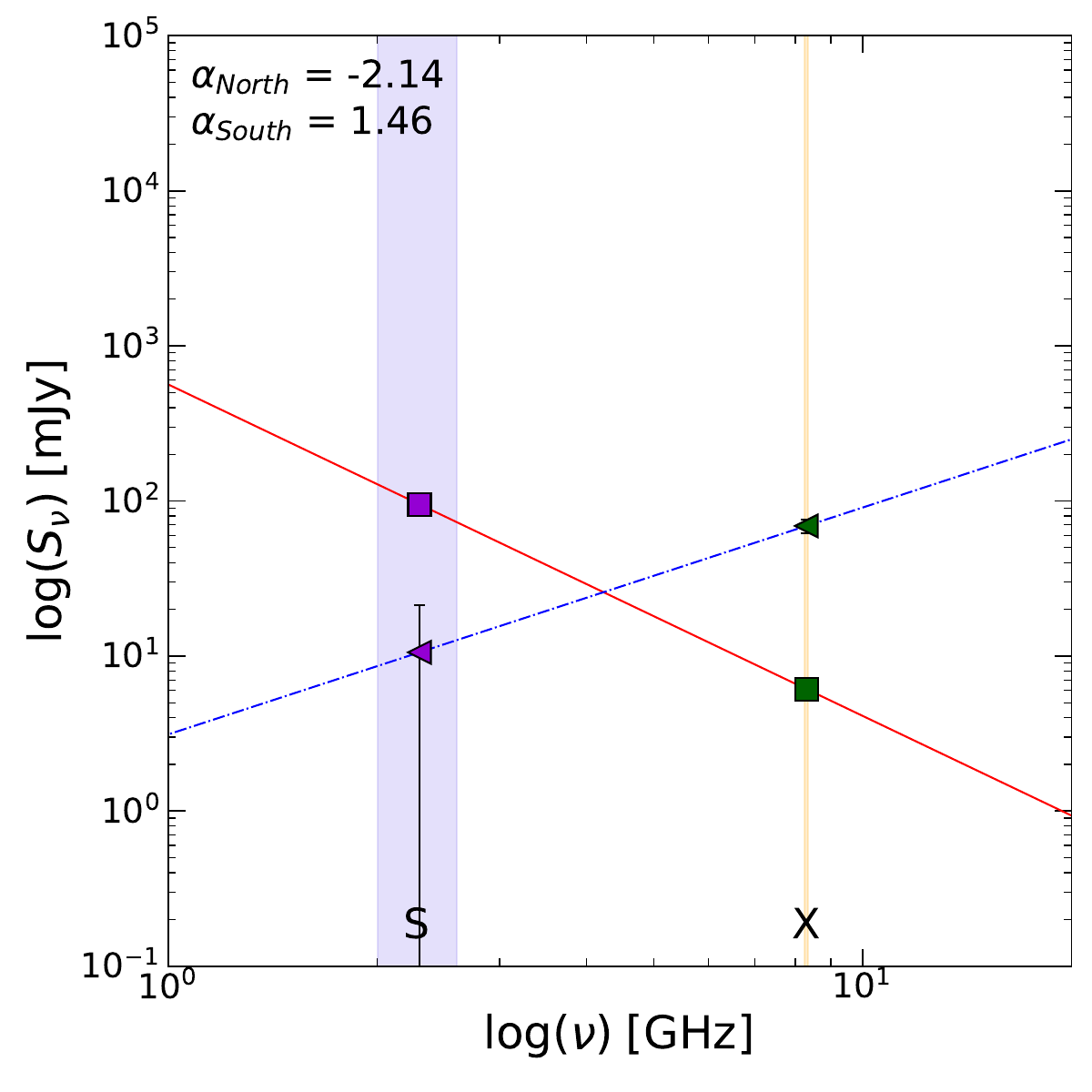}
     \includegraphics[width=8cm]{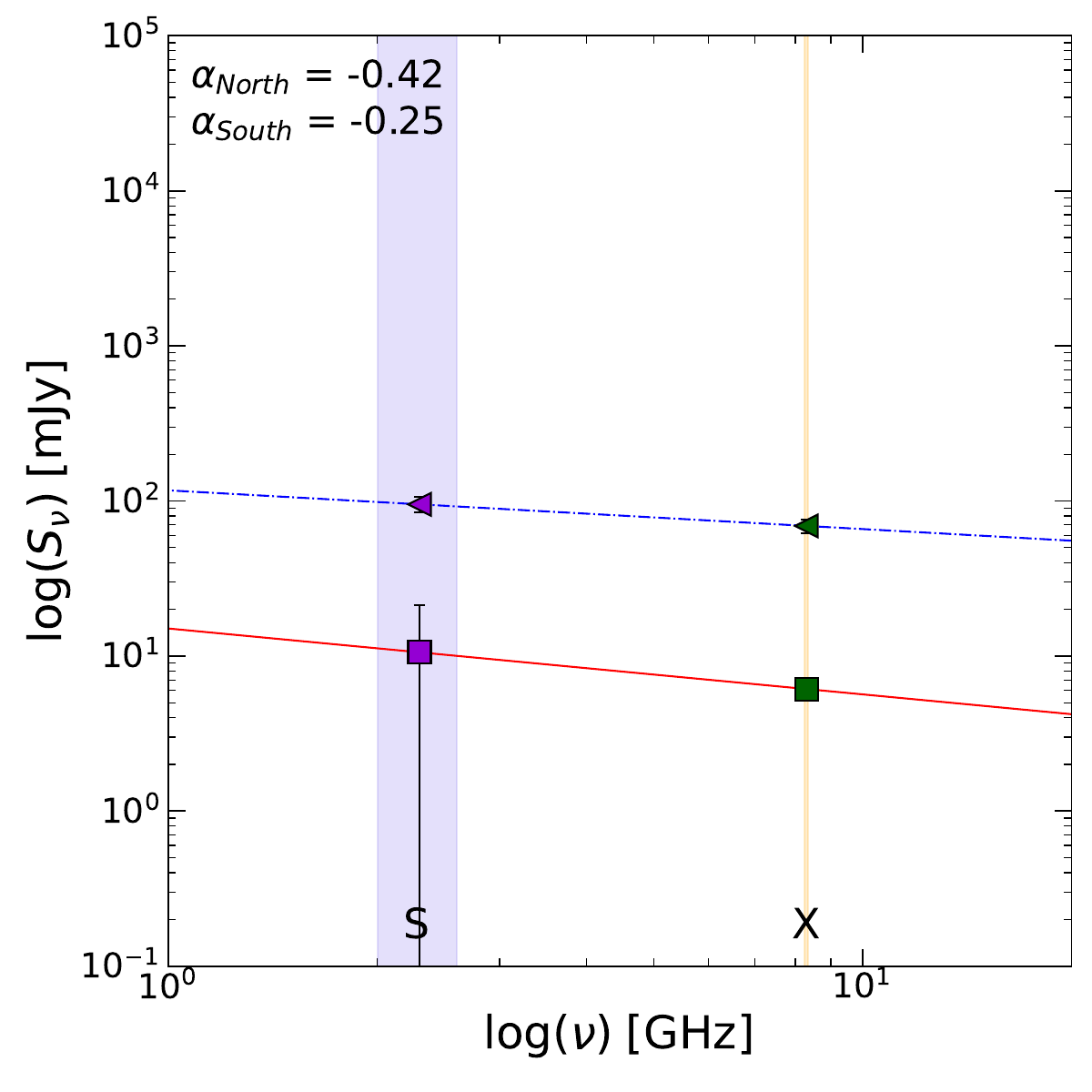}\\
     \includegraphics[width=8cm]{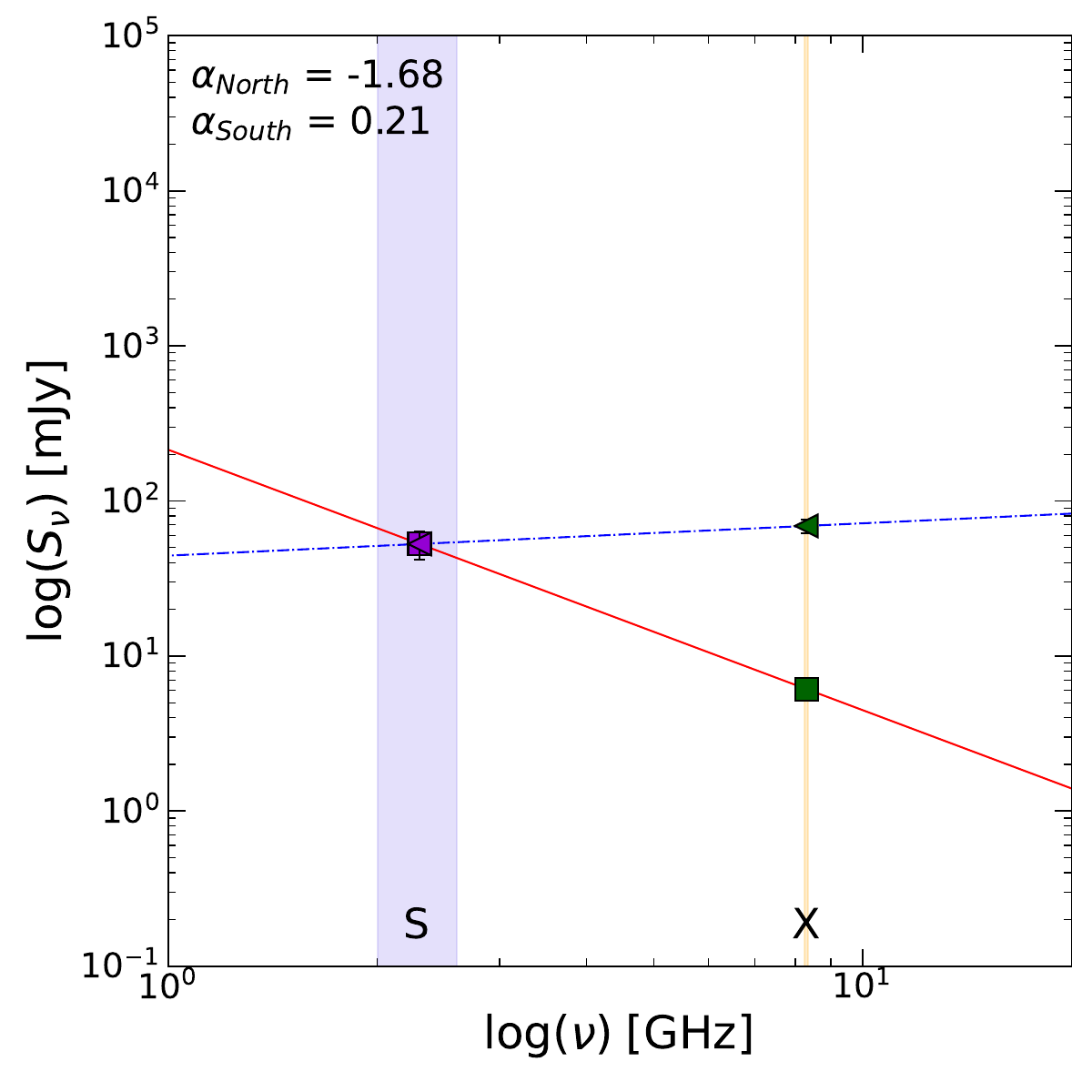}
        \vspace*{-3mm}
          \caption{Radio spectra for J0216-0105 showing the multi-frequency VLBA data. Due to the limited S-band resolution, three configurations have been shown. In all configurations, the X-band flux densities remain constant for both components. At S-band, they have been varied: 10\% South/90\% North (Upper Left), 90\% South/10\% North (Upper Right), and 50\% South/50\% North. Components are differentiated by shape (squares, triangles, and circles), while colors are attributable to observational frequencies: S-band in purple, C-band in green, X-band in brown, and U-band in grey. Error bars have been included, but for the majority of the flux densities, are too small to be visible. The standard power and curved power law fits are shown for each component. Non-physical fits have been excluded. Resultant spectral indices are included in the upper left hand corner. The superscript $C$ denotes a curved power law spectral index. The solid red, dot-dash blue, and dotted green lines each represent the standard and/or curved power laws for different components.}
   \label{fig:0216spec}
\end{figure*}

\begin{figure*}[ht!]
    \centering
    \includegraphics[width=8cm]{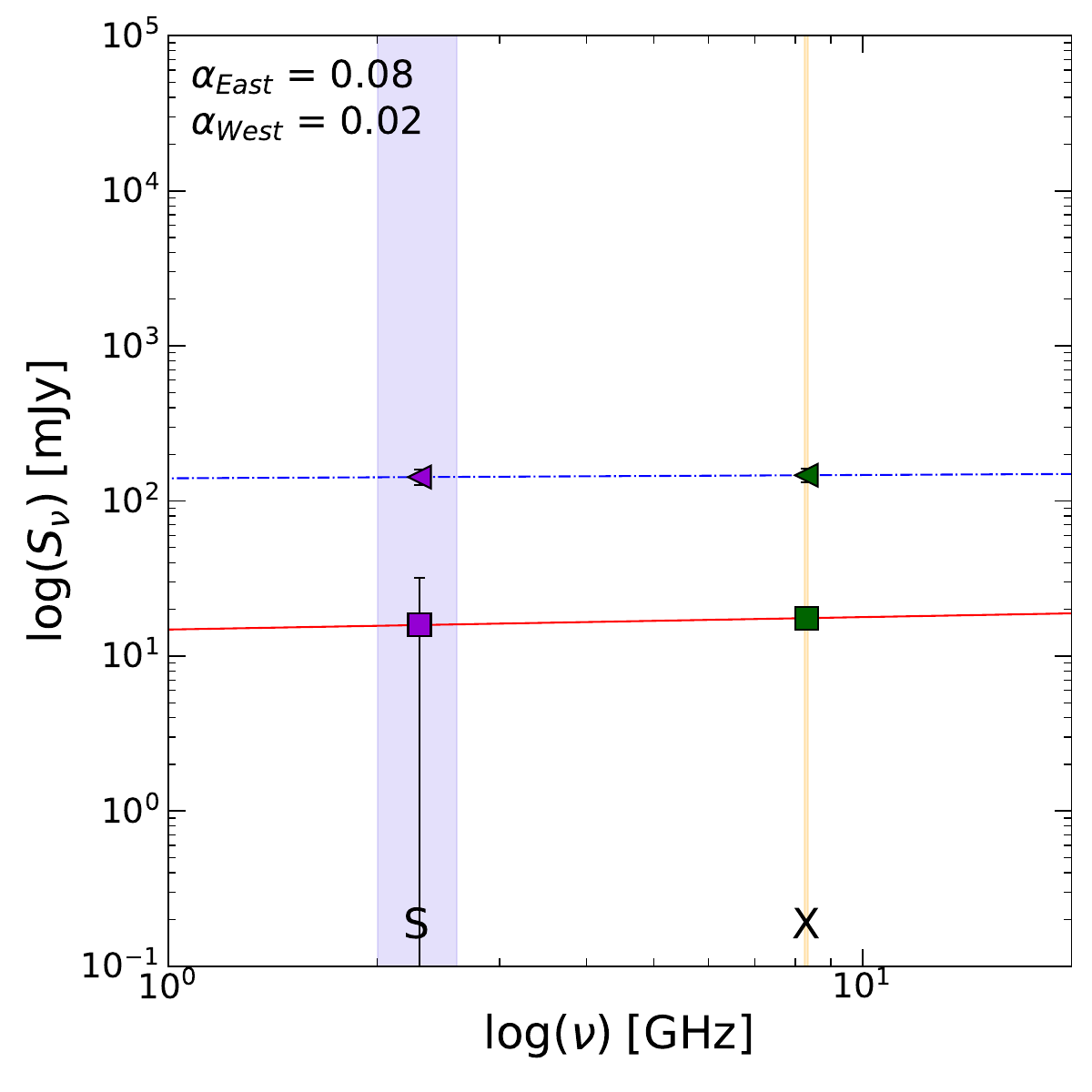}
     \includegraphics[width=8cm]{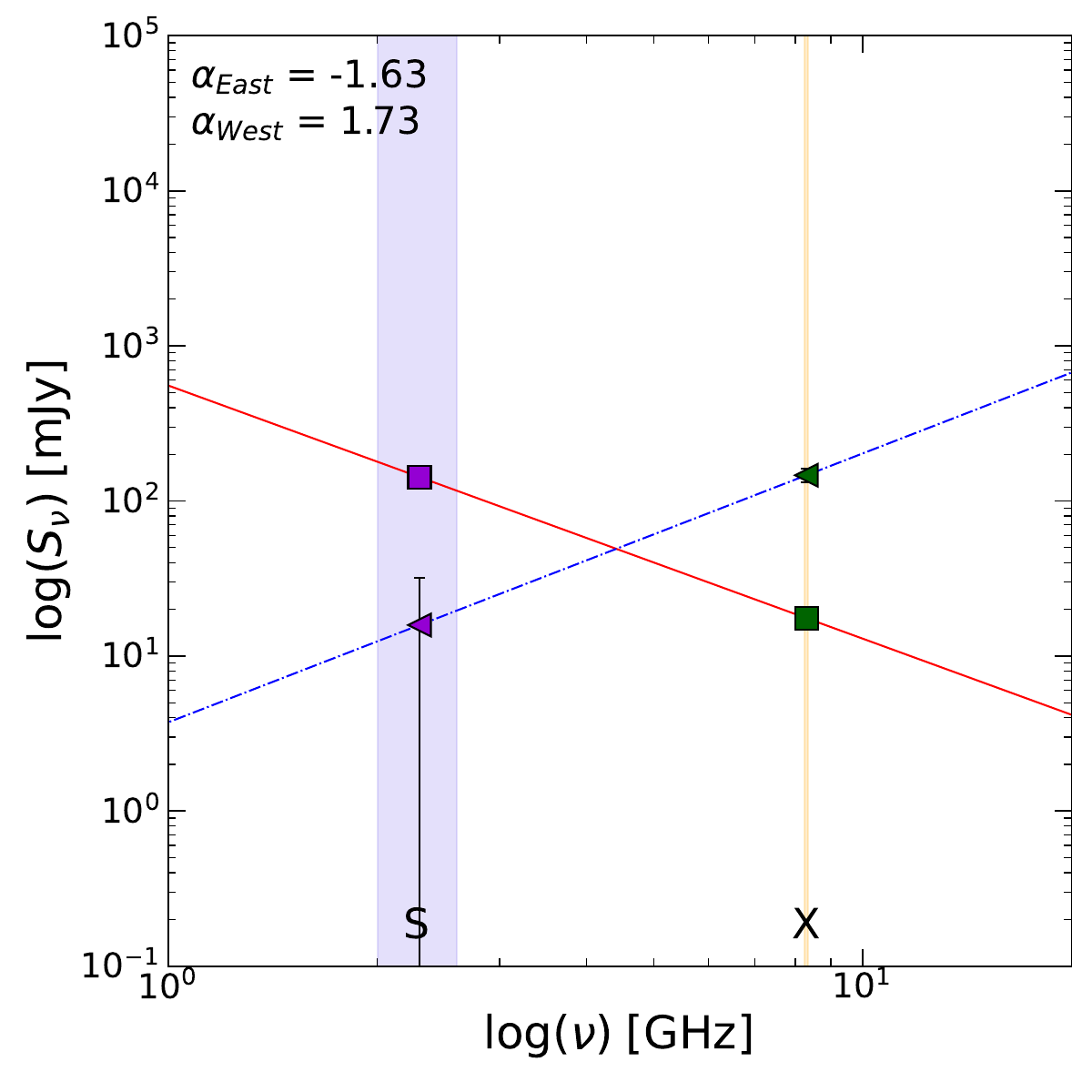}\\
     \includegraphics[width=8cm]{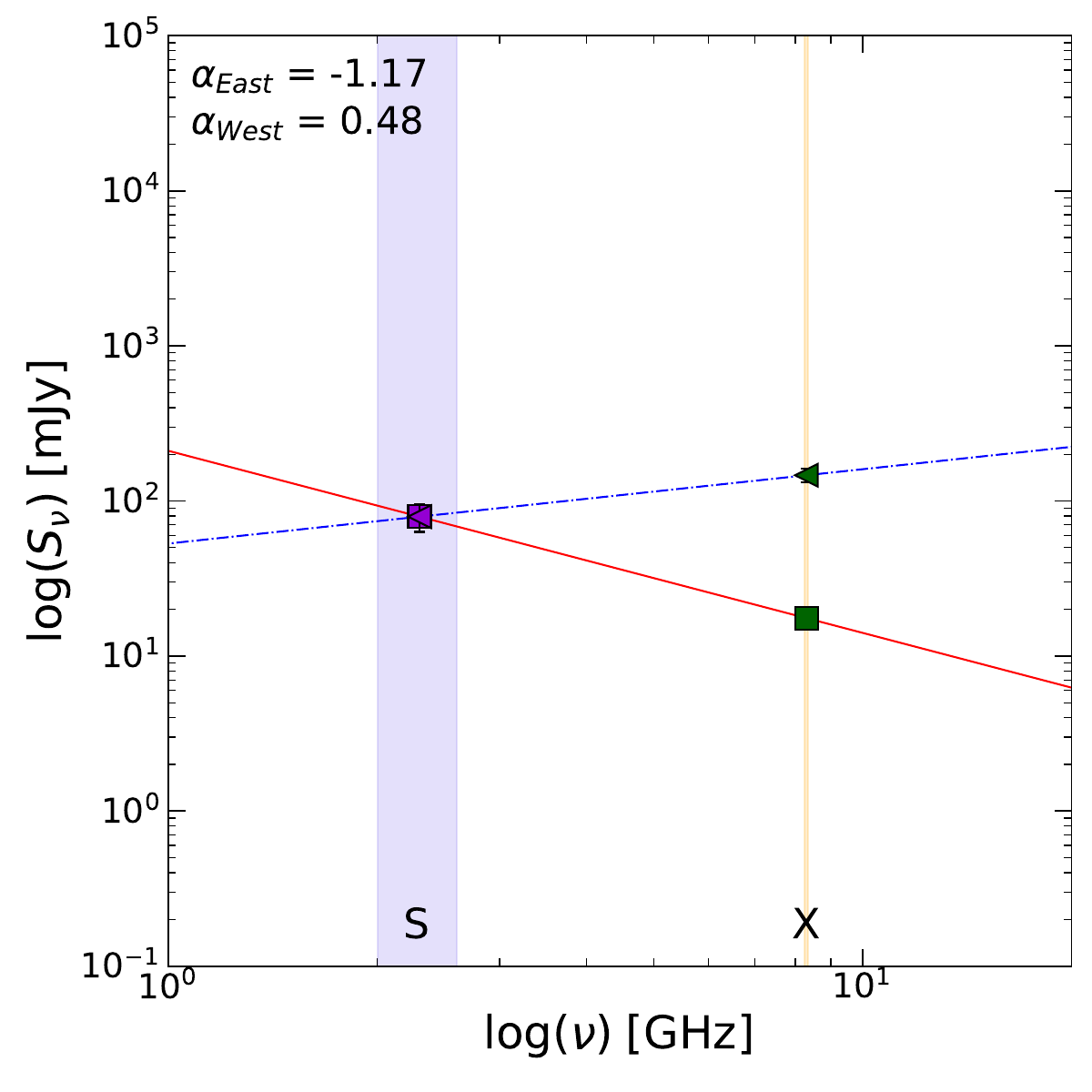}
        \vspace*{-3mm}
          \caption{Radio spectra for J0729-1320 showing the multi-frequency VLBA data. Due to the limited S-band resolution, three configurations have been shown. In all configurations, the X-band flux densities remain constant for both components. At S-band, they have been varied: 10\% East/90\% West (Upper Left), 90\% East/10\% West (Upper Right), and 50\% East/50\% West. For plot description, see Figure \ref{fig:0216spec}.}
   \label{fig:0729spec}
\end{figure*}

\begin{figure*}[ht!]
    \centering
    \includegraphics[width=8cm]{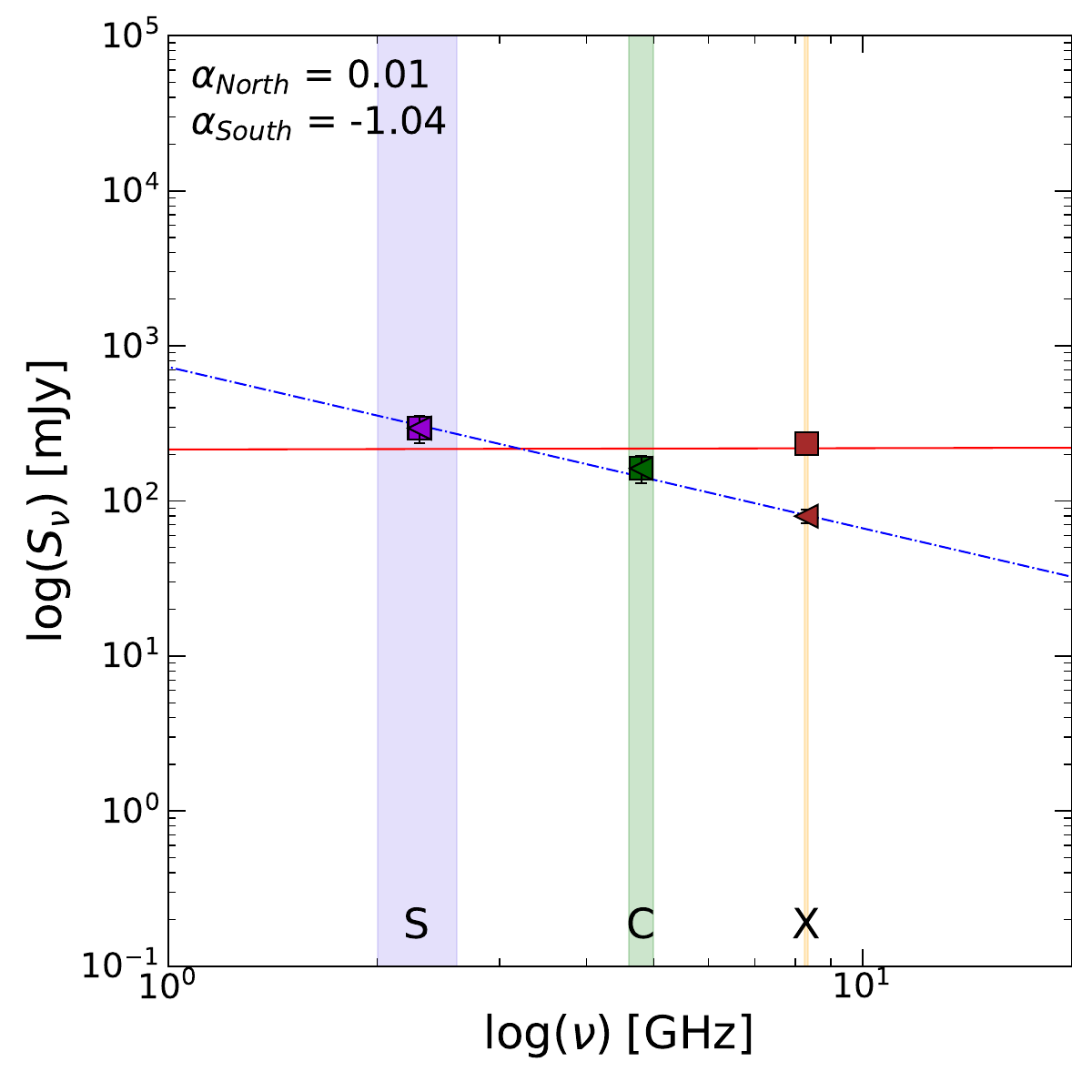}
     \includegraphics[width=8cm]{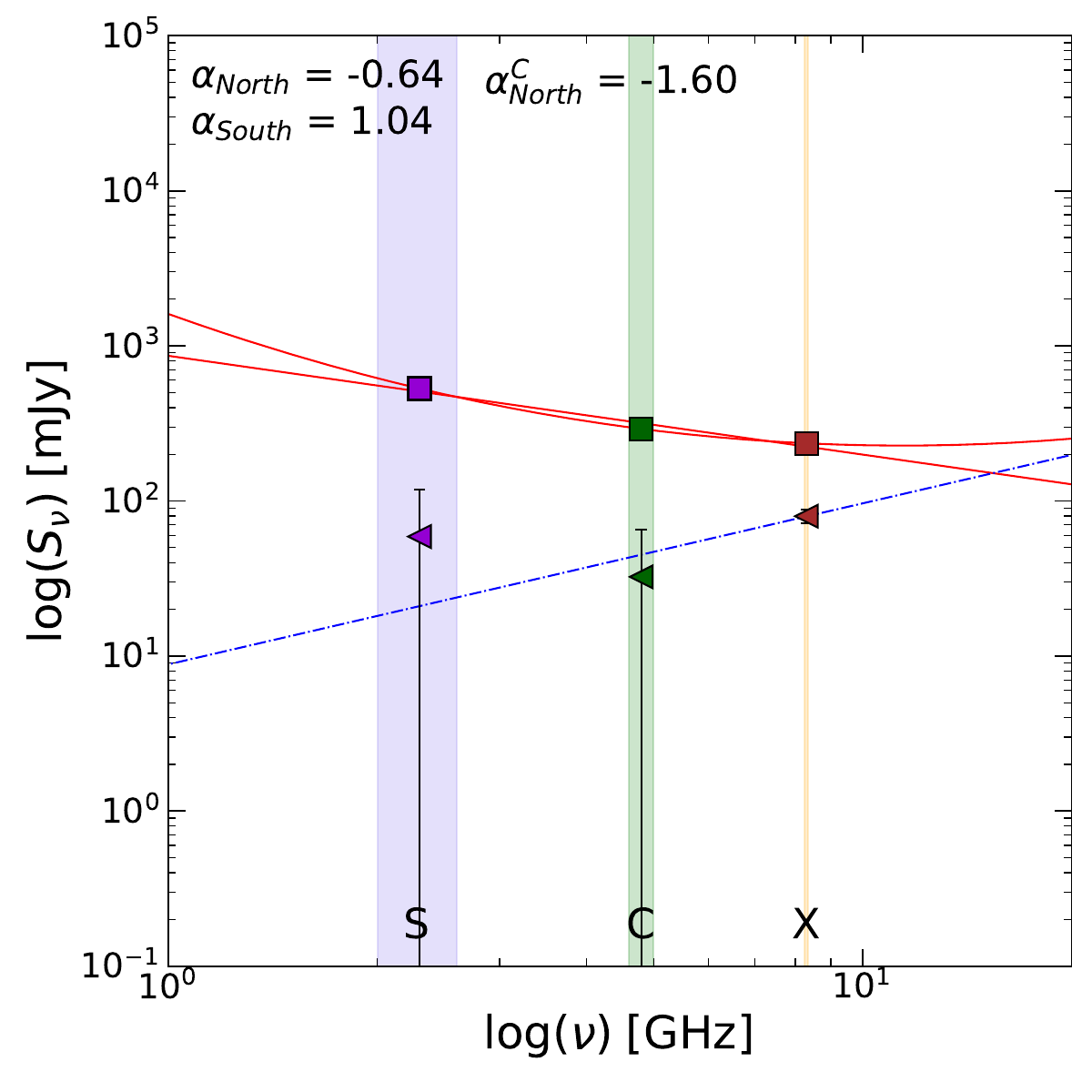}
        \vspace*{-3mm}
          \caption{Radio spectra for J0745+3142 showing the multi-frequency VLBA data. Due to the limited S and C-band resolution, two configurations have been shown. In all configurations, the X-band flux densities remain constant for both components. At S-band, they have been varied: 50\% North/50\% South (Left) and 90\% North/10\% South (Right). At C-band, they have been varied: 50\% North/50\% South (Left) and 90\% North/10\% South (Right). For plot description, see Figure \ref{fig:0216spec}.}
   \label{fig:0745spec}
\end{figure*}

\begin{figure*}[ht!]
    \centering
    \includegraphics[width=8cm]{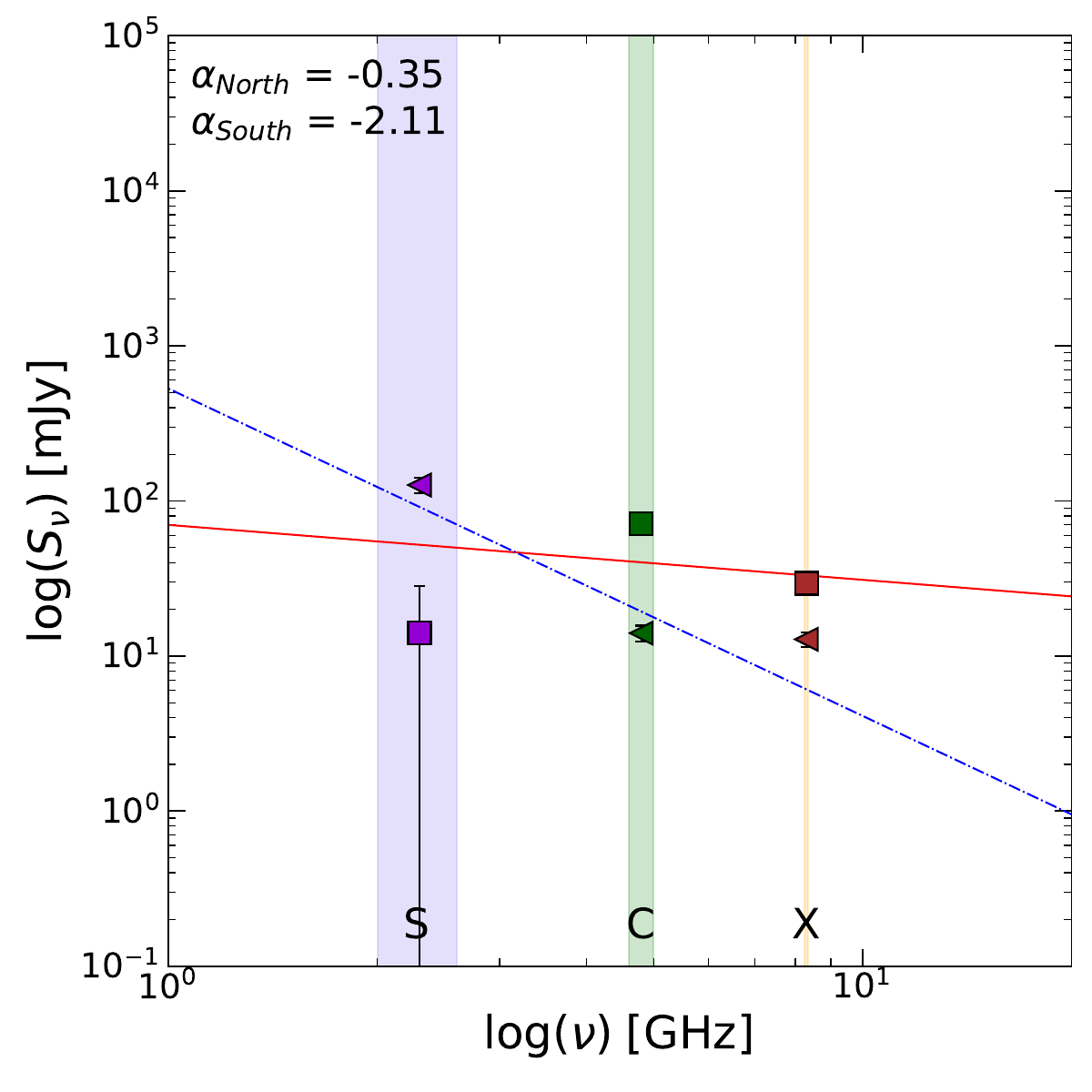}
     \includegraphics[width=8cm]{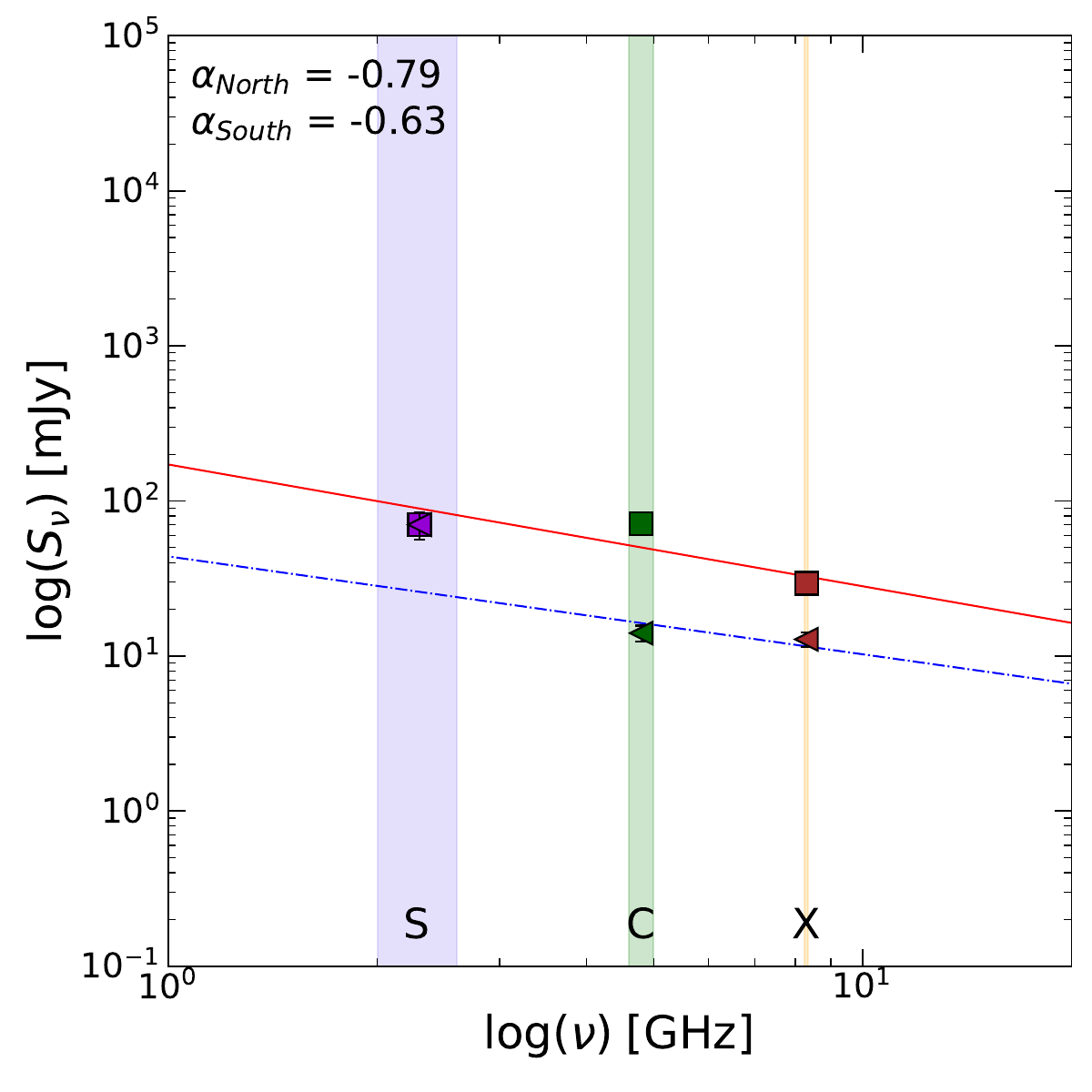}
        \vspace*{-3mm}
          \caption{Radio spectra for J0843+4537 showing the multi-frequency VLBA data. Due to the limited S-band resolution, two configurations have been shown. In all configurations, the C and X-band flux densities remain constant for both components. At S-band, they have been varied: 10\% South/90\% North (Left) and 50\% South/50\% North (Right). For plot description, see Figure \ref{fig:0216spec}.}
   \label{fig:0843spec}
\end{figure*}

\begin{figure*}[ht!]
    \centering
     \includegraphics[width=8cm]{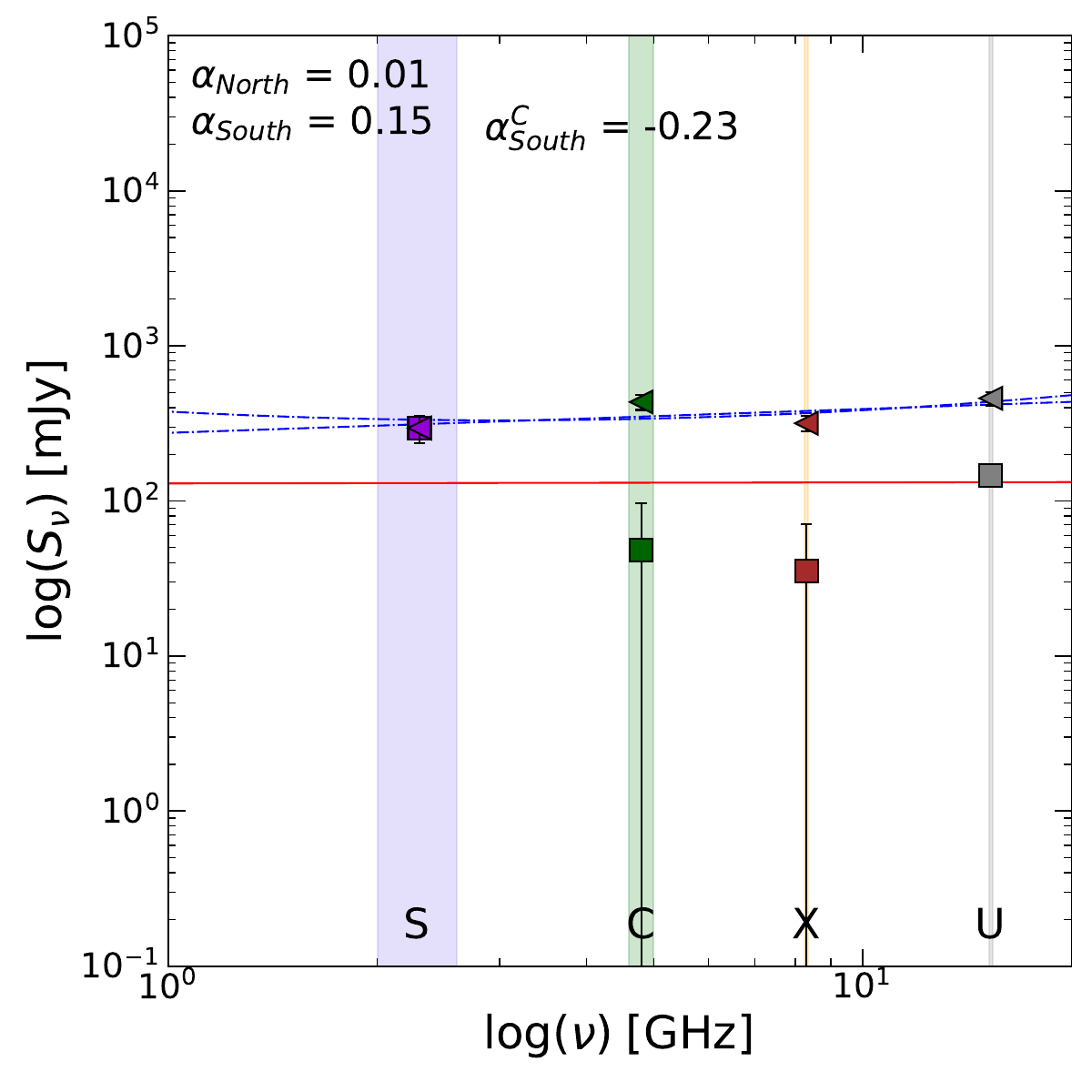}
    \includegraphics[width=8cm]{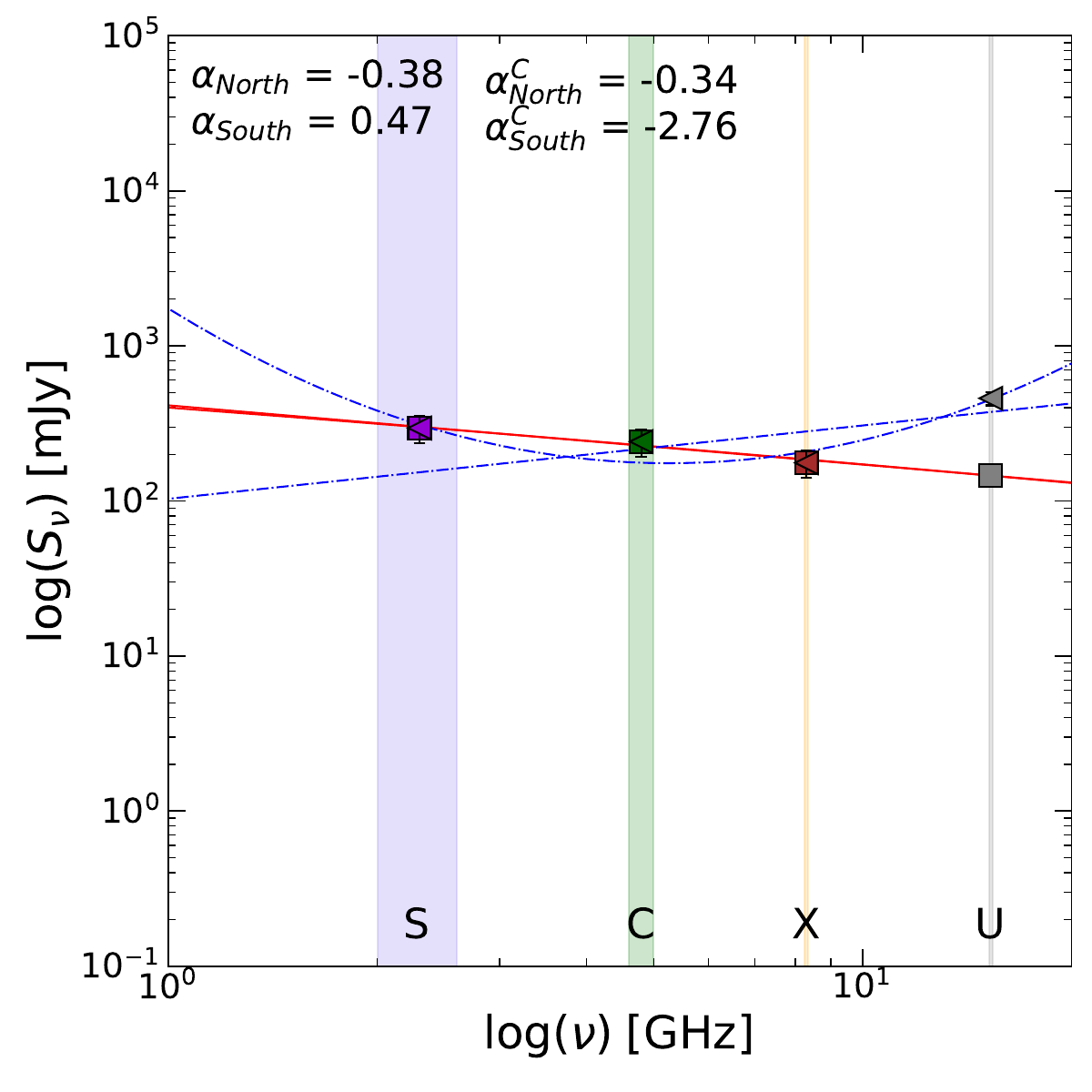}\\
        \vspace*{-3mm}
          \caption{Radio spectra for J1305-1033 showing the multi-frequency VLBA data. Due to the limited S, C, and X-band resolution, two configurations have been shown. In both configurations, the U-band flux densities remain constant for both components. At S-band, they have been varied: 50\% North/50\% South (Left) and 50\% North/50\% South (Right) . At C-band, they have been varied: 10\% North/90\% South (Left) and 50\% North/50\% South (Right). At X-band, they have been varied: 10\% North/90\% South (Left) and 50\% North/50\% South (Right). For plot description, see Figure \ref{fig:0216spec}.}
   \label{fig:1305spec}
\end{figure*}

\begin{figure*}[ht!]
    \centering
    \includegraphics[width=8cm]{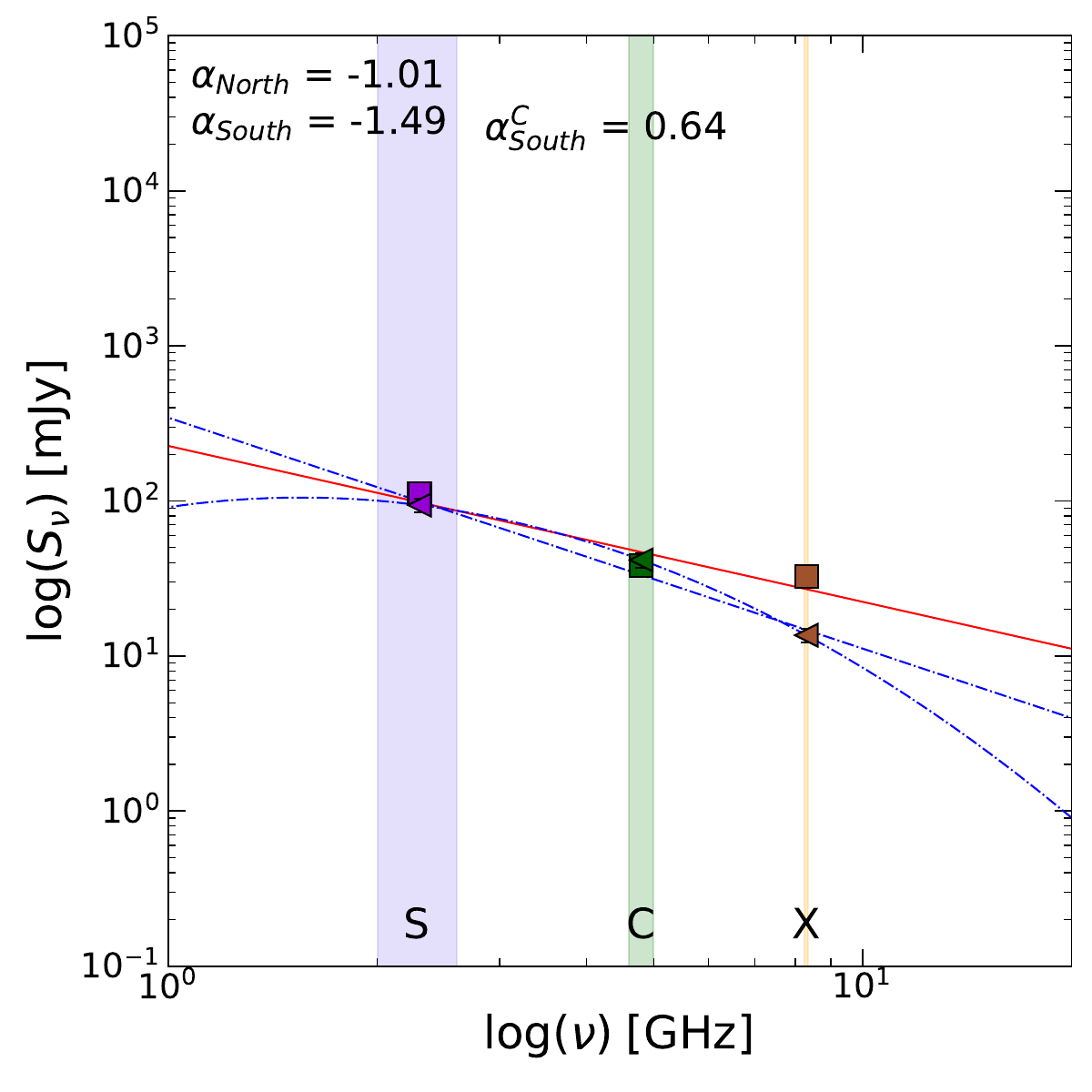}
        \vspace*{-3mm}
          \caption{Radio spectra for J1414+4554 showing the multi-frequency VLBA data. For plot description, see Figure \ref{fig:0216spec}.}
   \label{fig:1414spec}
\end{figure*}

\begin{figure*}[ht!]
    \centering
    \includegraphics[width=8cm]{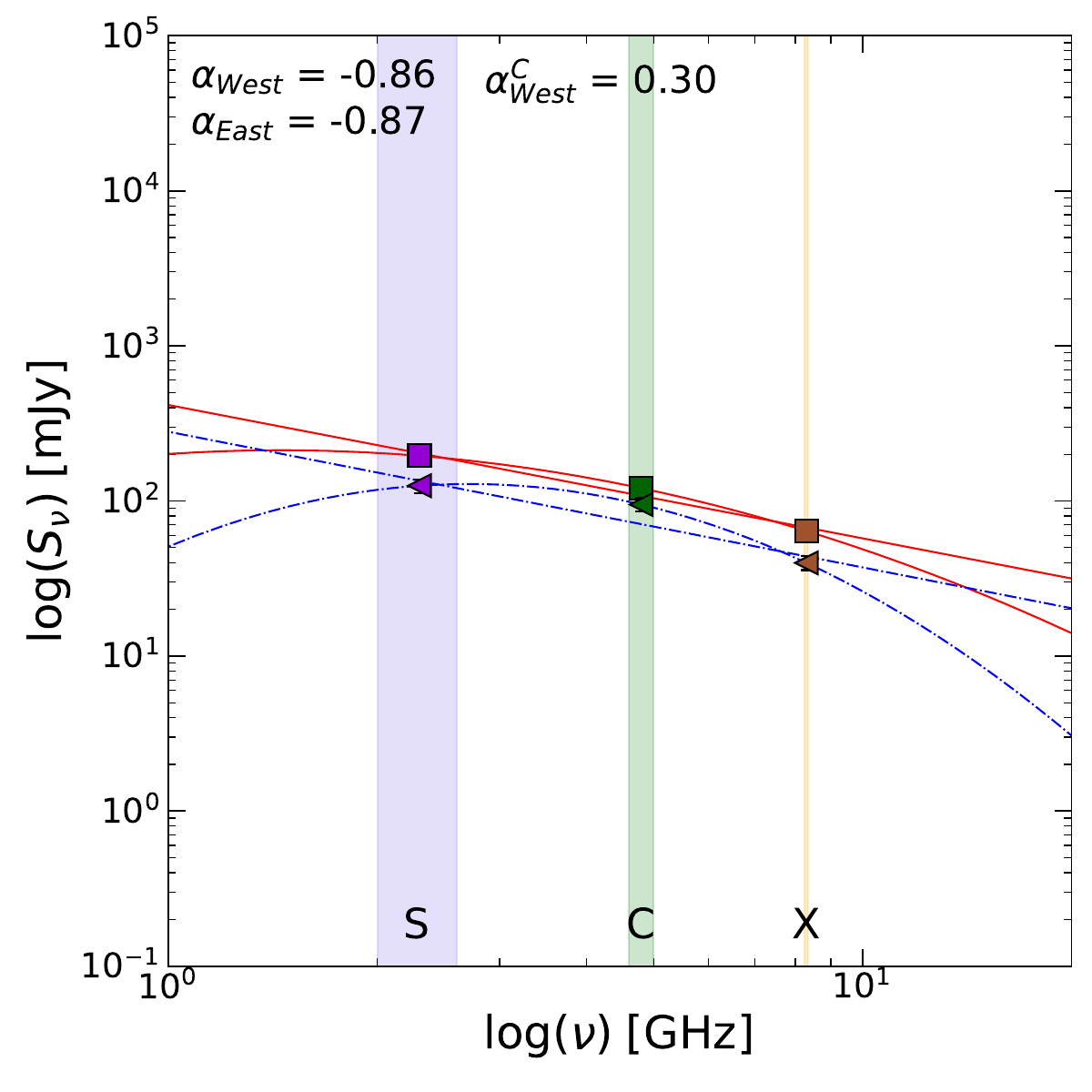}
        \vspace*{-3mm}
          \caption{Radio spectra for J1451+1343 showing the multi-frequency VLBA data. For plot description, see Figure \ref{fig:0216spec}.}
   \label{fig:1451spec}
\end{figure*}

\begin{figure*}[ht!]
    \centering
    \includegraphics[width=8cm]{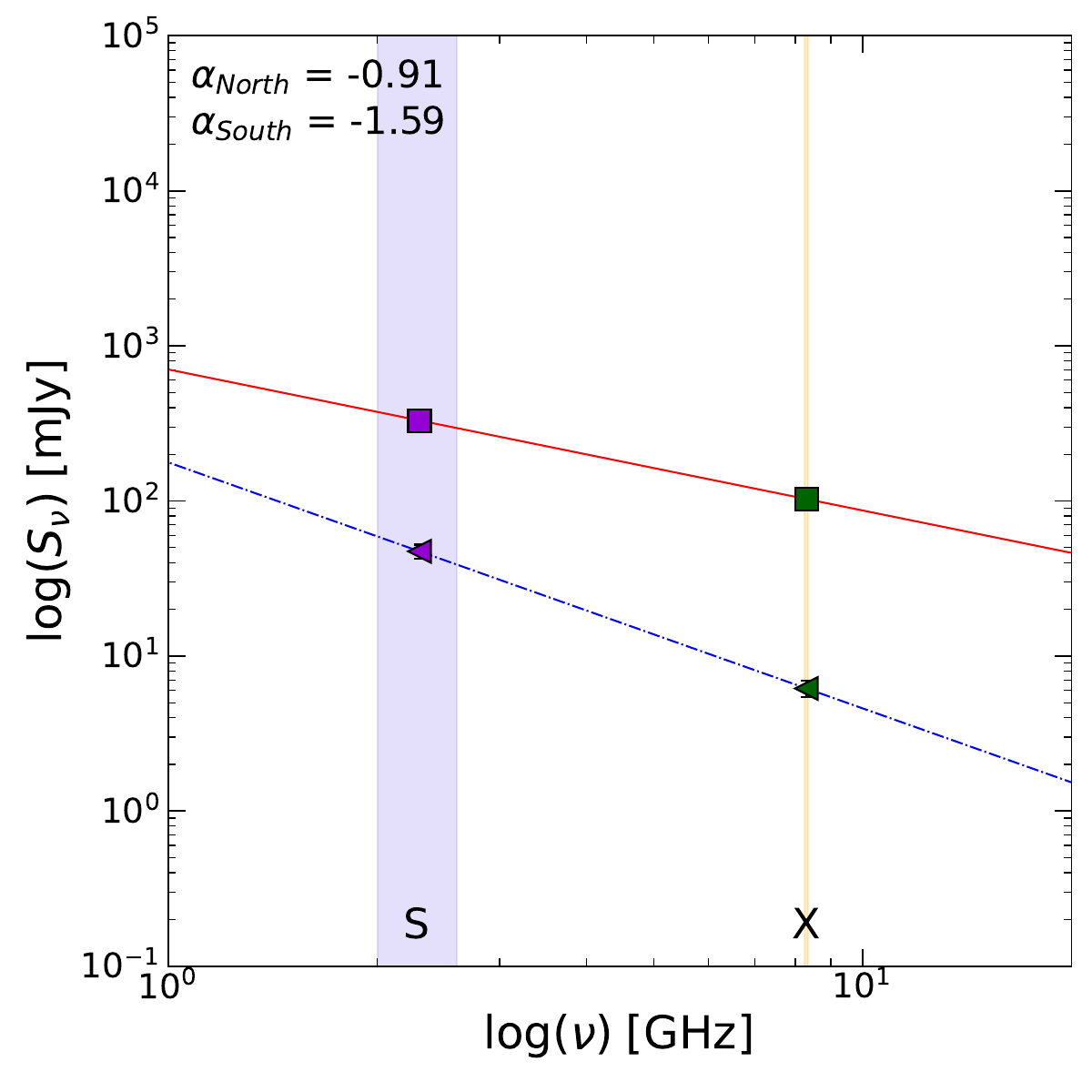}
        \vspace*{-3mm}
          \caption{Radio spectra for J1503+0917 showing the multi-frequency VLBA data. For plot description, see Figure \ref{fig:0216spec}.}
   \label{fig:1503spec}
\end{figure*}

\begin{figure*}[ht!]
    \centering
    \includegraphics[width=8cm]{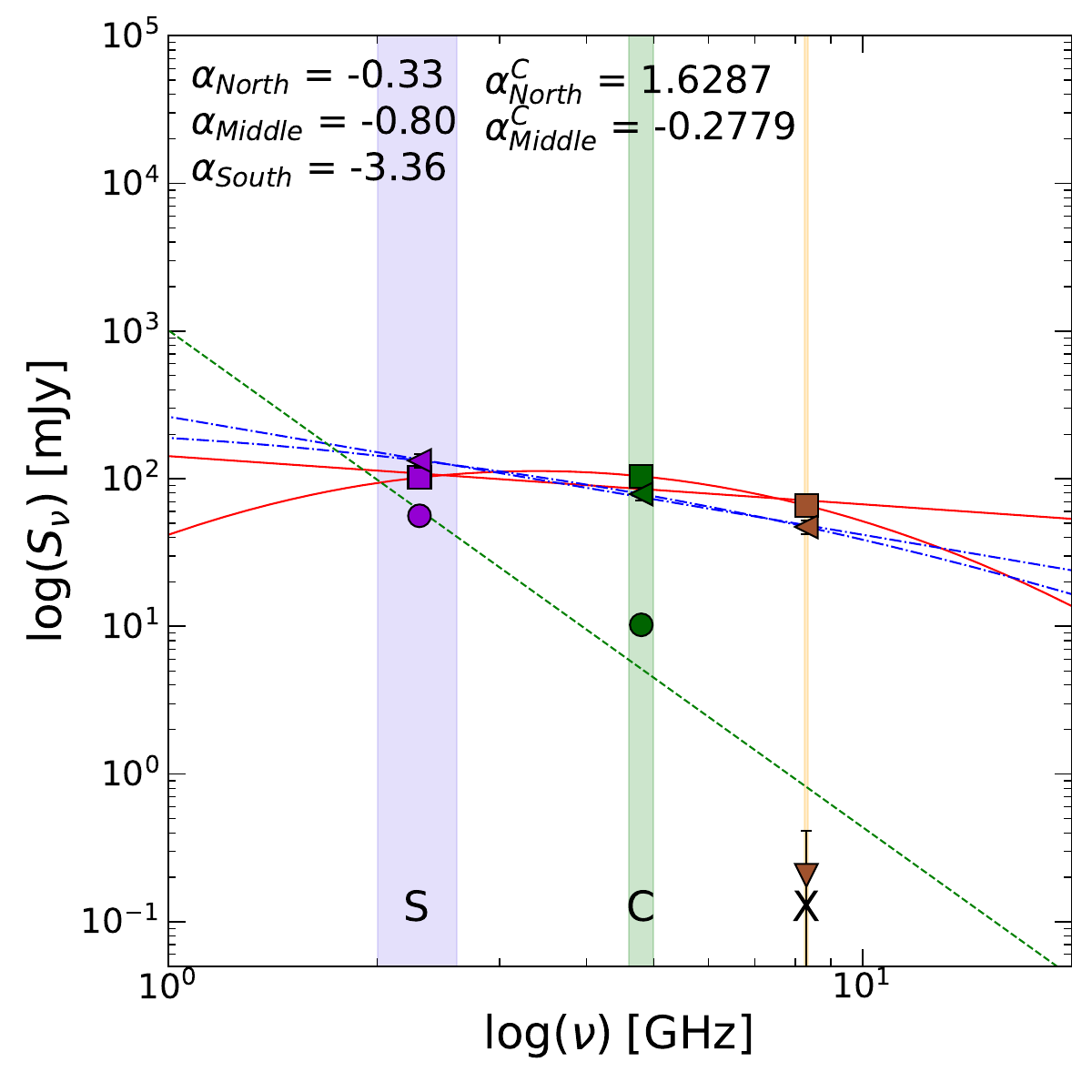}
        \vspace*{-3mm}
          \caption{Radio spectra for J2346+3011 showing the multi-frequency VLBA data. For plot description, see Figure \ref{fig:0216spec}.}
   \label{fig:2346spec}
\end{figure*}

\begin{figure*}[ht!]
    \centering
    \includegraphics[width=8cm]{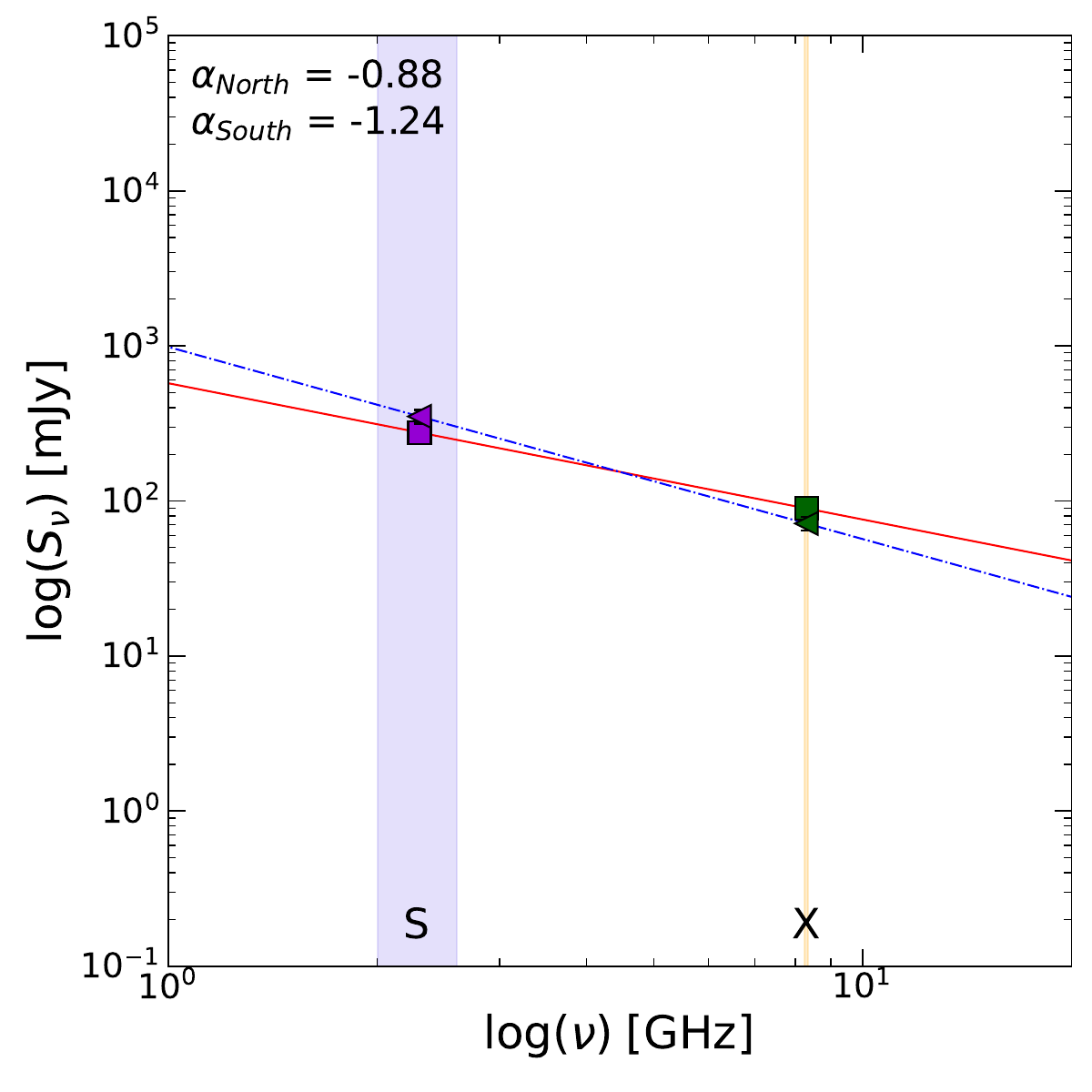}
        \vspace*{-3mm}
          \caption{Radio spectra for J2347-1856 showing the multi-frequency VLBA data. For plot description, see Figure \ref{fig:0216spec}.}
   \label{fig:2347spec}
\end{figure*}

\clearpage

\section{Appendix C: Optical Imaging} \label{sec:decals}

\begin{figure*}
\centering
\begin{minipage}{0.25\textwidth}
    \includegraphics[width=\linewidth]{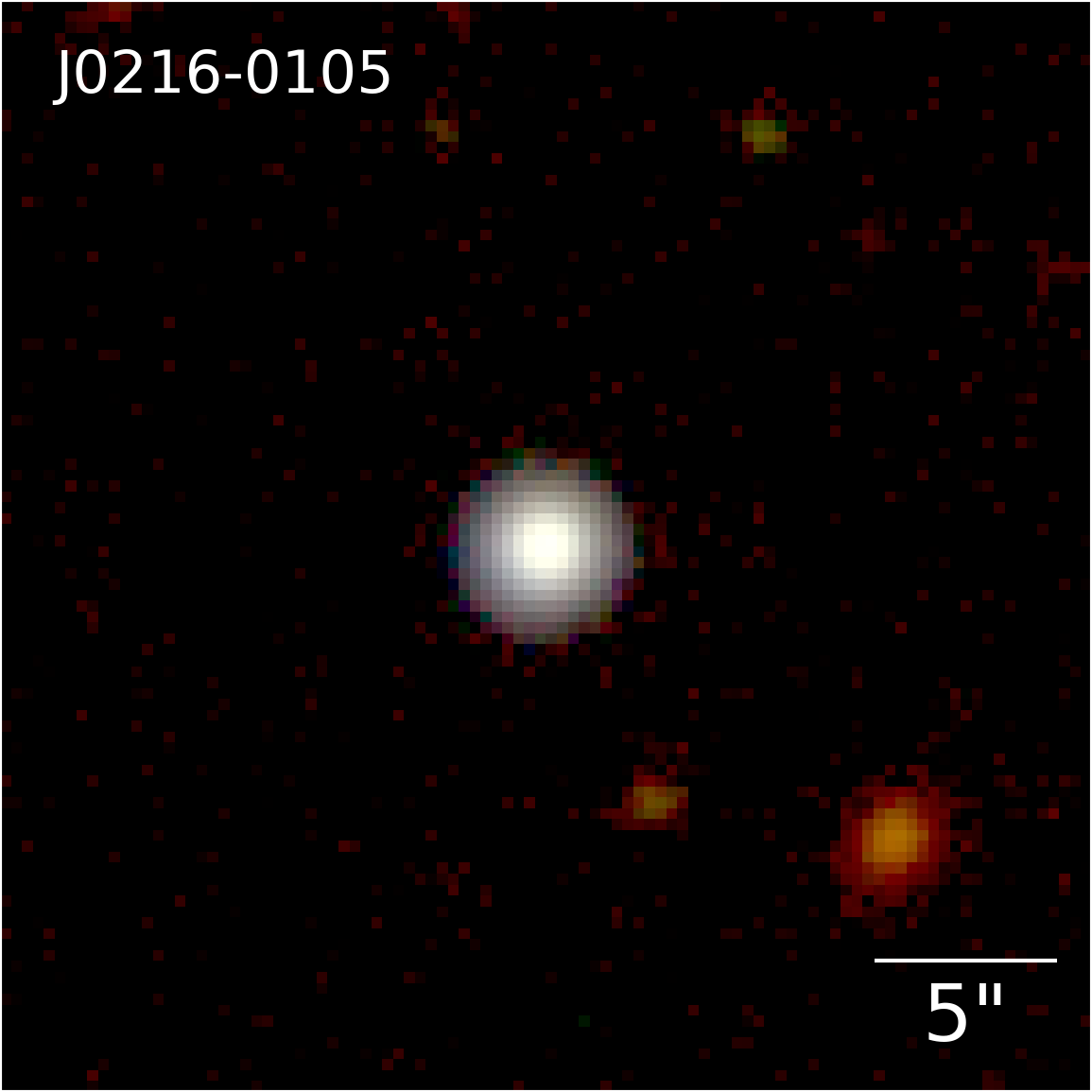}
\end{minipage}
\begin{minipage}{0.25\textwidth}
    \includegraphics[width=\linewidth]{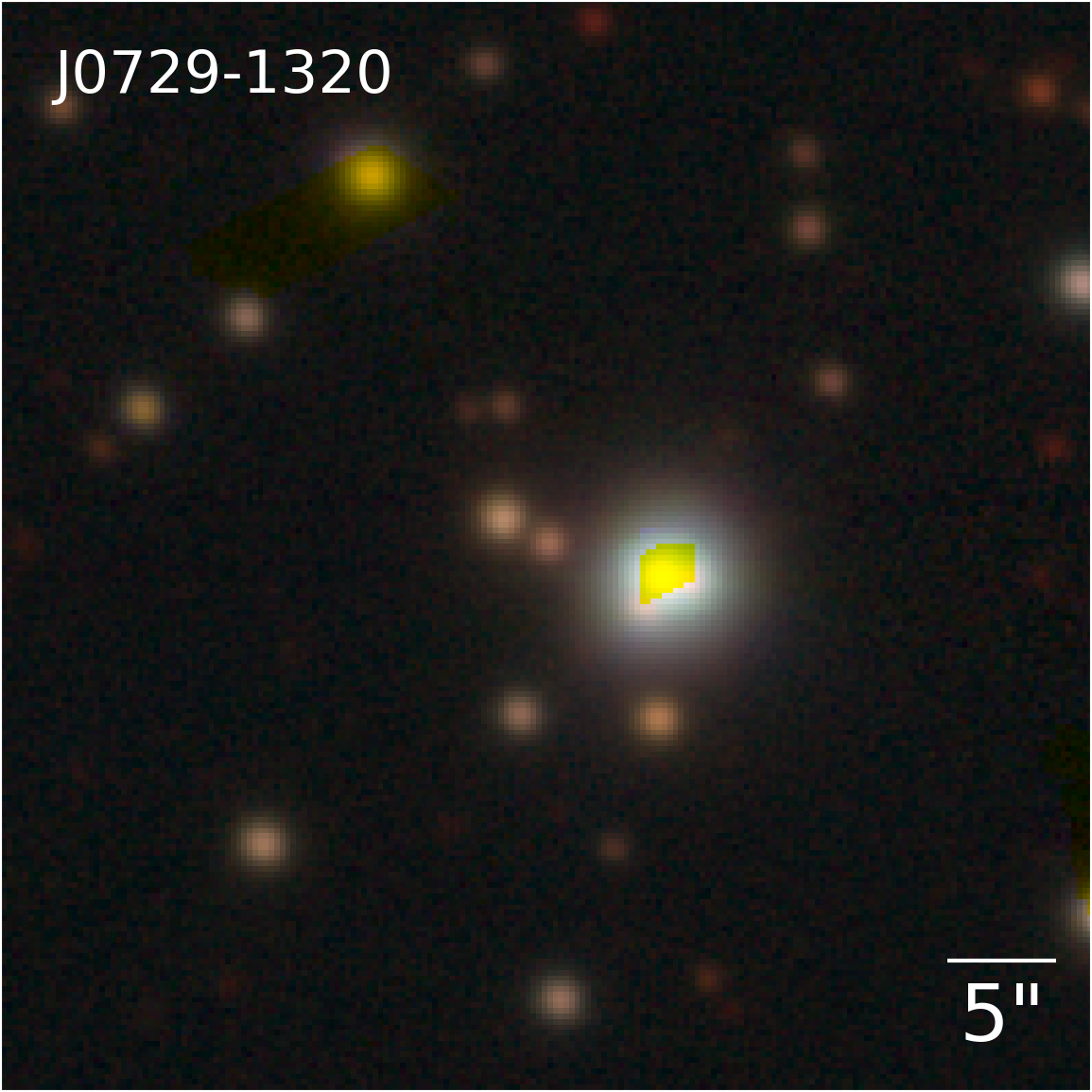}
\end{minipage}
\begin{minipage}{0.25\textwidth}
    \includegraphics[width=\linewidth]{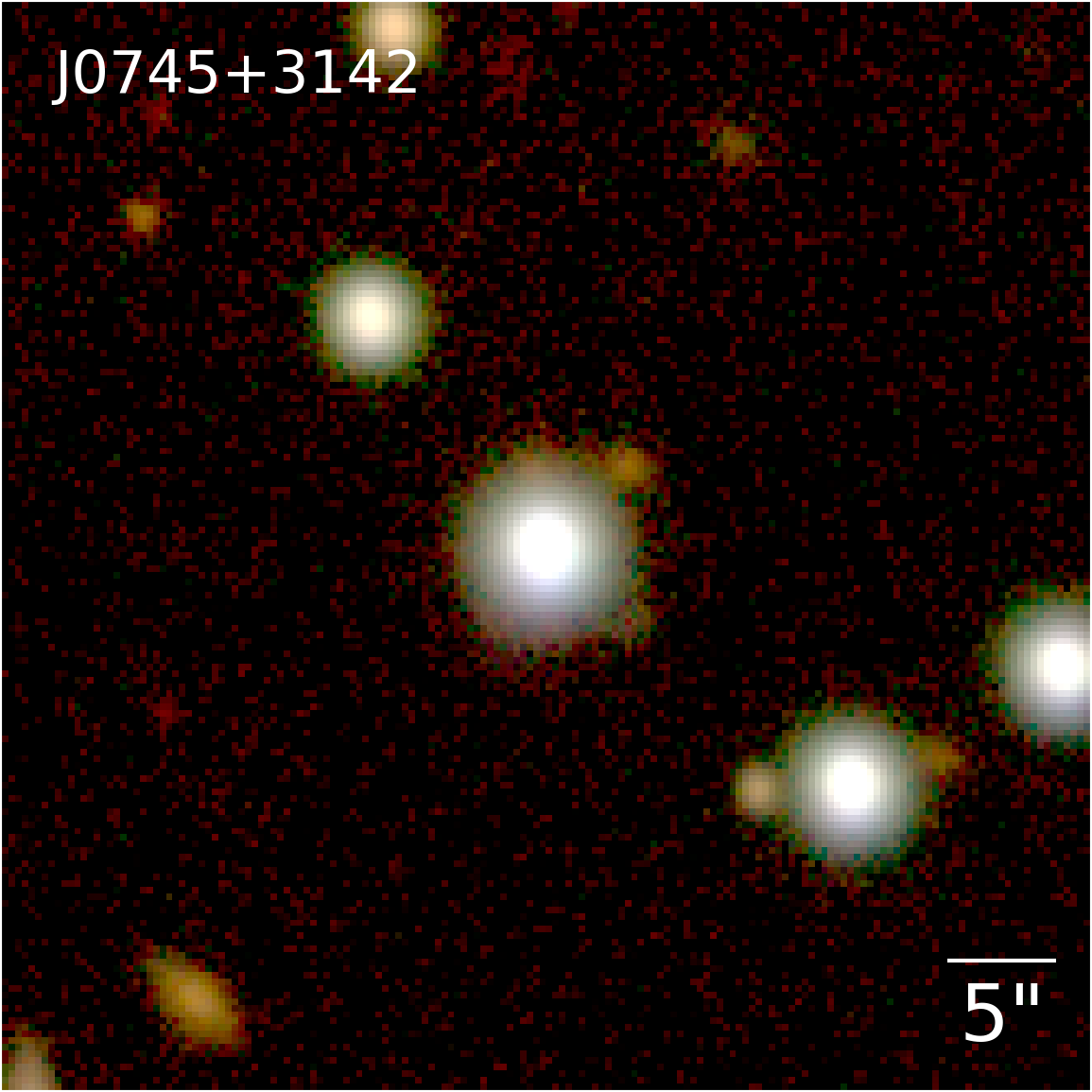}
\end{minipage}

\begin{minipage}{0.25\textwidth}
    \includegraphics[width=\linewidth]{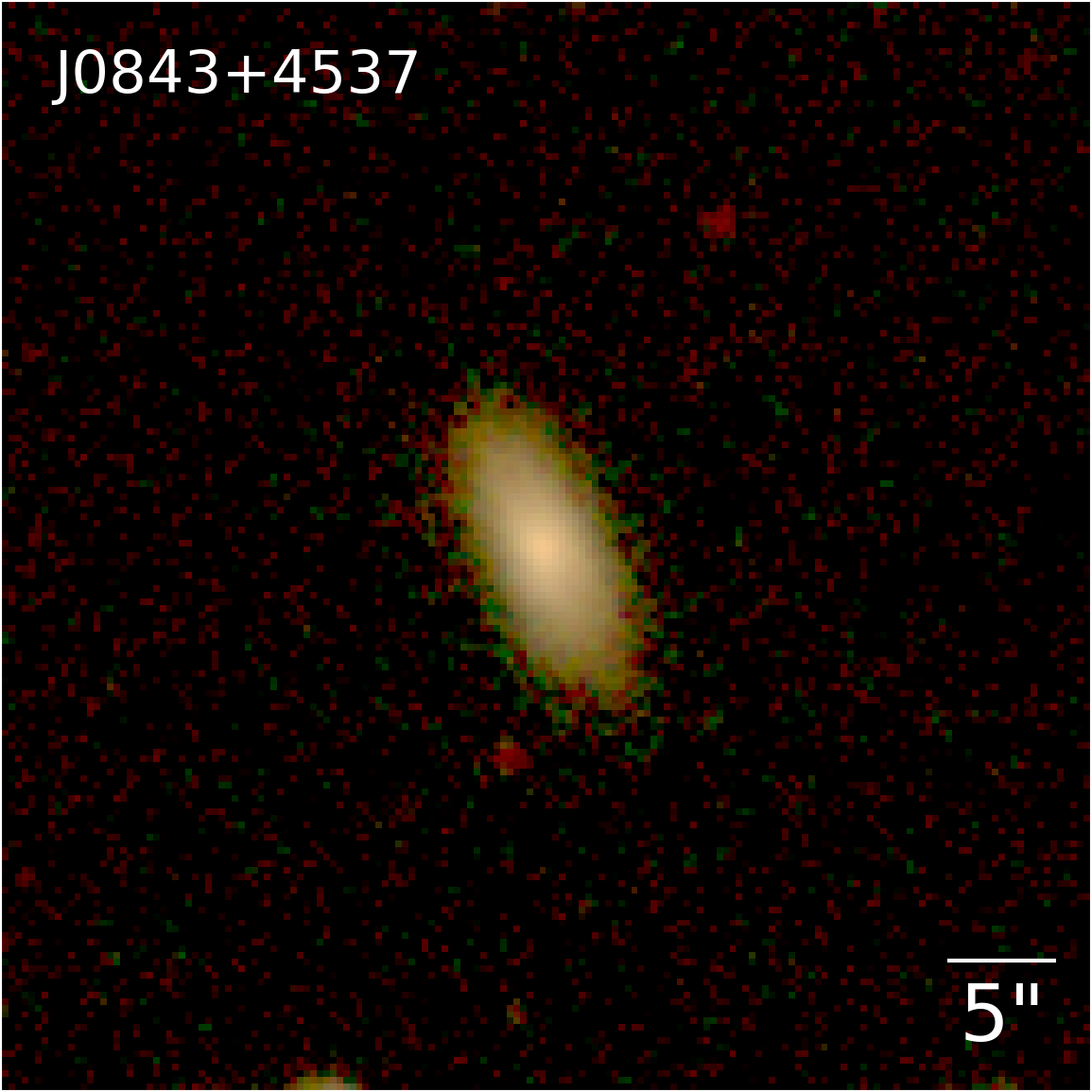}
\end{minipage}
\begin{minipage}{0.25\textwidth}
    \includegraphics[width=\linewidth]{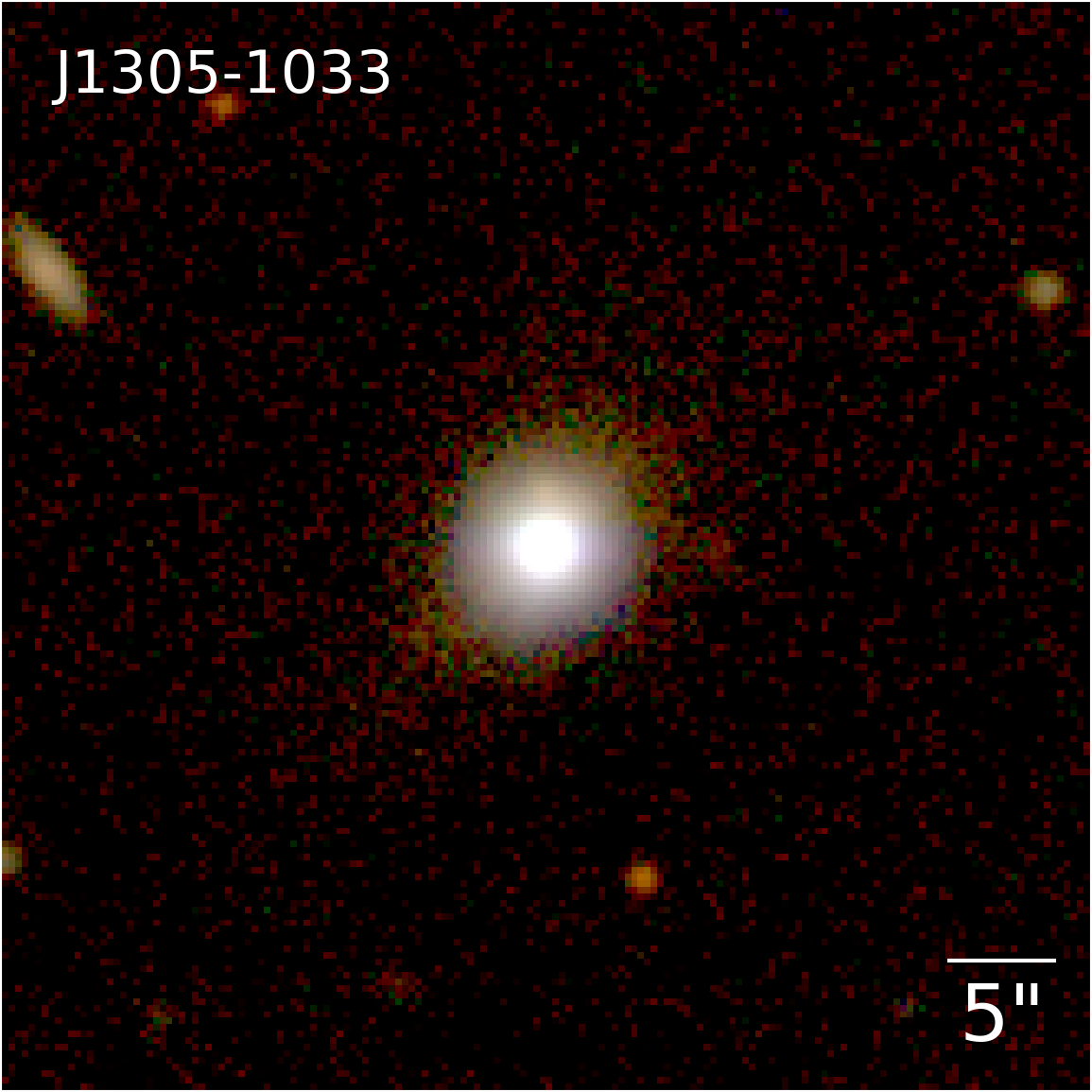}
\end{minipage}
\begin{minipage}{0.25\textwidth}
    \includegraphics[width=\linewidth]{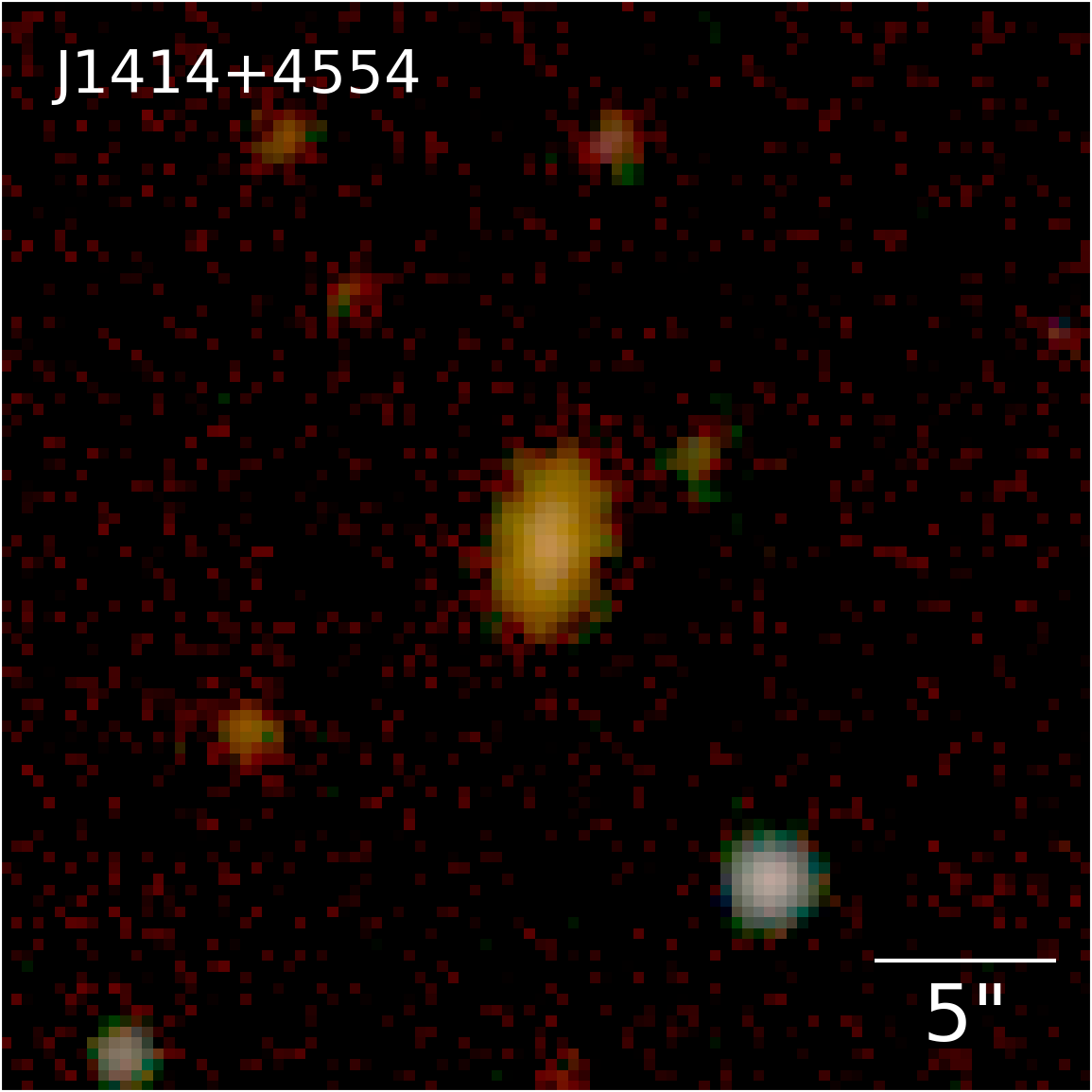}
\end{minipage}

\begin{minipage}{0.25\textwidth}
    \includegraphics[width=\linewidth]{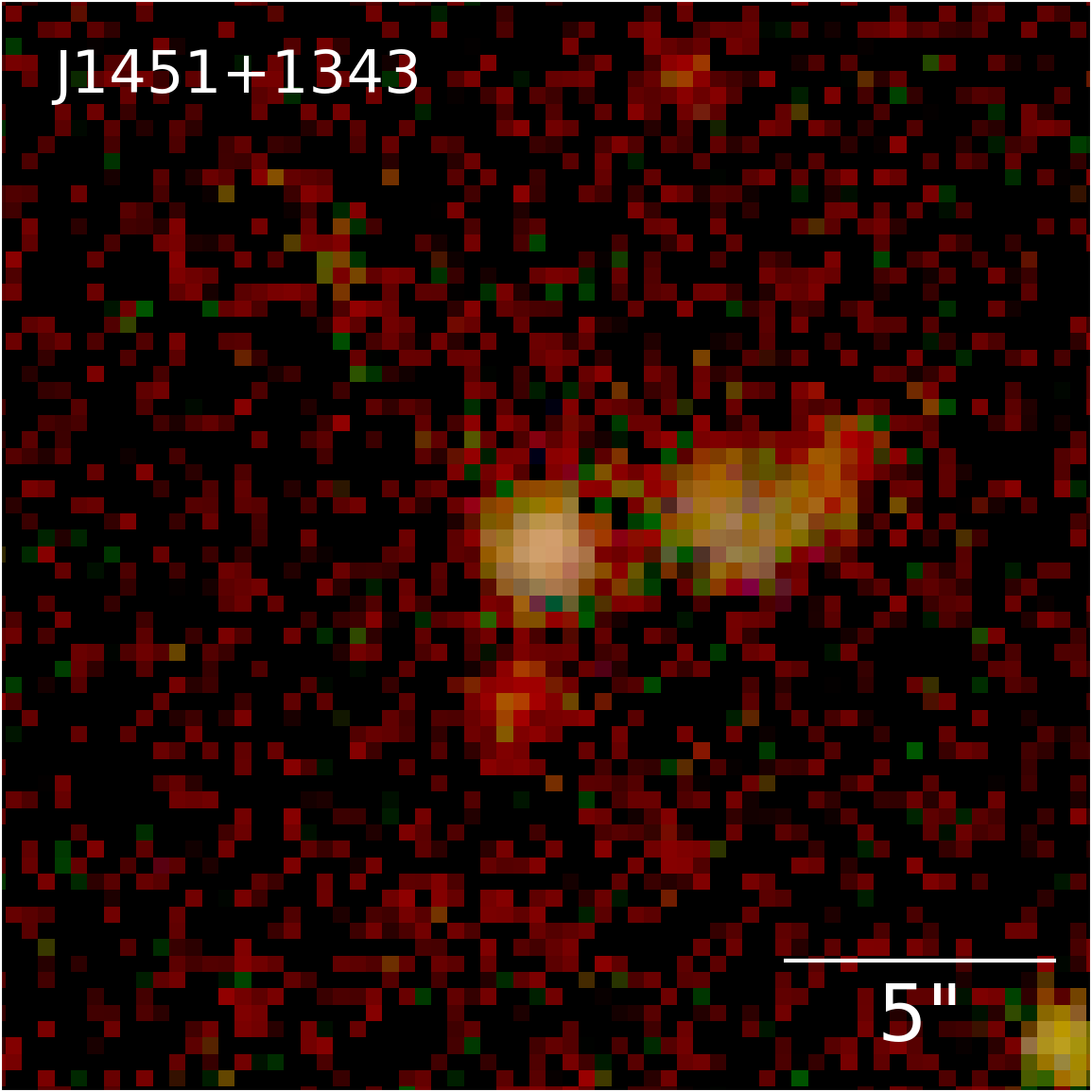}
\end{minipage}
\begin{minipage}{0.25\textwidth}
    \includegraphics[width=\linewidth]{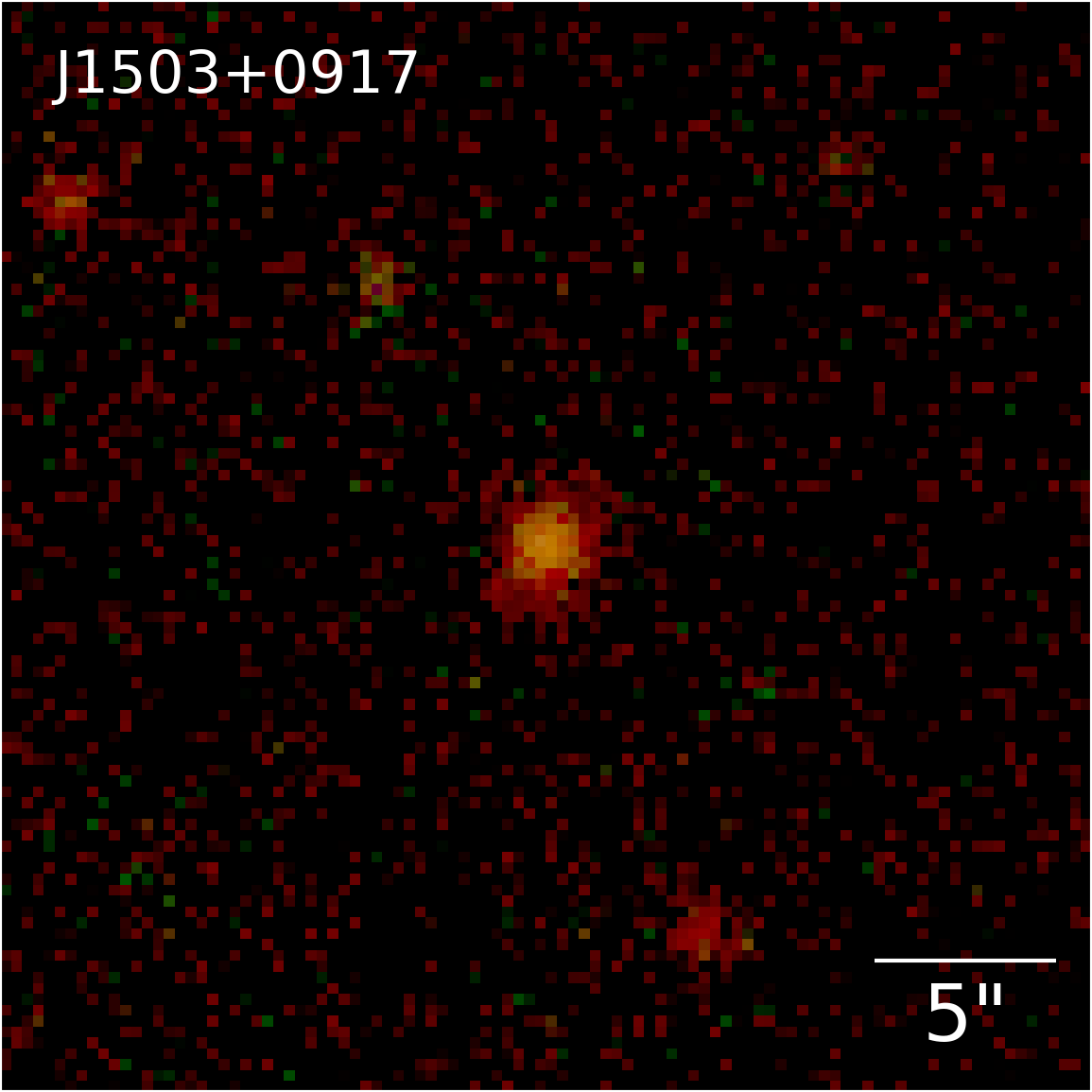}
\end{minipage}
\begin{minipage}{0.25\textwidth}
    \includegraphics[width=\linewidth]{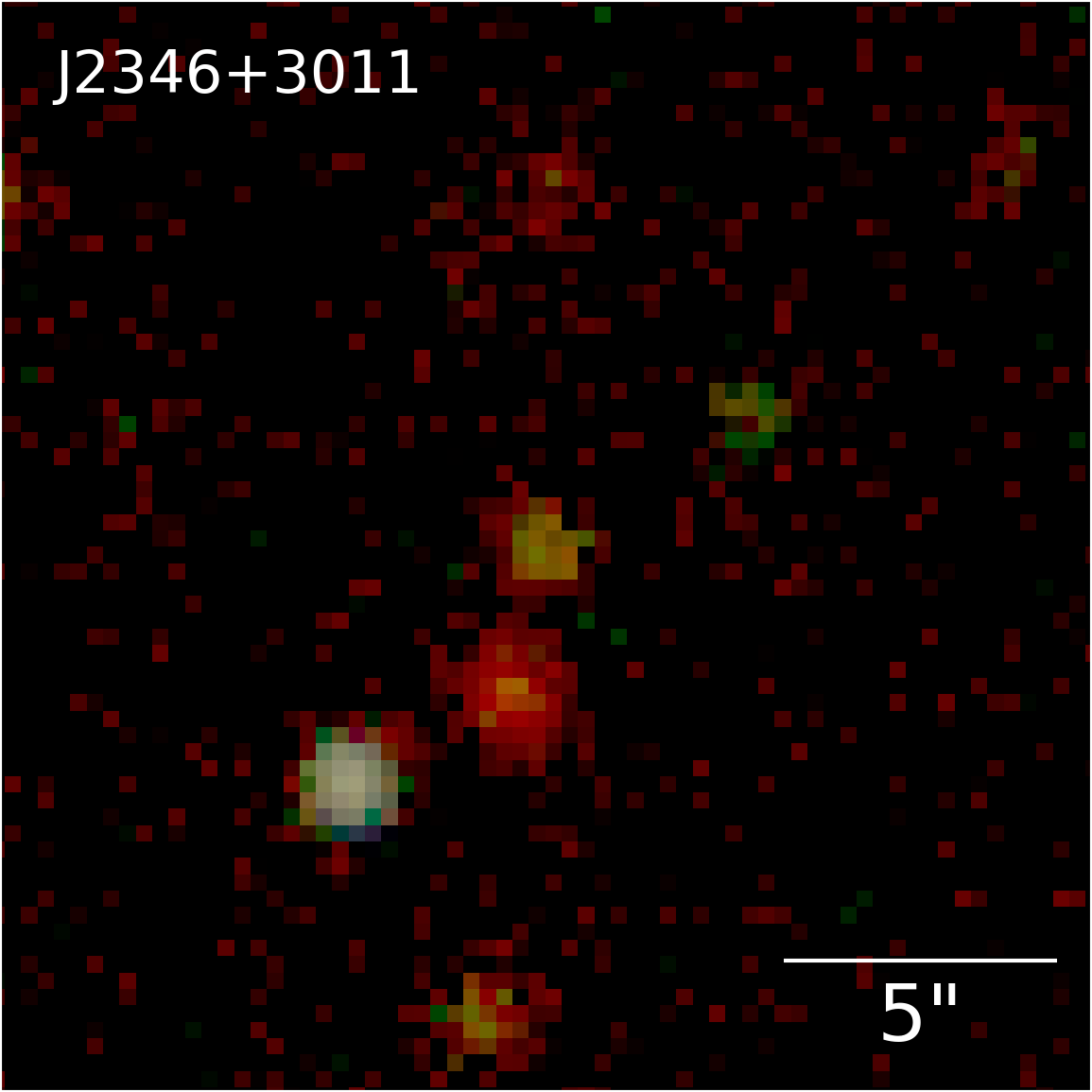}
\end{minipage}

\begin{minipage}{0.25\textwidth}
    \includegraphics[width=\linewidth]{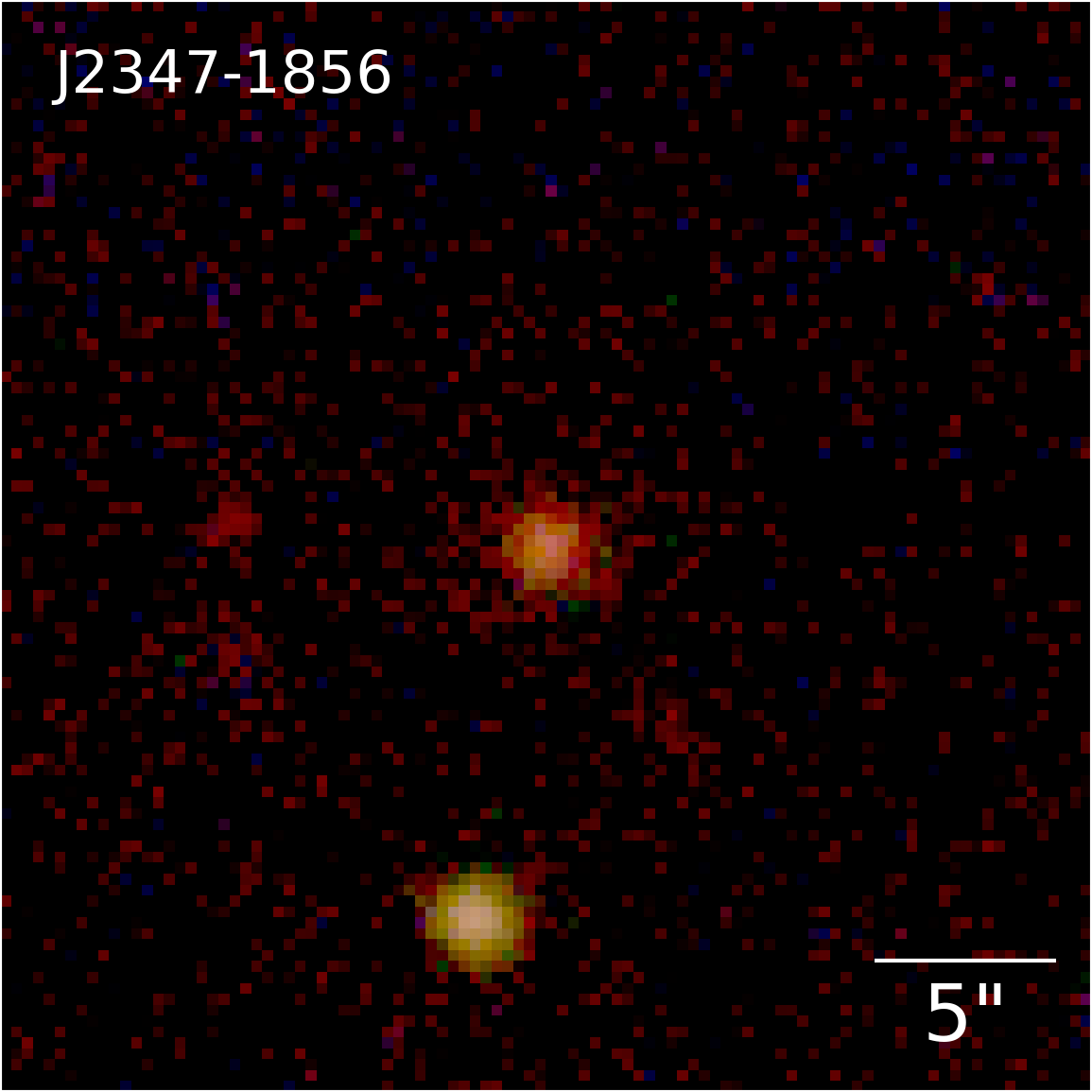}
\end{minipage}

\caption{Optical tricolor ugz images drawn from the DESI LS Legacy Viewer for the full sample. The tricolor ugz image for J0729-1320 uses the PanSTARRS observations, as it is outside the DESI LS sky coverage. The scale bar in the bottom right corner indicates 5''. Notice in several cases that the sources appear elongated rather than pointlike, which may potentially point to underlying complex host morphologies and/or a multiplicity of sources, neither of which are resolvable in these images.}
\label{fig:decals}
\end{figure*}

\end{document}